\documentclass[a4paper,clock]{article}
\usepackage[dvips]{epsfig}
\usepackage{amssymb}
\usepackage{bm}
\usepackage{dcolumn}

\catcode`\@=11
\def\marginnote#1{}
\newcount\hour
\newcount\minute
\newtoks\amorpm
\hour=\time\divide\hour by60
\minute=\time{\multiply\hour by60 \global\advance\minute
by-\hour}\edef\standardtime{{\ifnum\hour<12
\global\amorpm={am}%
        \else\global\amorpm={pm}\advance\hour by-12 \fi
        \ifnum\hour=0 \hour=12 \fi
        \number\hour:\ifnum\minute<10
0\fi\number\minute\the\amorpm}}
\edef\militarytime{\number\hour:\ifnum\minute<10
0\fi\number\minute}

\def\draftlabel#1{{\@bsphack\if@filesw {\let\thepage\relax
   \xdef\@gtempa{\write\@auxout{\string
      \newlabel{#1}{{\@currentlabel}{\thepage}}}}}\@gtempa
   \if@nobreak \ifvmode\nobreak\fi\fi\fi\@esphack}
        \gdef\@eqnlabel{#1}}
\def\@eqnlabel{}
\def\@vacuum{}
\def\draftmarginnote#1{\marginpar{\raggedright\scriptsize\tt#1}}
\def\draft{\oddsidemargin -.5truein
        \def\@oddfoot{\sl preliminary draft \hfil
        \rm\thepage\hfil\sl\today\quad\militarytime}
        \let\@evenfoot\@oddfoot \overfullrule 3pt
        \let\label=\draftlabel
        \let\marginnote=\draftmarginnote

\def\@eqnnum{(\theequation)\rlap{\kern\marginparsep\tt\@eqnlabel}%
\global\let\@eqnlabel\@vacuum}  }

\def\numberbysection{\@addtoreset{equation}{section}
        \def\theequation{\thesection.\arabic{equation}}}

\def\underline#1{\relax\ifmmode\@@underline#1\else
 $\@@underline{\hbox{#1}}$\relax\fi}

\catcode`@=12
\relax

\numberbysection

\advance \topmargin by -\headheight
\advance \topmargin by -\headsep

\evensidemargin \oddsidemargin
\def\nonu{\nonumber}
\def\br{\begin{eqnarray}}
\def\er{\end{eqnarray}}
\def\be{\begin{equation}}
\def\ee{\end{equation}}

\def\({\left(}
\def\){\right)}

\relax

\def\a{\alpha}

\def\b{\beta}

\def\g{\gamma}

\def\l{\lambda}

\def\pa{\partial}

\def\tp0{\Theta_{+}^{(0)}}
\def\tm0{\Theta_{-}^{(0)}}

\def\vp{\varphi}

\def\f#1#2#3 {f^{#1#2}_{#3}}

\def\win1{{\sf w_{1+\infty}}}

\def\Win1{{\sf W_{1+\infty}}}

\def\rlx{\relax\leavevmode}
\def\inbar{\vrule height1.5ex width.4pt depth0pt}
\def\IZ{\rlx\hbox{\sf Z\kern-.4em Z}}
\def\IR{\rlx\hbox{\rm I\kern-.18em R}}
\def\IC{\rlx\hbox{\,$\inbar\kern-.3em{\rm C}$}}
\def\IN{\rlx\hbox{\rm I\kern-.18em N}}
\def\IO{\rlx\hbox{\,$\inbar\kern-.3em{\rm O}$}}
\def\IP{\rlx\hbox{\rm I\kern-.18em P}}
\def\IQ{\rlx\hbox{\,$\inbar\kern-.3em{\rm Q}$}}
\def\IF{\rlx\hbox{\rm I\kern-.18em F}}
\def\IG{\rlx\hbox{\,$\inbar\kern-.3em{\rm G}$}}
\def\IH{\rlx\hbox{\rm I\kern-.18em H}}
\def\II{\rlx\hbox{\rm I\kern-.18em I}}
\def\IK{\rlx\hbox{\rm I\kern-.18em K}}
\def\IL{\rlx\hbox{\rm I\kern-.18em L}}
\def\one{\hbox{{1}\kern-.25em\hbox{l}}}
\def\0#1{\relax\ifmmode\mathaccent"7017{#1}%
B        \else\accent23#1\relax\fi}

\def\RQE#1#2#3{{\sl Radiophysics and Quantum Electronics} {\bf #1} (#2) #3}
\def\PRL#1#2#3{{\sl Phys. Rev. Lett.} {\bf#1} (#2) #3}
\def\NPB#1#2#3{{\sl Nucl. Phys.} {\bf B#1} (#2) #3}

\def\PRE#1#2#3{{\sl Phys. Rev.} {\bf E#1} (#2) #3}
\def\PRA#1#2#3{{\sl Phys. Rev.} {\bf A#1} (#2) #3}

\def\PLA#1#2#3{{\sl Phys. Lett.} {\bf #1A} (#2) #3}

\def\PLB#1#2#3{{\sl Phys. Lett.} {\bf #1B} (#2) #3}
\def\JMP#1#2#3{{\sl J. Math. Phys.} {\bf #1} (#2) #3}
\def\PTP#1#2#3{{\sl Prog. Theor. Phys.} {\bf #1} (#2) #3}

\def\PR#1#2#3{{\sl Phys. Reports} {\bf #1} (#2) #3}

\def\IJMPA#1#2#3{{\sl Int. J. Mod. Phys.} {\bf A#1} (#2) #3}

\def\JPA#1#2#3{{\sl J. Physics} {\bf A#1} (#2) #3}
\def\JPAMT#1#2#3{{\sl J. Physics A: Math. Theor.} {\bf A#1} (#2) #3}

\def\JHEP#1#2#3{{\sl JHEP} {\bf #1} (#2) #3}

\def\JCP#1#2#3{{\sl Journal of Computational Physics} {\bf #1} (#2) #3}
\def\SAM#1#2#3{{\sl Stud. Appl. Math.} {\bf #1} (#2) #3}
\def\MAA#1#2#3{{\sl Methods and  Applications of Analysis} {\bf #1} (#2) #3}
\def\OL#1#2#3{{\sl Optics Letters} {\bf #1} (#2) #3}
\def\Nonl#1#2#3{{\sl Nonlinearity} {\bf #1} (#2) #3}
\def\CJP#1#2#3{{\sl Canadian Journal of Physics} {\bf #1} (#2) #3}
\def\OC#1#2#3{{\sl Optics Communications} {\bf #1} (#2) #3}
\def\JPAMG#1#2#3{{\sl J. Physics A: Math. Gen.} {\bf A#1} (#2) #3}
\def\JGP#1#2#3{{\sl Journal of Geometry and Physics} {\bf #1} (#2) #3}
\def\EPL#1#2#3{{\sl Europhysics Letters} {\bf #1} (#2) #3}
\def\ScR#1#2#3{{\sl Sci. Rep.} {\bf #1} (#2) #3}

\begin{document}

\begin{titlepage}

\vspace{.2in}
\begin{center}
{\large\bf Modified non-linear Schr\"odinger models, ${\cal C}{\cal P}_s{\cal T}_d$ invariant $N-$bright solitons and infinite towers of anomalous charges}\cal 
\end{center}

\vspace{.2in}

\begin{center}

H. Blas$^{(a)}$,  M. Cerna Magui\~na$^{(b)}$ and  L.F. dos Santos$^{(c)}$

\par \vskip .2in \noindent

$^{(a)}$Instituto de F\'{\i}sica\\
Universidade Federal de Mato Grosso\\
Av. Fernando Correa, $N^{0}$ \, 2367\\
Bairro Boa Esperan\c ca, Cep 78060-900, Cuiab\'a - MT - Brazil. \\
$^{(b)}$ Departamento de Matem\'atica\\
Universidad Nacional Santiago Ant\'unez de Mayolo\\
Campus Shancay\'an, Av. Centenario 200, Huaraz - Per\'u\\
$^{(c)}$ Centro Federado de Educa\c c\~ao Tecnologica-CEFET-RJ\\
Campus Angra dos Reis, Rua do Areal, 522, Angra dos Reis- RJ -Brazil

\normalsize
\end{center}

\vspace{.3in}

\begin{abstract}
\vspace{.3in}

Modifications of the non-linear Schr\"odinger model (MNLS) $ i \partial_{t} \psi(x,t) + \partial^2_{x} \psi(x,t) -  [\frac{\delta V}{\delta |\psi|^2} ] \psi(x,t) =  0,$ where $\psi \in \IC$  and $V: \IR_{+} \rightarrow \IR$, are considered.  We show that the MNLS models possess infinite towers of quasi-conservation laws for soliton-type configurations with a special complex conjugation, shifted parity and delayed time reversion (${\cal C}{\cal P}_s{\cal T}_d$) symmetry. Infinite towers of anomalous charges appear even in the standard NLS model for ${\cal C}{\cal P}_s{\cal T}_d$ invariant $N-$bright solitons. The true conserved charges emerge through some kind of anomaly cancellation mechanism. A dual Riccati-type pseudo-potential formulation is introduced for a modified AKNS system (MAKNS) and infinite towers of novel anomalous conservation laws are uncovered. In addition, infinite towers of exact non-local conservation laws are uncovered in a linear system formulation. Our analytical results are supported by numerical simulations of $2-$bright-soliton scatterings with potential $ V = - \frac{ 2\eta}{2+ \epsilon} ( |\psi|^2 )^{2 + \epsilon}, \epsilon \in \IR, \eta>0$. Our numerical simulations show the elastic scattering of bright solitons for a wide range of values of the set $\{\eta, \epsilon\}$ and a variety of amplitudes and relative velocities. The AKNS-type system is quite ubiquitous, and so, our results  may find potential applications in several areas of non-linear physics, such as Bose-Einstein condensation, superconductivity, soliton turbulence and the triality among gauge theories, integrable models and gravity theories. 
\end{abstract}

\end{titlepage}

\section{Introduction}

Integrable partial differential equations such as sine-Gordon (SG), Korteweg-de Vries (KDV) and nonlinear Schr\"odinger (NLS) systems are regarded as universal models of nonlinear phenomena. They can describe, at leading order, several nonlinear systems and can be integrated using the inverse scattering transform. Soliton type solutions and the infinite number of conserved charges are among the distinguishing features of the integrable models \cite{das, faddeev, abdalla}. However, some non-linear field theory models with important physical applications and solitary wave solutions are not integrable. Recently, some deformations of integrable models, which exhibit soliton-type properties, have been put forward. Various quasi-integrability properties of the deformations of the integrable models, such as SG, NLS, Toda, KdV, Boullogh-Dodd and SUSY-SG have recently been examined in the frameworks of the anomalous zero-curvature formulations \cite{jhep1, jhep2, jhep6, jhep3, toda, cnsns, npb, jhep4, jhep5, epl} and the deformations of the Riccati-type pseudo-potential approach \cite{npb1, jhep33}. 
 
The dynamics of the soliton-like configurations in the quasi-integrable models are, so far, largely unknown. We summarize the known results. First, the one-soliton sectors exhibit infinite number of anomalous charges, since the space-time integration of the so-called anomalies vanish. Second, the space-time integration of the anomalies also vanish for configurations such that the one-soliton like solutions are located far away from each other. The anomalies are appreciable around the space-time regions of their mutual interaction. Third, a sufficient condition for the vanishing of the space-time integrated anomaly is that the $N-$soliton configurations possess definite parity, either odd or even, under a special shifted parity and delayed time reversion (${\cal P}_s{\cal T}_d$) symmetry. When the anomaly densities are odd under this symmetry the space-time integration of them vanish, which imply the existence of asymptotically (anomalous) conserved charges. Fourth, the existence of several towers of infinitely many anomalous charges. Several new towers of anomalous charges have been uncovered in \cite{npb1, jhep33}, in this way adding a new list to the ones presented in \cite{jhep1, cnsns, jhep4, jhep5} for deformations of SG and KdV models. Fifth, some deformed models possess a subset of infinite number of anomalous charges for soliton eigenstates simply of the shifted space-reflection symmetry ${\cal P}_s$. The deformed defocusing (focusing) NLS model for a variety of two-soliton configurations \cite{jhep4, jhep5} and the deformed sine-Gordon model \cite{cnsns} for two-kink-type and breather-type solutions have been shown to exhibit this property. The above results have been obtained by combining analytical and numerical methods.   

Remarkably, it has been observed that even the standard KdV model possesses several towers of quasi-conservation laws with anomalous charges for analytical $N-$ soliton configurations satisfying the special  ${\cal P}_s{\cal T}_d$ symmetry properties \cite{jhep33}.  For the standard SG theory this property has also been discussed for the 2-soliton sector of the theory \cite{npb1}.  So, one can argue that the truly  integrable systems inherit certain properties to their deformed counterparts. 

Moreover, in the context of the Riccati-type pseudo-potential approach to quasi-integrability, there have been shown that the deformed SG and KdV models \cite{npb1, jhep33} can be formulated as the compatibility condition of certain linear systems of equations and that they possess infinite towers of exact non-local conservation laws.   

The new properties mentioned above have been examined for the deformations of the relativistic invariant sine-Gordon model with topological solitons and the non-relativistic KdV model with non-topological and unidirectional solitons, respectively. Conventionally, both of  them are defined for real scalar fields. So, it would be desirable to examine those properties for NLS-type models defined for a complex field with envelope solitons.  In general, those integrable models can be formulated in the framework of  the AKNS system, from which they can be obtained through relevant reduction processes. The NLS-type model stands in the same level of importance as the KdV-type and SG-type  models in their potential applications, since they are ubiquitous in all areas of nonlinear physics, such as  Bose-Einsten condensation and superconductivity \cite{frantzeskakis, tanaka1}, soliton gas and soliton turbulence in fluid dynamics \cite{gauss, pla1, turbu, prlgas}, the Alice-Bob physics \cite{alice}, and the understanding of a kind of triality among the gauge theories, integrable models and gravity theories (see \cite{nian} and references therein).    

In the first part of this paper we search for additional quasi-conservation laws, different from  the ones related to the anomalous zero-curvature approach in \cite{jhep3}, and study the role played by them in the phenomenon of quasi-integrability. A special complex conjugation, shifted parity and delayed time reversion (${\cal C}{\cal P}_s{\cal T}_d$) symmetry will play a central role in our constructions. Using a direct constructive method, starting from the equation of motion, it is shown that for each monomial or polynomial of homogeneous degree and even parity under ${\cal P}_s{\cal T}_d$ and formed by the NLS field and its derivatives one can construct an infinite tower of quasi-conservation laws with anomalous charges with successively increasing degrees and odd parities under ${\cal P}_s{\cal T}_d$. In addition, it is shown that some exactly conserved charges (e.g. the energy) emerge through some kind of anomaly cancellation mechanism. In fact, an exact conservation law arises when a suitable linear combination of relevant quasi-conservation laws leads to the vanishing of the combined anomalies.

As a byproduct we will show that even the standard NLS model exhibits infinite towers of  infinitely many anomalous conservation laws with analogous properties to their counterparts in the quasi-integrable NLS theory. It will be shown analytically the quasi-conservation of the infinite towers of anomalous charges for $N-$soliton solution satisfying the  ${\cal C}{\cal P}_s{\cal T}_d$ symmetry since their relevant anomalies possess odd parities under the ${\cal P}_s{\cal T}_d$ symmetry transformation. The lowest order anomalous charges can be identified to the so-called  statistical characteristics of the NLS model and their ensemble-averaged values  for large number of $N-$solitons have been examined \cite{turbu} and it has been argued to play a fundamental role in the understanding of the phenomena of soliton gas and soliton turbulence in integrable models, see e.g. \cite{pla1, prlgas} and references therein. 

Remarkably, in the sector described by $N-$bright solitons possessing ${\cal C}{\cal P}_s{\cal T}_d$  symmetry invariance and broken space-time translation symmetry the standard NLS model equation of motion can be rewritten as the one belonging to the list studied in \cite{ablowitz, reverse}, i.e. the so-called reverse space-time nonlocal NLS \cite{reverse}. Thus, the usual NLS, in the ${\cal C}{\cal P}_s{\cal T}_d$ symmetric $N-$soliton sector,  belongs to the family of non-local generalization of the NLS model considered in the recent literature. 

We numerically simulate various two-soliton configurations of the deformed model by numerically evolving
linear superpositions of two (initially located far away) solitary-wave exact solutions of a particular deformation of the NLS model.  The numerical simulations will allow us to verify the analytical expectations for the novel set of quasi-conservation laws, and check the  behavior of some of the lowest order anomalies associated to the relevant new towers of anomalous conservation laws of the deformed model for a variety of values of the coupling constant and deformation parameters  $\{\eta, \epsilon\}$.  In order to perform the numerical simulations we will use  the time-splitting cosine pseudo-spectral finite difference method (TSCP) \cite{bao1, bao2}, which is a suitable method in order to control the highly oscillatory background. The collisions would be elastic if the solitons preserved their initial profiles and velocities and there was no noticeable production of radiation. These properties hold for truly integrable models and the fact that they would also hold for the deformed NLS equation, which is not integrable, will be the distinguishing features of  the quasi-integrable deformed NLS model.

In the second part of this paper we perform a particular deformation of the Riccati-type pseudo-potential approach related to the AKNS system \cite{nucci},  from which the NLS-like modified model is obtained through a particular reduction process. A modified  AKNS system (MAKNS) is defined by introducing a deformed potential $V$ and some auxiliary fields into the pair of Riccati-type system of equations and   another system of equations for the set of auxiliary fields, such that the compatibility condition applied to the extended system gives rise to the modified AKNS model equations of motions. Then, it is constructed a set of infinite number of quasi-conservation laws order by order in powers of the spectral parameter. A dual system of the Riccati-type pseudo-potential approach is introduced and novel infinite set of anomalous conservation laws are derived, which encompasses, as a subset, the ones obtained by a direct constructive method as mentioned above. 

In addition, in the framework  of the pseudo-potential approach, it is  proposed a linear system of equations whose compatibility condition gives rise to the MAKNS equations of motion. As an application of the linear system formulation of the modified AKNS model, it is obtained a pair of infinite towers of exact non-local conservation laws.  A particular reduction $MAKNS \rightarrow MNLS$ allows one to reproduce the relevant quantities of the MNLS model out of the ones constructed for  the MAKNS system.   
    
This paper is organized as follows: In the next section we define a modification of the NLS model as a quasi-integrable theory and the special parity symmetry, i.e. a complex conjugation, a shifted space-reflection and time-delayed inversion $({\cal C}{\cal P}_s{\cal T}_d)$ is discussed. In sec. \ref{secmnlsano} we find novel towers of quasi-conservation laws of the modified NLS. In particular, we introduce the anomaly cancellation mechanism in order to get the exact conservation laws. In  section \ref{sec:stNLS} we show by direct construction that even the standard NLS model possesses infinite towers of infinite number of anomalous conservation laws. In sec. \ref{sec:PCTsol} it is shown analytically the quasi-conservation of the infinite number of anomalous charges for $N-$bright soliton solution satisfying the symmetry ${\cal C}{\cal P}_s{\cal T}_d$. The section \ref{simul} presents the results of our numerical simulations. The time evolution of  $2-$bright soliton collisions for the modified NLS are performed and then it is verified whether the observed results supported the vanishing of the space and space-time integrated anomalies of the quasi-conservation laws for several values of the deformation parameters and relative velocities and amplitudes. Our numerical results are shown in the Figs. 1,  2,...,15.

The section \ref{sec:riccati} considers a particular deformation of the $sl(2)$ AKNS model in the context of the Riccati-type pseudo-potential approach, and it discusses a particular reduction to the modified NLS model. An infinite set of quasi-conservation laws are constructed. In sec. \ref{sec:dual1} a dual Riccati-type formulation and novel anomalous charges are presented. 
In sec. \ref{linear} it is found a linear system formulation of the deformed AKNS model and constructed an infinite set of non-local conservation laws. In sec. \ref{ap:conclu} we present our conclusions and discussions. The appendices \ref{fsca1} , \ref{ap:chi} and \ref{app:uchid} present the components of the expansions in power of $\zeta$ of the Riccati-type pseudo-potentials. Finally, the appendix \ref{ap:3sol} presents the relevant parameters in the construction of the ${\cal C}{\cal P}_s{\cal T}_d-$symmetric $3-$bright soliton.

\section{The model}
\label{sec:themodel}

We will consider non-relativistic models   in $(1+1)-$dimensions with  equation of motion given by
\br
\label{nlsd}
i \frac{\pa}{\pa t} \psi(x,t) + \frac{\pa^2}{\pa x^2} \psi(x,t) -  V^{(1)}(|\psi(x,t)|^2)\psi(x,t) &=&  0,\\
V^{(n)}(I)&\equiv &\frac{d^n}{d I^n} V(I),\,\,\,\, I \equiv \bar{\psi} \psi, \er 
where $\psi$ is a complex scalar field  and the potential $V: \IR_{+} \rightarrow \IR$.

The model (\ref{nlsd}) defines a modified non-linear Schr\"odinger model (MNLS) and it supports bright and dark soliton type solutions in analytical form for some special functions $V[I]$. The potential  $V[I] = \eta I^2,\,(\eta < 0)$, corresponds to the integrable focusing NLS model and supports N-bright soliton solutions; whereas, the case $\eta > 0$ defines the integrable defocusing NLS model and supports N-dark soliton solutions. The potential  $V[I] = \eta  I^2-\epsilon I^3/6$ defines the non-integrable cubic-quintic NLS model (CQNLS) which possesses analytical bright ($\eta < 0 $) and dark ($\eta > 0$) type solitons \cite{cowan, crosta}. In \cite{sombra, cowan} the bright solitary waves of the cubic-quintic focusing NLS have been regarded as quasi-solitons presenting partially inelastic collisions in certain region of parameter space. Among the models with saturable non-linearities \cite{kivshar}, the case $V[I] = \frac{1}{2} \rho_s (I+\frac{\rho_s^2}{I+ \rho_s})$ also exhibits analytical dark solitons \cite{kroli}. The deformed NLS model with $V^{(1)}(I) = 2 \eta I - \epsilon \frac{I^q}{1+ I^q}$ passes the Painlev\'e test for arbitrary positive integers  $q\in \IZ_{+}$ and $\epsilon =1$ \, \cite{enns}. 

We will consider the solutions of (\ref{nlsd}) satisfying the boundary conditions
\br
\label{bcs}
|\psi|_{x=-\infty} = |\psi|_{x=+\infty} = 0,\,\,\,\, \pa_x \psi =0,\,\,\, \mbox{as}\,\,\,  |x| \rightarrow  +\infty. 
\er

As a particular example, we will consider the deformed (focusing)  NLS equation  with potential 
\br
\label{pot10}
V(I) = - \frac{ 2\eta }{2+ \epsilon} ( I )^{2 + \epsilon},  \,\, \epsilon \in \IR 
\er
which implies the following equation of motion
\begin{eqnarray}
\label{mnls}
i \frac{\partial}{\partial t} \psi(x,t) +   \frac{\partial^2}{\partial x ^2} \psi(x,t) + 2 \eta \Big[ |\psi(x,t)|^2\Big]^{(1+ \epsilon)} \psi(x,t) =  0,\,\,\,\,\psi(x,t) \in C;\,\,\,\, \eta > 0,\end{eqnarray}
where $\epsilon$ is a deformation parameter. Notice  that in the limit $\epsilon  = 0$ one has the standard (focusing) NLS model. The model (\ref{mnls}) has recently been considered in \cite{jhep3, jhep5} in order to study  the concept of quasi-integrability for bright soliton collisions. An analytical solitary wave solution with vanishing boundary condition (bright soliton) for this potential is well known in the literature (see for example \cite{jhep3})
\br 
\label{solitary}
\psi(x,t) = \Big[\frac{2 + \epsilon}{2} \frac{\rho^2}{\eta} \frac{1}{\cosh^2{[(1+ \epsilon) \rho (x - v t -x_0)]}}\Big]^{\frac{1}{2(1+ \epsilon)}} \,\, e^{i [(\rho^2-\frac{v^2}{4})t + \frac{v}{2} x]}.
\er
In our numerical simulations for two-soliton collisions for deformed NLS  we will consider the potential of type (\ref{pot10}) corresponding to the model  (\ref{mnls}) and as an initial condition two solitary waves of type (\ref{solitary}) sewed together conveniently and located far away.   
  
Next, let us consider a special space-time symmetry related to soliton-type solutions of the model. So, consider a reflection around a fixed point $(x_{\Delta},t_{\Delta})$
\br
\label{par1}
\widetilde{{\cal P}}:  (\widetilde{x},\widetilde{t}) \rightarrow (-\widetilde{x},-\widetilde{t});\,\,\,\,\,\,\,\,\widetilde{x} = x - x_{\Delta},\,\,\widetilde{t} = t- t_{\Delta}. 
\er 
The transformation $\widetilde{{\cal P}}$ defines a shifted parity ${\cal P}_{s}$ for the spatial variable $x$  and a delayed time reversal ${\cal T}_d$ for the time variable $t$. Notice that, when $x_{\Delta}=0$ ($t_\Delta=0$), ${\cal P}_{s}$ (${\cal T}_d$) is reduced to the pure parity ${\cal P}$ (pure time reversal ${\cal T}$).  
  
In the quasi-integrability approach \cite{jhep3} it is assumed that the  $\psi$ solution of the deformed NLS model  possesses the following property under the transformation (\ref{par1})
\br
\label{par2}
\widetilde{{\cal P}} \equiv {\cal P}_{s}{\cal T}_{d},\,\,\,\,\,\,\,\,\widetilde{{\cal P}}(\psi) = e^{i \delta} \bar{\psi},\,\,\,\widetilde{{\cal P}}(\bar{\psi}) = e^{-i \delta} \psi,\,\,\,\,\bar{\psi} \equiv \psi^{\star},\,\,\,\,\delta = \mbox{constant.}
\er
The model (\ref{mnls}) with $\epsilon=0$ becomes the standard NLS
\br
\label{nls0}
i \frac{\partial}{\partial t} \psi(x,t) +   \frac{\partial^2}{\partial x ^2} \psi(x,t) + 2 \eta   |\psi(x,t)|^2  \psi(x,t) =  0,\,\,\,\,\psi(x,t) \in C;\,\,\,\, \eta > 0.
\er
Let us examine the scaling dimensions  of the NLS model. So,  it enjoys the scale-invariance 
\br
\label{sca1}
\psi  \rightarrow \frac{1}{\l} \psi(\frac{t}{\l^2}, \frac{x}{\l}).
\er
So, inspecting the scaling (inverse length dimension) of the $\psi$ and $\psi^{\star}$ fields in the eq. (\ref{nls0}) one notices that the fields and derivatives can be associated with the scale dimensions\footnote{Notice that all terms  in the  standard NLS equation should be of the same scale dimension with $\pa_x$ being $\l^{-1}$, which we define as $\deg(\pa_x)=1$.}
\br
\label{sca2}
\deg(\pa_t) = 2;\,\,\,\,\deg(\pa_x) = 1;\,\,\,\, \deg(\psi)= \deg(\bar{\psi})=1. 
\er

Next, let us perform the complex conjugation of the equation (\ref{nls0}) 
\begin{eqnarray}
\label{nls0c}
	-i \frac{\partial}{\partial t} \bar{\psi}(x,t)+   \frac{\partial^2}{\partial x ^2} \bar{\psi}(x,t) + 2 \eta  |\psi(x,t)|^2\, \bar{\psi}(x,t) =  0, \end{eqnarray}
where we have considered the charge conjugation operator
\br
\label{chargeop}
{\cal C} (\psi) = \bar{\psi}.
\er 
In addition,  consider the space and time reflections around the origin
\br
\label{ref0}
{\cal P}: x \rightarrow -x,\,\,\,\,\,\,{\cal T}: t \rightarrow -t.
\er 
Let us apply the above reflection transformations to the eq.  (\ref{nls0}). So, one has
\br
\label{nls0xt}
-i \frac{\partial}{\partial t} \psi(-x,-t) +   \frac{\partial^2}{\partial x ^2} \psi(-x,-t) + 2 \eta   |\psi(-x,-t)|^2  \psi(-x,-t) =  0,\,\,\,\,\psi(-x,-t) \in C;\,\,\,\, \eta > 0.
\er
Then, comparing the eqs.  (\ref{nls0c}) and (\ref{nls0xt}) the next relationships follow
\br
\label{inv}
\psi(-x, -t) = e^{i \delta} \bar{\psi}(x,t),\,\,\,\bar{\psi}(-x, -t) = e^{-i \delta} \psi(x,t),\,\,\,\,\delta = \mbox{constant.}
\er
So, we define the special  transformation as the product of complex conjugation ${\cal C}$, shifted space-reflection ${\cal P}_d$ and time-delayed inversion ${\cal T}_d$ as
\br
\label{cpstd}
 {\cal C}{\cal P}_s{\cal T}_d :   \psi(\widetilde{x},\widetilde{t}) \rightarrow \bar{\psi}(-\widetilde{x},-\widetilde{t}), 
\er
for a complex field $\psi$, such that $\bar{\psi}$ stands for complex conjugation of the field. Therefore, the condition (\ref{par2}) satisfied by the  field of a  quasi-integrable model can be rewritten as
\br
\label{cpstd1}
 {\cal C}{\cal P}_s{\cal T}_d (\psi) = e^{-i \delta}\, \psi.
\er 

Moreover, the NLS equation exhibits the space-time translation invariance, i.e. 
\br
\label{trxt}
x \rightarrow x + x_0,\,\,\,t \rightarrow  t + t_0. 
\er

This implies that for the N-soliton solution its j-soliton component can be located
anywhere $x_{j}=\eta_{0j}$, for $\eta_{0j}$ being arbitrary constants.  Notice that the shifted parity and delayed time inversion symmetry (\ref{par1}) becomes the usual space-time reflection symmetry (\ref{ref0}) if one sets $x_\Delta=t_\Delta= 0$ in (\ref{par1}), i.e. (\ref{ref0}) is a particular case of (\ref{par1}).  

So, the standard NLS equation is symmetric under scaling transformation, therefore its conservation laws, generalized symmetries, and recursion operator inherit the same scaling property \cite{anco}. Below, we will systematically use this concept in order to identify the higher scale dimensions of different quantities, such as the relevant charge densities in the both deformed and standard NLS models, respectively. In fact, the single tower of NLS-type asymptotically conserved charges defined in the previous references \cite{jhep3, jhep4, jhep5} for deformations of  NLS is composed by anomalous charge densities with homogeneous scale dimensions; i.e. each charge density is a homogeneous polynomial  which exhibits the  degree $n$; in their notation $\deg{\{a_x^{(3, -n)}\}} = n,\,\,n=0,1,2,...$.  

In recent papers \cite{jhep4, jhep5}, there have been shown analytically and numerically that the quasi-integrable defocusing and focusing non-linear Schr\"odinger models of type (\ref{mnls}), respectively,  support a tower of infinite number of exactly conserved charges for two-soliton (dark-dark and bright-bright) configurations possessing definite parity simply under the space-reflection symmetry, for a wide range of values of the deformation parameter $\epsilon$. The associated charges exhibit the same form as the ones for the standard NLS model. Moreover, in the both types (focusing and defocusing) of deformed NLS models there have been reported that for various two-soliton configurations without parity symmetry, the first nontrivial fourth order charge, which presents an `anomalous' term in the quasi-integrability formulation, is exactly conserved, within numerical accuracy; i.e. both the space and space-time integrals of its associated anomaly vanishes.  

In the following we will tackle the problem of finding novel  towers of quasi-conservation laws of the deformed NLS model (\ref{nlsd}) and show that some even parity monomials and polynomials with homogeneous degrees can be regarded as building blocks in order to construct the first exact conservation laws of the deformed model, provided that certain combinations of the relevant anomalies vanish.  In the next sections we construct  towers of infinite number of  anomalous conservation laws of the deformed model (\ref{nlsd}) and then we discuss the presence of relevant quasi-conservation laws  even for the standard NLS (\ref{nls0}).  

\section{Modified NLS and anomalous charges}
\label{secmnlsano}

In the context of deformed sine-Gordon and KdV models there have been analyzed the behavior of novel infinite towers of  asymptotically conserved (anomalous) charges in \cite{npb1}  and \cite{jhep33}, respectively. Those results uncovered new anomalous charges and extended the earlier results on SG model in \cite{jhep1, jhep2, cnsns} and KdV model in \cite{npb}, respectively.  Remarkably, the standard SG and KdV models exhibit some sets of quasi-conservation laws, such that the space-time integrals of the relevant anomalies vanish for 
N-soliton solutions with special space-time symmetries. In those developments the novel anomalous charges and related anomalies, associated to standard and modified integrable models, have been obtained through a direct construction method starting from the relevant equations of motion. In the section \ref{sec:riccati} of this paper we will provide an unified and rigorous approach in order to construct the new quasi-conservation laws based on the Riccati-type pseudopotential method for the standard and modified NLS cases. However, in this section we follow the direct construction method in order to generate those types of charges by analyzing the behavior of each monomial and polynomial of increasing number of degrees in the scaling dimensions defined in (\ref{sca2}).  

\subsection{First order charge and its generalization}

Next we construct a conservation law with a charge density whose relevant terms possess scaling dimensions of order one, according to  the definition (\ref{sca1})-(\ref{sca2})\footnote{Formally, one can assume $\deg{(\psi^{-1})} = \deg{(\bar{\psi}^{-1})}=  -1$.}. From the eqs. of motion (\ref{nlsd}) and its complex conjugate one can write the exact conservation law
\br
\label{topod}
\pa_t [i \frac{\bar{\psi}\pa_x \psi - \psi \pa_x \bar{\psi}}{\bar{\psi} \psi}] + \pa_x [ \frac{\pa_x^2 \psi}{\psi}+\frac{\pa_x^2 \bar{\psi}}{\bar{\psi}} -2 V^{(1)}] =0.
\er
The relevant charge density possesses even parity under (\ref{par2}) and it is associated  to the phase difference (or the phase jump) of the solutions. In particular, it represents the topological charge related to the dark solitons of the defocusing NLS \cite{jhep4, jpa1}.
 
A direct generalization of the above conservation law is provided by the next quasi-conservation law
\br
\label{topoFd}
\pa_t \Big[i F(I) \frac{\bar{\psi}\pa_x \psi - \psi\pa_x \bar{\psi}}{\bar{\psi}\psi}\Big] + \pa_x \Big[ F(I) \left(\frac{\pa_x^2 \psi}{\psi}+\frac{\pa_x^2 \bar{\psi}}{\bar{\psi}} -2 V^{(1)}\) - 2 G(I)\Big] &=& \hat{\alpha}_1,
 \er
with
\br
\hat{\alpha}_1 &\equiv & 2 F^{(1)}(I)\, \pa_x [\pa_x \bar{\psi} \pa_x \psi], \label{an1d}\\ 
G(I) &\equiv& - \int^{I} dI'\,V^{(1)}(I')\, F^{(1)}(I'),\nonumber\\
F^{(1)}(I) &\equiv&  \frac{d}{dI} F(I),\nonumber
\er
where $I$ has been defined in (\ref{nlsd}) and $F(I)$  is an arbitrary function. In fact, this quasi-conservation law reduces to the exact conservation law (\ref{topod}) for $F=1$. The r.h.s. of (\ref{topoFd}) defines the anomaly $\hat{\alpha}_1$, which is an odd function under the space-time transformation defined in (\ref{par2}).  Notice that in the generalized expression (\ref{topoFd}) the even parity condition imposed in the construction of  the charge density implied the desired odd parity of  the anomaly $\hat{\alpha}_1$.

So, one can define the asymptotically conserved charge
\br
\label{q1d}
\frac{d  }{dt}  Q_{1}(t) &=&  \int \, dx\, \hat{\alpha}_1\\
Q_{1} (t) &=& \int_{-\infty}^{\infty} \, dx\, [i F(I) \frac{\bar{\psi}\pa_x \psi - \psi \pa_x \bar{\psi}}{\bar{\psi}\psi}].
\er
For the special solutions satisfying the parity property (\ref{par1})-(\ref{par2}) one must have the vanishing of the space-time integral of the anomaly $\hat{\alpha}_1$, i.e.
\br
\int^{\widetilde{t}}_{-\widetilde{t}} dt \, \int^{\widetilde{x}}_{-\widetilde{x}} dx \,  \hat{\alpha}_1=0,\,\,\,\, \mbox{for} \,\,\,\widetilde{t} \rightarrow \infty,\,\,\,\widetilde{x} \rightarrow  \infty.
\er
Therefore, integrating in $t$ on the b.h.s.'s of (\ref{q1d}) one can get
\br
Q_{1} (t \rightarrow \infty) =Q_{1} (t \rightarrow - \infty) .
\er

Below we will compute the $x-$integral and $(x,t)-$integral, respectively,  of the anomaly $\hat{\alpha}_1$ in (\ref{an1d}) for a special case of $F (I)$, through numerical simulation of two-bright soliton collisions. In fact, we will use the exponential distribution $F(I) = e^{-I^2}$ which has been used in order to compute numerically the ensemble-averaged values of the first integrals in the study of soliton-gas and integrable turbulence in the focusing NLS model \cite{turbu, gauss}.

\subsection{Second order and related higher order tower}

From eq. (\ref{nlsd}) and its complex conjugate one can write the conservation law
\br
\label{rho1}
\pa_t (\bar{\psi}\psi) - \pa_x \Big[i (\bar{\psi} \pa_x \psi - \psi \pa_x \bar{\psi}) \Big] =0.
\er
The even parity charge density $\bar{\psi}\psi$ possesses scaling dimension $2$. The relevant charge $\widetilde{Q}_1 = \int\, dx \, \frac{1}{2}(\bar{\psi}  \psi)$ defines the normalization of the solution $\psi$ and it is related to the internal symmetry of the model: $\psi \rightarrow e^{i\delta} \psi, \delta = const.$

Multiplying by $(\bar{\psi} \psi)^{n-1}$ on the both sides of eq. (\ref{rho1}) and making use of the eq. of motion (\ref{nlsd}) one can rewrite  (\ref{rho1}) as
\br
\label{rho1t}
\pa_{t} [\frac{1}{2n} (\bar{\psi}  \psi)^n] -\pa_x [ \frac{1}{2n} i  (\bar{\psi} \psi)^{n-1} (\bar{\psi} \pa_x \psi - \psi \pa_x \bar{\psi} ) ] & = &\hat{\beta}_n, \,\,\,\,\,\,n=1,2,3...\\
\hat{\beta}_n &\equiv &-\frac{1}{2n} \pa_x [(\bar{\psi} \psi)^{n-1}] i (\bar{\psi} \pa_x \psi - \psi \pa_x \bar{\psi} ).
\label{rho1tan}
\er
Notice that for $n=1$ it is the first exact conservation law (\ref{rho1}). Let us write the asymptotically conservation laws as
\br
\label{q2d}
\frac{d}{dt} {\cal Q}_{n} &=&\int dx\, \hat{\beta}_n;\,\,\,\,\,n=2,3,...\\
{\cal Q}_{n} &=& \frac{1}{2n}\int\, dx \, (\bar{\psi}  \psi)^n .\label{beta2d}
\er
For the field $\psi$ satisfying (\ref{par1})-(\ref{par2}) the anomaly density  $\hat{\beta}_n$ possesses an odd parity for any $n$. Notice that the even parity condition imposed in the construction of  the monomials $(\bar{\psi}\psi)^{n}$ of  scaling dimension $2n$ in (\ref{rho1t}) implied the odd parities of  the   anomalies $\hat{\beta}_n$  which appear in the r.h.s of the generalized quasi-conservation laws (\ref{rho1t}).

Therefore, one must have the vanishing of the space-time integral of the anomaly $\hat{\beta}_n$, i.e.
\br
\int^{\widetilde{t}}_{-\widetilde{t}} dt \, \int^{\widetilde{x}}_{-\widetilde{x}} dx \,  \hat{\beta}_n =0,\,\,\,\, \mbox{for} \,\,\,\widetilde{t} \rightarrow \infty,\,\,\,\widetilde{x} \rightarrow  \infty,\,\,\,\,\,n=2,3,...
\er
So,   the asymptotically conserved charges satisfy 
\br
{\cal Q}_{n}(t \rightarrow \infty) = {\cal Q}_{n} (t \rightarrow - \infty),\,\,\,\, n=2,3,...
\er

Below we will compute the $x-$integral and $(x,t)-$integral, respectively,  of the anomaly $\hat{\beta}_2$  through numerical simulation of two-bright soliton collisions.  

\subsection{Third order and related higher order tower}

The conservation law related to the space translation symmetry of the model can be written as
\br
\label{momd}
\pa_t \Big[i (\bar{\psi}\pa_x \psi - \psi \pa_x \bar{\psi})\Big] - \pa_x \Big[ 2 \pa_x \psi \pa_x \bar{\psi} -\psi \pa^2_x \bar{\psi} - \bar{\psi} \pa_x^2 \psi + 2  V^{(1)}  |\psi|^2 - 2 V\Big] =0.  
\er
The relevant charge is associated to the momentum conservation. Notice that the relevant charge density is a third order homogeneous polynomial according to the scaling (\ref{sca2}) and possesses even parity under (\ref{par1})-(\ref{par2}). Next, multiplying by 
 $\frac{(i)^n}{n} (\bar{\psi}\pa_x \psi - \psi \pa_x \bar{\psi}) ^{n-1}$ on the both sides of the above eq. and  conveniently  rewritten it one has
\br \nonumber
\pa_t \Big[\frac{(i)^n}{n} (\bar{\psi}\pa_x \psi - \psi \pa_x \bar{\psi}) ^n\Big]& -&\\
 \pa_x \Big[\frac{(i)^n}{n} (\bar{\psi}\pa_x \psi - \psi \pa_x \bar{\psi}) ^{n-1} (2 \pa_x \psi \pa_x \bar{\psi} -\psi \pa^2_x \bar{\psi} - \bar{\psi} \pa_x^2 \psi + 2  V^{(1)}  |\psi|^2 - 2 V)\Big]&=&\hat{\gamma}_n \label{gam3d}\\
\hat{\gamma}_n  \equiv  - \frac{(i)^n}{n} \pa_x [(\bar{\psi}\pa_x \psi - \psi \pa_x \bar{\psi}) ^{n-1}]\Big[ 2 \pa_x \psi \pa_x \bar{\psi} -\psi \pa^2_x \bar{\psi} - \bar{\psi} \pa_x^2 \psi + 2  V^{(1)}  |\psi|^2 - 2 V\Big].&& \label{gam33d}\\
n=2,3,4,...&&\nonumber
\er
Let us write the asymptotically conservation laws as
\br
\label{q3d}
\frac{d}{dt} \widetilde{Q}_{n}  &=& \int \, dx\, \hat{\gamma}_n ;\,\,\,\,\,n=2,3,...\\
\widetilde{Q}_{n} (t) &=& \int\, dx \,\frac{(i)^n}{n} (\bar{\psi}\pa_x \psi - \psi \pa_x \bar{\psi}) ^n.
\er
Notice that the anomaly density $\hat{\gamma}_n$ (\ref{gam33d})  possesses odd parity under the  transformation $\widetilde{{\cal P}}$ defined in (\ref{par1})-(\ref{par2}). Therefore, the  integral $\int \, dt\,\int \, dx \,\hat{\gamma}_n$ will vanish providing the asymptotically conserved charges
\br
\widetilde{Q}_{n}(t=+\infty) &=& \widetilde{Q}_{n}(t=-\infty),\,\,\,\,n=2,3,...
\er

As in the previous constructions,  the even parity condition imposed in the construction of  the  charge density in (\ref{gam3d}) ($n^{th}$ power of the charge density in (\ref{momd})) implied the odd parity of  the anomalies $\hat{\gamma}_n$ in the generalized expression (\ref{gam3d}).

Below we will compute the $x-$integral and $(x,t)-$integral, respectively,  of the anomaly $\hat{\gamma}_2$  through numerical simulation of two-bright soliton collision.  

\subsection{Fourth order and related higher order towers}

From the eqs. of motion one can write the next conservation law
\br
\label{energy1}
\pa_t [\pa_x \bar{\psi} \pa_x \psi + V(|\psi|^2) ] -i \pa_x\Big[ (\bar{\psi}\pa_x \psi - \psi \pa_x \bar{\psi}) V^{(1)}+ \pa_x\bar{\psi}\pa^2_x \psi - \pa_x \psi \pa^2_x \bar{\psi}  )  \Big]=0.\er
The relevant charge of this the conservation law   is related to the energy of the system
\br
\label{chenergy}
H = \int \, dx [\pa_x \bar{\psi} \pa_x \psi + V(|\psi|^2)].
\er
Notice that for the standard NLS the relevant energy charge density is composed of terms with even parity under (\ref{par2}) and homogeneous scalings of degree $4$ according to the definition (\ref{sca1})-(\ref{sca2}), i.e. $\deg{(\pa_x \bar{\psi} \pa_x \psi)}=4,\,\deg{(V_{NLS})}= \deg{(-\eta I^2)} = 4$; whereas, the deformed NLS  is composed of terms with mixed degrees, i.e. the deformed potential $V(|\psi|^2)$ would carry scaling dimension different from $4$.
  
Associated to this conservation law one can write the following pair of quasi-conservation laws
\br
\label{kinetic1}
\pa_t [\pa_{x} \bar{\psi} \pa_x{\psi}] + i \pa_x [ \pa_x^2\bar{\psi} \pa_x \psi - \pa_x^2\psi \pa_x \bar{\psi}] &=& \hat{\delta}_{1}\\
\pa_t V(|\psi|^2) - i \pa_x[ (\bar{\psi}\pa_x \psi - \psi \pa_x \bar{\psi})V^{(1)}] &=&-\hat{\delta}_{1}. \label{potential1}\\
\hat{\delta}_{1} \equiv i [(\bar{\psi} \pa_x \psi)^2-(\psi \pa_x\bar{\psi})^2] V^{(2)}. \label{ano4}
\er 
In fact, the corresponding anomalous charges correspond to the kinetic and potential terms, respectively, associated to the enegy charge (\ref{chenergy}). Notice that the r.h.s.'s of the above eqs. possess odd parities under the symmetry transformation (\ref{par1})-(\ref{par2}).   

Remarkably, the relevant anomalies possess opposite signs, and so, upon adding the l.h.s.'s of (\ref{kinetic1}) and (\ref{potential1}) one gets the exact conservation law (\ref{energy1}), since the corresponding anomalies in the r.h.s.'s cancel to each other. So, we have shown a first instance in which an anomaly cancellation mechanism of two quasi-conservation laws  gives rise to an exact conservation law for a deformed NLS model with an arbitrary  potential of type $V(I)$.

From the relationships (\ref{kinetic1}) and (\ref{potential1}) one can construct their related higher order quasi-conservation laws. So, multiplying by $\frac{1}{n} (\pa_{x} \bar{\psi} \pa_x{\psi})^{n-1}$ on the both hand sides of   (\ref{kinetic1}) and rewritten conveniently one has
\br
 \label{kinetic11}
\pa_t [\pa_{x} \bar{\psi} \pa_x{\psi}]^{n} & + & \frac{i}{n} \pa_x[(\pa_{x} \bar{\psi} \pa_x{\psi})^{n-1} (\pa_x^2\bar{\psi} \pa_x \psi - \pa_x^2\psi \pa_x \bar{\psi})] =\hat{\delta}_n\\
 \hat{\delta}_n &\equiv & \frac{i}{n} [(\bar{\psi} \pa_x \psi)^2-(\psi \pa_x\bar{\psi})^2] (\pa_{x} \bar{\psi} \pa_x{\psi})^{n-1} V^{(2)}  +\nonumber \,\,\,\,\,\,\,\,\,\, \\
&& \frac{i}{n}\pa_x [\pa_{x} \bar{\psi} \pa_x{\psi}]^{n-1} (\pa_x^2\bar{\psi} \pa_x \psi - \pa_x^2\psi \pa_x \bar{\psi}),\,\,\,\,n=1,2,3...\label{an4d}
\er
The case $n=1$ reduces to the quasi-conservation law (\ref{kinetic1}).  Moreover, the anomalies $ \hat{\delta}_n$   possess odd parities under the symmetry transformation (\ref{par1})-(\ref{par2}). Therefore, one can define the tower of anomalous charges 
\br
\label{q4kd} 
\frac{d}{dt} K_n &=& \int dx\, \hat{\delta}_n,\,\,\,\,\,\,\,\,\, n=1,2,...
\\
\label{q4d1}
K_n &=& \int \, dx \, [\pa_{x} \bar{\psi} \pa_x{\psi}]^{n}.
\er

Likewise, the generalization of the quasi-conservation law (\ref{potential1}) becomes
\br
\label{potential11}
\pa_t [ G(|\psi|^2) V(|\psi|^2)] &-& i \pa_x[ (\bar{\psi}\pa_x \psi - \psi \pa_x \bar{\psi})V^{(1)}(|\psi|^2) G(|\psi|^2)] = \hat{\sigma}_{G}\\
 \hat{\sigma}_{G}&\equiv&
-i [(\bar{\psi} \pa_x \psi)^2-(\psi \pa_x\bar{\psi})^2] V^{(2)}(|\psi|^2)G(|\psi|^2)+ V(|\psi|^2)\pa_t G(|\psi|^2) -\nonumber
\\ && i (\bar{\psi}\pa_x \psi - \psi \pa_x \bar{\psi})V^{(1)}(|\psi|^2) \pa_x G.
\label{sig}
\er
In fact, the special case $G(I)=1$ of (\ref{potential11}) reduces to the quasi-conservation law  (\ref{potential1}). In addition, the anomaly $\hat{\sigma}_{G}$ possesses odd parity under  the symmetry transformation (\ref{par1})-(\ref{par2}) for any function $G(I)$, provided that $\widetilde{{\cal P}} (G) = G$. So, one can define a family of anomalous charges
\br
\label{q4d2}
V_{G} = \int\,  dx \,  G(|\psi|^2) V(|\psi|^2). 
\er

Below we will compute the $x-$integral and $(x,t)-$integral  of the anomaly $\widetilde{\delta}_1$ in (\ref{ano4}) or (\ref{kinetic11}), which corresponds to  $\hat{\sigma}_{G=1} = - \widetilde{\delta}_1$ in (\ref{potential11}),  through numerical simulation of two-bright soliton collision.   

So, we have shown that the deformed NLS (\ref{nlsd}) model  exhibits four exact conservation laws for any deformed potential of type $V(I)$.  They are  the first order topological charge  density (\ref{topod}), second order normalization (\ref{rho1}), and the third order momentum (\ref{momd}).  Whereas, the energy density charge (\ref{chenergy}), even though it satisfies an exact conservation law, possesses mixed scale dimension; i.e. the fourth order monomial $\pa_{x} \bar{\psi} \pa_x{\psi}$ and the deformation dependent term $V(I)$, which can take an arbitrary dimension. For example, for the case (\ref{mnls}) it takes the degree $2(2+ \epsilon)$.  

It is well known that each conserved charge  of the standard NLS model is composed by homogeneous degree polynomials on the field $\psi (\bar{\psi})$, and their $x-$derivatives, such that the higher order charges are organized as polynomials with increasing scaling dimensions. The natural question is whether there exist exact conservation laws of higher order (beyond fourth order) for the deformations of NLS of type (\ref{nlsd}).  Next, we will look for new charges, at least for the polynomial types,  in the context of the quasi-conservation law formulation and following the anomaly cancellation mechanism discussed above. 

\subsection{Fifth order monomials and related higher order towers}

An exact conservation law with fifth order charge density is not available for the deformed NLS (\ref{nlsd}). In fact,  in the anomalous zero-curvature \cite{jhep3, jhep4, jhep5} and in the Riccati-type pseudo-potential approaches, which we will present below, the fifth order  exact conservation law is absent. Let us write the quasi-conservation laws for the monomials of fifth order. They become 
\br\nonumber
\pa_t \Big[i (\bar{\psi}\pa^3_x \psi - \psi \pa^3_x \bar{\psi})\Big] &-& \pa_x \Big[\pa^2_x  \bar{\psi}\pa^2_x \psi -  \bar{\psi}\pa^2_x (\psi V^{(1)})-\psi\pa^2_x (\bar{\psi} V^{(1)}) + 4 \pa_x\bar{\psi} \pa_x\psi V^{(1)} (I)+ \pa_x I \pa_x  V^{(1)}(I)-\\
&& V^{(1)}(I)\pa^2_x I +\bar{\psi}\pa^4_x \psi  + \psi \pa^4_x\bar{\psi} - \pa_x\bar{\psi}\pa^3_x \psi-\pa_x \psi\pa^3_x\bar{\psi}  \Big] =\hat{\rho} \label{fifth1}\\ 
\hat{\rho}&\equiv &-2 V^{(1)} (\bar{\psi}\pa^3_x \psi + \psi \pa^3_x \bar{\psi})
\er
and 
\br
\pa_t \Big[i I (\bar{\psi}\pa_x \psi - \psi \pa_x \bar{\psi})\Big] &+& \pa_x\Big[\frac{1}{2}(\bar{\psi}\pa_x \psi - \psi \pa_x \bar{\psi})^2 -H(I) -\frac{1}{3} I \pa^2_x I +\frac{1}{6}(\pa_x I)^2\Big] =\hat{\zeta} \label{fifth2}\\
 \hat{\zeta} &\equiv & -\frac{4}{3} I (\bar{\psi}\pa^3_x \psi + \psi \pa^3_x \bar{\psi}),\,\,\,\,\,\, H^{(1)}(I) \equiv 2 I^2  V^{(2)}(I).
\er
Moreover, one considers the following combination of the charge densities in  the l.h.s's of (\ref{fifth1}) and (\ref{fifth2})
\br
\label{q5nlsd}
\hat{Q}_5 = \int \, dx\,  i \Big[ 3 \eta  I (\bar{\psi}\pa_x \psi - \psi \pa_x \bar{\psi}) +  (\bar{\psi}\pa^3_x \psi - \psi \pa^3_x \bar{\psi})\Big],
\er
and the relevant  combination of their anomalies 
\br
\label{ancom}
3 \eta   \hat{\zeta} +  \hat{\rho} = -2 (V^{(1)} + 2 \eta I)  \Big[\bar{\psi}\pa^3_x \psi + \psi \pa^3_x \bar{\psi}\Big].
\er
So, for the deformed NLS one gets a quasi-conservation law
\br
\label{Q5an}
\frac{d}{dt} \hat{Q}_5 &=& \hat{\xi}\\
\label{5an}
\hat{\xi} &\equiv& \int \, dx\, [ -2 (V^{(1)} + 2 \eta I)  \Big[\bar{\psi}\pa^3_x \psi + \psi \pa^3_x \bar{\psi}\Big] ].
\er 
The anomalies $\hat{\rho}$ in (\ref{fifth1}) and $ \hat{\zeta}$ in (\ref{fifth2}) exhibit odd parities  under  the symmetry transformation (\ref{par1})-(\ref{par2}), and then, as in the previous constructions, one has the asymptotically conserved charge
\br
\hat{Q}_{5}(t=+\infty) &=& \hat{Q}_{5}(t=-\infty).
\er
Remarkably, the combination of the anomalies (\ref{ancom}) vanishes for the potential of type $V(I) = - \eta I^2$. Therefore, adding the l.h.s. of  (\ref{fifth1}) and (\ref{fifth2}) one gets an exact  conservation law and the associated fifth order conserved charge of the standard NLS. Notice that the fifth order charges of the standard NLS and its deformed counterpart are similar in form. So, at this level the anomaly cancellation mechanism does not provide a new exactly conserved charge of the deformed NLS.

Similarly, one can construct the infinite number of  higher order quasi-conservation laws associated to the identities  (\ref{fifth1})  and  (\ref{fifth2}), respectively. Although we do not present them explicitly, let us emphasize that their anomalies will exhibit odd parities and the relevant asymptotically conserved charges would be constructed following analogous steps as above.

\subsection{Sixth order monomials and related higher order towers}

Following similar constructions as above one can write the next quasi-conservation laws for the relevant sixth order three monomials and one polynomial
\br
\label{q61}
\frac{d}{dt} \int dx \, [2(\bar{\psi} \psi)^3] &=&\int dx \,  {\cal A}_1\\
{\cal A}_1 &\equiv& -12 i I \pa_x I ( \bar{\psi} \pa_x \psi - \psi \pa_x \bar{\psi}),\label{a1}\\
\label{q62}\frac{d}{dt}  \int dx \, [\pa_x^2\bar{\psi} \pa_x^2\psi] &=&\int dx \,  {\cal A}_2\\
{\cal A}_2 &\equiv&  i V^{(1)} ( \bar{\psi} \pa^4_x \psi - \psi \pa^4_x \bar{\psi}),\label{a2}\\
\label{q63}\frac{d}{dt} \int dx \,  [ I (\bar{\psi} \pa_x^2\psi + \psi \pa_x^2\bar{\psi})] &=&\int dx \,  {\cal A}_3, \\
\nonumber
{\cal A}_3 &\equiv&  i [( \bar{\psi} \pa^2_x \psi)^2 - (\psi \pa^2_x \bar{\psi})^2] + i I ( \bar{\psi} \pa^4_x \psi - \psi \pa^4_x \bar{\psi})-\\
&&2i I V^{(2)} \pa_x I ( \bar{\psi} \pa_x \psi - \psi \pa_x \bar{\psi}),\label{a3}\\
\label{q64}\frac{d}{dt} \int dx \,  [ I \pa_x \bar{\psi} \pa_x \psi] &=&\int dx \,  {\cal A}_4\\
\nonumber
{\cal A}_4 &\equiv&  i ( \bar{\psi} \pa^2_x \psi - \psi \pa^2_x \bar{\psi}) \pa_x \bar{\psi} \pa_x \psi - i I ( \pa_x \psi \pa^3_x \bar{\psi} - \pa_x\bar{\psi} \pa^3_x \psi)+\\
&&i I V^{(2)} \pa_x I ( \bar{\psi} \pa_x \psi - \psi \pa_x \bar{\psi})\label{a4}.
\er
Notice that the relevant anomalies ${\cal A}_j \, (j=1,...,4)$ in  (\ref{a1}), (\ref{a2}), (\ref{a3}) and (\ref{a4}), possess odd parities under the symmetry transformation (\ref{par1})-(\ref{par2}), and then, as in the previous constructions, one has the relevant charge density monomials in (\ref{q61}), (\ref{q62}) and  (\ref{q64}) and charge density polynomial in  (\ref{q63})  which will give rise to  asymptotically conserved charges, respectively.

One can show the following identities
\br
\label{id1}
3i ( \bar{\psi} \pa^2_x \psi - \psi \pa^2_x \bar{\psi}) \pa_x \bar{\psi} \pa_x \psi &\dot{\equiv} & i I ( \pa_x \psi \pa^3_x \bar{\psi} - \pa_x\bar{\psi} \pa^3_x \psi)\\
i I ( \bar{\psi} \pa^4_x \psi - \psi \pa^4_x \bar{\psi})  &\dot{\equiv} & 8  i ( \bar{\psi} \pa^2_x \psi - \psi \pa^2_x \bar{\psi}) \pa_x \bar{\psi} \pa_x \psi+ i [( \bar{\psi} \pa^2_x \psi)^2 - (\psi \pa^2_x \bar{\psi})^2],\label{id2}
\er
where $\dot{\equiv} $ stands for an equivalence relationship, up to a term of type $\pa_x [...]$, between the expresions on the both sides of the relevant identity.

The linear combination of the charge density monomials (\ref{q61}), (\ref{q62}) and (\ref{q64}) and polynomial (\ref{q63}) provides the next total charge of the sixth order
\br
\label{q6lc}
\hat{Q}_6 = \int \, dx [2\l_1  (\bar{\psi} \psi)^3+ \l_2 \pa_x^2\bar{\psi} \pa_x^2\psi+ \l_3 I (\bar{\psi} \pa_x^2\psi + \psi \pa_x^2\bar{\psi})+ \l_4  I \pa_x \bar{\psi} \pa_x \psi],
\er   
where the $\l_j's$ are arbitrary constant parameters.
 
Next,  using the above identities (\ref{id1})-(\ref{id2}) into the anomalies (\ref{a1}), (\ref{a2}), (\ref{a3}) and (\ref{a4}), one can write a linear combination of the anomalies as follows
\br
\nonumber \sum_{j=1}^{4} \lambda_j {\cal A}_j &=& i[ V^{(2)}(\l_4-2\l_3) -12 \l_1] I \pa_x I ( \bar{\psi} \pa_x \psi - \psi \pa_x \bar{\psi})+ i [\l_2 V^{(1)}+ 2 \l_3 I] ( \bar{\psi} \pa^4_x \psi - \psi \pa^4_x \bar{\psi}) -\\
&&2i [\l_4 + 4 \l_3]  ( \bar{\psi} \pa^2_x \psi - \psi \pa^2_x \bar{\psi}) \pa_x \bar{\psi} \pa_x \psi.\label{lc1}
\er
So, one has the quasi-conservation law
\br
\label{q66}
\frac{d}{dt} \hat{Q}_6 = \int dx\,  \sum_{j=1}^{4} \lambda_j {\cal A}_j.
\er
In general, for an arbitrary  set of values $\l_j$ and an arbitrary potential $V(I)$, the sum $\sum_{j=1}^{4} \lambda_j {\cal A}_j$ does not vanish, giving rise to the asymptotically conserved total charge $\hat{Q}_6$. However, for the special case of the standard NLS, $[V(I)= - \eta I^2,\,\,\,V^{(1)}(I) =- 2\eta I \,\,\, \mbox{and}\,\,\,\, V^{(2)}(I)= - 2\eta]$, one can choose the set of values: $\l_1= \eta\l_3,\,\l_2= \frac{\l_3}{\eta},\,\l_4= -4 \l_3,\,$ such that  the summation $\sum_{j=1}^{4} \lambda_j {\cal A}_j$ identically vanishes. So, one can define the exactly conserved charge  
\br
\label{q6exact}
Q_{6}^{NLS} = \l_3\, \int \, dx [ \frac{1}{\eta} \pa_x^2\bar{\psi} \pa_x^2\psi+  I (\bar{\psi} \pa_x^2\psi + \psi \pa_x^2\bar{\psi})+ 2 \eta (\bar{\psi} \psi)^3 - 4 I \pa_x \bar{\psi} \pa_x \psi].
\er 
As we will show below, this anomaly free charge is an exact  conserved charge of the usual NLS, up to an overall constant $\l_3$.  Then, this is another example of constructing an exact conservation law through the anomaly cancellation mechanism, provided that the potential $V$ takes a special form.

Similarly, one can construct  infinite number of  higher order quasi-conservation laws associated to the monomial charge densities of order $6$  in (\ref{q61}),   (\ref{q62})  and   (\ref{q64}), and the polynomial charge density of order $6$ in  (\ref{q63}), respectively. We do not present them explicitly here, but their anomalies will exhibit odd parities and their anomalous charges would be constructed following the procedure above. 	

Some comments are in order here. First, the anomalous charge densities above are composed of monomials or polynomials of homogeneous degree under the scaling defined in  (\ref{sca2}) and even parity under the symmetry transformation ${\cal P}_s{\cal T}_d$ in (\ref{par1})-(\ref{par2}). Second, associated to each lowest order monomial or polynomial of this type one can construct an increasingly higher order infinite tower of quasi-conservation laws with even anomalous charge densities  and odd parity anomaly densities. Third, it is shown that the exactly conserved charges (e.g. the energy $H$ and the  charges $\hat{Q}_5$ and $\hat{Q}_6$ in (\ref{chenergy}), (\ref{q5nlsd}) and (\ref{q6lc}), respectively ) emerge through some kind of anomaly cancellation mechanism. So, an exact conservation law arises when a suitable linear combination of relevant quasi-conservation laws leads to the vanishing of the combined anomalies. In fact, for the charge $H$ in (\ref{chenergy}) it happens for any potential $V(I)$, whereas for the charges $\hat{Q}_5$ and $\hat{Q}_6$ this mechanism holds for a special type of potential $V(I)= - \eta I^2$.

\section{Standard NLS: Exactly conserved and anomalous charges}
 \label{sec:stNLS}

The exactly conserved charges of the standard NLS can be found recursively as follows \cite{faddeev}
\br
w_1 &=& \psi\\
w_{n+1} &=& -i \frac{\pa}{\pa x} w_n - \eta \bar{\psi} \sum_{k=1}^{n-1} w_k w_{n-k}.
\er
Then the conserved quantities are defined as
\br
\label{qnw}
q_n  = \int_{-\infty}^{+\infty} dx \, \bar{\psi} w_n, \,\,\,\,\,n=1,2,3,...
\er
Therefore, the first five lowest order conserved charges become
\br
\label{q11}
q_1  &=& \int_{-\infty}^{+\infty} dx \, I,\,\,\,\,\,\,\,\,\,\,\,\,I \equiv \bar{\psi} \psi,\\
\label{q22}
q_2  &=& -\frac{1}{2}\int_{-\infty}^{+\infty} dx \,  i(\bar{\psi} \pa_x \psi - \psi \pa_x \bar{\psi}),\\
\label{q33}
q_3  &=& -\int_{-\infty}^{+\infty} dx \, [ \bar{\psi} \pa_x^2 \psi + \eta I^2 ],\\
\label{q44}
q_4  &=& \frac{1}{2}\int_{-\infty}^{+\infty} dx \, i[3 \eta I(\bar{\psi} \pa_x \psi - \psi \pa_x \bar{\psi}) + (\bar{\psi} \pa^3_x \psi - \psi \pa^3_x \bar{\psi})],\\
\label{q55}
q_5  &=& \int_{-\infty}^{+\infty} dx \, \{\pa^2_x\bar{\psi} \pa^2_x \psi +  \eta I (\bar{\psi} \pa^2_x \psi + \psi \pa^2_x \bar{\psi}) + 2 \eta^2 I^3 - 4  \eta I \pa_x\bar{\psi} \pa_x \psi \}.\label{q56}
\er
Notice that according to the scaling dimension defined in (\ref{sca2}) the relevant charge densities above (\ref{qnw}) possess degree $n+1$, i.e. $\deg{(\bar{\psi} w_n)} =  n+1$. 

The  sequence of charges (\ref{q11})-(\ref{q55}) can be reproduced from the relevant charges of the deformed model discussed in the last section, provided that the potential $V$ takes the form $V = -\eta I^2$. In fact, the charges $q_1$, $q_2$ and $q_3$ correspond to the exact conservation laws (\ref{rho1}), (\ref{momd}) and (\ref{energy1}), respectively.  The charge (\ref{q5nlsd}), up to the overall factor $\frac{1}{2}$, corresponds to $q_4$ provided that the anomaly in  (\ref{Q5an})-(\ref{5an}) vanishes in the NLS case. Similarly, the anomaly free charge in (\ref{q6exact}) ($\lambda_ 3 = \eta$) corresponds to the charge $q_5$ in the standard NLS case.

\subsection{Standard NLS and infinite towers of quasi-conservation laws}
\label{sec:NLSqc}

New towers of infinite number of anomalous charges have also been uncovered in the standard SG and KdV models in \cite{npb1} and  \cite{jhep33}, respectively. In the standard KdV case a particular tower of anomalous charges comprises the so-called statistical moments defined by the integrals of type $M_n(t) = \int_{\infty}^{+\infty} u^n \, dx,\,\,n=1,2,3,....$, where $u$ is the KdV field. It has been examined the behavior of these charges for two-soliton interactions, which are thought to play an important role in the study of soliton gas and turbulence in integrable systems \cite{pla1}. In the quasi-integrable KdV model the charges of type $M_n$ remain also as asymptotically conserved ones \cite{jhep33}.  

In the following, taking into account the symmetry transformations and the scaling dimension argument discussed in section \ref{sec:themodel}\, we uncover, through a direct construction,  novel towers of infinite number of anomalous conservation laws of the standard NLS model. In particular, we will construct  the relevant anomalies possessing odd parities under the special symmetry transformation (\ref{par2}), such that the space-time integration of them vanish for the N-soliton solutions .

The constructions follow similar steps as in the last section. So, we specialize to the standard NLS model the generalized quasi-conservation laws provided for the deformed NLS in sec. \ref{secmnlsano}.  So, in the corresponding equations of sec. \ref{secmnlsano} the potential and its derivatives must be taken as
\br \label{potder}
V(I)= - \eta I^2,\,\,\,V^{(1)}(I) =- 2\eta I \,\,\, \mbox{and}\,\,\,\, V^{(2)}(I)= - 2\eta.
\er
 Next, we describe order by order the corresponding charges and their generalizations.

${\bf 1^{st}}$ order and generalization

The exact first order conservation law  (\ref{topod}) and its generalized quasi-conservation  law  (\ref{q1d}) maintain the same form, provided that $V(I)$ satisfies (\ref{potder}).

${\bf 2^{nd}}$ order and higher order generalized tower of quasi-conservation laws

The second order exact conservation law (\ref{rho1})  and the tower of infinite number of generalized quasi-conservation laws (\ref{rho1t})-(\ref{q2d}), $n=2,3,...$, maintain the same form in the case of  the standard NLS.

${\bf 3^{rd}}$ order and higher order generalized tower

The exact third order conservation law  (\ref{momd}) and its generalized tower of quasi-conservation  laws  (\ref{gam3d})-(\ref{q3d}) maintain the same form, provided that  (\ref{potder}) is assumed.

${\bf 4^{th}}$ order and  higher order generalized tower

The exact  fourth order conservation law remain the same as in (\ref{energy1}) provided that $V$ satisfies  (\ref{potder}). In addition, one has the both  quasi-conservation laws (\ref{kinetic11}) and (\ref{potential11}) in which one must use (\ref{potder}). Then, one has the equations
\br
\frac{d}{dt} K_n &=& \delta_n,\,\,\,\,\,\,\,n=1,2,3,...;\\
\frac{d}{dt} V_{G} &=& \sigma_{G},
\er
where the anomalies $ \delta_n $ and $\sigma_{G}$ have been defined from their deformed counterparts (\ref{an4d}) and (\ref{sig}) by the replacement $V= -\eta I^2$; i.e. $ \delta_n \equiv  \hat{\delta}_n|_{V=- \eta I^2}$ and $\sigma_{G} \equiv  \hat{\sigma}_{G}|_{V=-\eta I^2}$.

${\bf 5^{th}}$ order and generalizations

In this case one has  the quasi-conservation laws  (\ref{fifth1}) and (\ref{fifth2}) such that   (\ref{potder}) is assumed. Notice that similar constructions as in the lower order cases can be made in order to get infinite towers of quasi-conservation laws even for the standard NLS.

${\bf 6^{th}}$ order and generalizations

In this case one has  the quasi-conservation laws  (\ref{q61})-(\ref{a4}) such that  (\ref{potder}) is assumed. So, similar constructions as above can be made in order to get infinite towers of quasi-conservation laws even for the standard NLS.

Notice that the potential $V(I)$ and its $n^{th}$-order derivative, $V^{(n)}(I)$, possess even parity under the parity transformation (\ref{par1})-(\ref{par2}), i.e. $\widetilde{{\cal P}}[V(I)] = V(I)$ and $\widetilde{{\cal P}}[V^{(n)}(I)] = V^{(n)}(I)$. So,
the main observation is that all the anomalies of the quasi-conservation laws for the standard NLS  (\ref{nls0}) and the relevant anomalies for  the deformed NLS models  of type (\ref{nlsd})  share the same parity under the space-time parity transformation (\ref{par1})-(\ref{par2}).  

Moreover, the statistical properties of nonlinear integrable systems, generally called integrable turbulence, is a rapidly developing
area of research. In particular, some of the lowest order anomalous charges presented above have appeared as statistical characteristics of the soliton gas in NLS systems, and their ensemble-averged values  for large number of $N-$solitons have been examined  (see e.g. \cite{turbu} and refs. therein).

\subsection{$ {\cal C} {\cal P}_s {\cal T}_d$ invariant N-bright solitons}
 \label{sec:PCTsol}

The explicit expression of the general N-soliton solution of the 
standard NLS model has been derived in the literature using a variety of methods, among them one can mention the direct method \cite{hirota} and the dressing method (see \cite{arxiv0} and references therein). The Hirota direct method provides \cite{hirota}
\br
\label{hirota}
\psi_N &=& \sqrt{\frac{2}{\eta}} \, \frac{f_N}{g_N}\\
f_N &\equiv &\sum_{\mu_{>}} \exp{(\sum_{j=1}^{2N}\mu_j \xi_j + \sum_{j<l}\mu_j \mu_l \theta_{jl})}\\
g_N &\equiv& \sum_{\mu_{=}} \exp{(\sum_{j=1}^{2N}\mu_j \xi_j + \sum_{j<l}\mu_j \mu_l \theta_{jl})}
\er
where 
\br
\xi_j &=&\Big\{ \begin{array}{ll}
k_j x + i k_j ^2 t + \xi_{0j}, & j= 1,2,...,N,\\
\bar{k}_j x - i \bar{k}_j^2 t + \bar{\xi}_{0j}, & j= N+1,...,2N;\,\,\,\,k_{N+j} \equiv \bar{k}_j,\,\,\,\,\xi_{0(N+j)} \equiv \bar{\xi}_{0j}
\end{array}
\\
\exp{(\theta_{jl})}&= &2 (k_j-k_l)^2,\,\,\,j,l=1,2,...,N\\
\exp{(\theta_{j(N+l)})}&= &\frac{1}{2 (k_j+\bar{k}_l)^2},\,\,\,j,l=1,2,...,N\\
\theta_{(N+j)(N+l)} &=&\bar{\theta}_{jl},
\er
such that the constant complex parameters $k_j, \, \xi_{0j},\,j=1,2,...,N$ are arbitrary.

The summations labeled by $\mu_{=}$ and $\mu_{>}$ must be done for all permutations $\mu_j = \{0 , 1\}$ such that
\br
\sum_{j=1}^{N} \mu_j =  \sum_{j=1}^{N} \mu_{N+j}
\er
and
\br
\sum_{j=1}^{N} \mu_j = 1 +  \sum_{j=1}^{N} \mu_{N+j},
\er
respectively.

The space-time reflection around the origin (\ref{ref0}) and the space-time translation (\ref{trxt})  symmetries, respectively, must be broken in order to construct a subset of solutions possessing the special space-time symmetry (\ref{par1})-(\ref{par2}). The two-soliton solutions satisfying (\ref{par1})-(\ref{par2}) have been constructed in \cite{jhep3, jhep5} for bright solitons and in \cite{jhep4} for dark solitons.  Below we will construct the N-bright soliton solution possessing the above symmetry (\ref{par2}) or (\ref{cpstd1}), i.e.
\br
\label{paritypsi}
\widetilde{{\cal P}} (\psi_{N-sol})= e^{i \delta_N}\, \bar {\psi}_{N-sol},\,\,\,\,\delta_{N} = 0.
\er
So, in this case the $N-$soliton solution $\psi_{N-sol}$ in (\ref{paritypsi}) would satisfy the special space-time reflection symmetry (\ref{par1})-(\ref{par2}) for $\delta = 0$.

The construction proceeds by introducing a set of new functions $\eta_j\,(j=1,2,...N)$ \cite{lou1}, as well as the special point $(x_\Delta, t_{\Delta})$,  defined by 
\br
\label{etas1}
\xi_j &=& k_j (x-x_{\Delta}) + i k^2_j (t-t_{\Delta}) + \eta_{0j} -\frac{1}{2} \sum_{i=1}^{j-1} \theta_{ij} - \frac{1}{2} \sum_{i=j+1}^{2N} \theta_{ji} \\
 &\equiv & \eta_j -\frac{1}{2} \sum_{i=1}^{j-1} \theta_{ij} - \frac{1}{2} \sum_{i=j+1}^{2N} \theta_{ji}.\label{etas2}
\er
Let us construct the ${\cal C}-{\cal P}_s-{\cal T}_d$ symmetric $N-$bright solitons order by order for the first three cases ($N=1,2,3$).
 
\subsubsection{1-bright soliton}
Let us write (\ref{hirota}) for the case $N=1$
\br
\label{1sol}
\psi_{1sol} &=& \sqrt{\frac{2}{\eta}} \, \frac{e^{\xi_1}}{1+ e^{\xi_1 + \xi_2 + \theta_{12}}}
\er
Next, introducing the new function $\eta_1$ through (\ref{etas1})-(\ref{etas2}) one has
\br
\xi_1 &=& k_1  \widetilde{x}  + i k_1^2  \widetilde{t}  + \eta_{01} - \frac{\theta_{12}}{2} \equiv \eta_1 - \frac{\theta_{12}}{2};\,\,\, \, \widetilde{x} = x-x_{\Delta},\, \widetilde{t} = t-t_{\Delta}\\
\xi_2 &=& \bar{k}_1 \widetilde{x} - i \bar{k}_1^2 \widetilde{t} + \bar{\eta}_{01} - \frac{\theta_{12}}{2} \equiv \bar{\eta}_1 - \frac{\theta_{12}}{2}.
\er
Then, (\ref{1sol}) can be written as
 \br
\psi_{1sol} &=& \sqrt{\frac{1}{2\eta}} \, e^{\frac{\eta_1-\bar{\eta}_1 -\theta_{12}}{2}} \, \mbox{sech}(\frac{\eta_1+\bar{\eta}_1}{2}).
\er
Taking  $k_1 = k_{1R} + i k_{1I}$, the last expression can be rewritten as 
\br
\psi_{1sol} = \sqrt{\frac{1}{2\eta}} \, e^{\frac{\eta_{01}-\bar{\eta}_{01} - \theta_{12}}{2}} e^{i[k_{1I}  \widetilde{x}  + (k_{1R}^2- k_{1I}^2)  \widetilde{t} ]}\, \mbox{sech}\Big[k_{1R}  \, \widetilde{x} - 2 k_{1I} k_{1R}  \, \widetilde{t}  + \frac{\eta_{01}+\bar{\eta}_{01}}{2}\Big].
\er
A relationship between the coordinates of the special point $(x_{\Delta}\,,\, t_{\Delta})$ can be determined by defining the initial position of the soliton, for   $\eta_{01}=\bar{\eta}_{01}=0$, to be located at $x_{01}$ for the initial time $t=0$. So the coordinates  $(x_{\Delta}\,,\, t_{\Delta})$ satisfy the relationship $x_{\Delta} = x_{01} + 2 k_{1I} t_{\Delta}$. 

Notice that for $\eta_{01}=\bar{\eta}_{01}=0$ one has that $\psi_{1sol}$ satisfies the special space-time symmetry property (\ref{par2}), i.e.  $\bar{\psi}_{1sol}(\widetilde{x}, \widetilde{t}) = \psi_{1sol}(-\widetilde{x}, -\widetilde{t})$.

\subsubsection{2-bright soliton}
The eq. (\ref{hirota}) for the case $N=2$ becomes
\br
\label{2sol}
\psi_{2sol} &=& \sqrt{\frac{2}{\eta}} \,\, \frac{e^{\xi_1}+e^{\xi_2} + e^{\xi_1+\xi_2+ \xi_3 + \theta_{12}+ \theta_{13}+\theta_{23}} + e^{\xi_1+\xi_2+ \xi_4 + \theta_{12}+ \theta_{14}+\theta_{24}} }{1+ e^{\xi_1 + \xi_3 + \theta_{13}} +e^{\xi_1 + \xi_4 + \theta_{14}} +e^{\xi_2 + \xi_4 + \theta_{24}} + e^{\xi_2 + \xi_3 + \theta_{23}} +  e^{\sum_{j=1}^4 \xi_j +  \sum_{j<l}^4 \theta_{jl}}}
\er
Next, introducing the new functions $\eta_1$\, and $\eta_2$ through (\ref{etas1})-(\ref{etas2}) one has
\br
\xi_1 &=& k_1  \widetilde{x}  + i k_1^2  \widetilde{t}  + \eta_{01} - \frac{\theta_{12}+\theta_{13}+\theta_{14}}{2} \equiv \eta_1 -  \frac{\theta_{12}+\theta_{13}+\theta_{14}}{2} ;\,\,\, \, \widetilde{x} = x-x_{\Delta},\, \widetilde{t} = t-t_{\Delta}\\
\xi_2 &=& k_2 \widetilde{x} + i k_2^2 \widetilde{t} + \eta_{02} - \frac{\theta_{12}+\theta_{23}+\theta_{24}}{2} \equiv \eta_2 - \frac{\theta_{12}+\theta_{23}+\theta_{24}}{2},\\
\xi_3 &=& \bar{k}_1  \widetilde{x}  - i \bar{k}_1^2  \widetilde{t}  + \bar{\eta}_{01} - \frac{\theta_{12}+\theta_{13}+\theta_{14}}{2} \equiv \bar{\eta}_1 -  \frac{\theta_{12}+\theta_{13}+\theta_{14}}{2} ,\\
\xi_4 &=& \bar{k}_2 \widetilde{x} - i \bar{k}_2^2 \widetilde{t} + \bar{\eta}_{02} - \frac{\theta_{12}+\theta_{23}+\theta_{24}}{2} \equiv \bar{\eta}_2 - \frac{\theta_{12}+\theta_{23}+\theta_{24}}{2},\,\,\,\,\,\, \theta_{34} = \bar{\theta}_{12}.
\er
Then, (\ref{2sol}) can be written as
\br
\psi_{2sol} &=& \sqrt{\frac{2}{\eta}} \times \nonumber
\\&&\Big\{\frac{\sqrt{R_2}\, e^{\frac{\eta_2-\eta_4}{2}} \, \cosh{(\frac{\eta_1+\eta_3}{2}-i\frac{\vp_2}{2})}+\sqrt{R_1} \, e^{\frac{\eta_1-\eta_3}{2}} \, \cosh{(\frac{\eta_2+\eta_4}{2}-i\frac{\vp_1}{2})} }{\cosh{(\frac{\eta_1+\eta_3+\eta_2+\eta_4}{2})} +\frac{|k_{1}+ \bar{k}_{2}|}{|k_1-k_2|^2}\cosh{(\frac{\eta_1+\eta_3-\eta_2-\eta_4}{2})} +\frac{4|k_{1R} k_{2R}|}{|k_1-k_2|^2} \cosh{(\frac{\eta_1-\eta_3-\eta_2+\eta_4}{2} )} }\Big\},\\
R_1 e^{i \vp_1} &\equiv& \exp{(\theta_{12}+\theta_{13}+\theta_{14})},\,\,R_2 e^{i \vp_2} \equiv \exp{(\theta_{12}+\theta_{23}+\theta_{24})}; \, R_1\in \IR_{+},\,R_2\in \IR_{+}.
\er
Taking  $k_j = k_{jR} + i k_{jI},\,\,\eta_{0j} = \eta_{0jR} + i \eta_{0jI}\,(j=1,2)$, the last expression becomes
\br
\psi_{2sol} &=& \sqrt{\frac{2}{\eta}} \,  \,\(\frac{F_2}{G_2} \)\\
 F_2 &\equiv & \sqrt{R_2} \, e^{i[k_{2I}\widetilde{x}+(k_{2R}^2-k_{2I}^2)\widetilde{t} +\eta_{02I}]} \, \cosh{(k_{1R}\widetilde{x}-2k_{1I} k_{1R} \widetilde{t}+\eta_{01R}-i\frac{\vp_2}{2})}+\\
&&\sqrt{R_1}  \, e^{i[k_{1I}\widetilde{x}+(k_{1R}^2-k_{1I}^2)\widetilde{t} +\eta_{01I}]} \, \cosh{(k_{2R}\widetilde{x}-2k_{2I} k_{2R} \widetilde{t}+\eta_{02R}-i\frac{\vp_1}{2})},\\
G_2 &\equiv& 
\cosh{[ (k_{1R}+k_{2R})\widetilde{x}-2(k_{1I} k_{1R} + k_{2I} k_{2R} )\widetilde{t}+\eta_{01R}+\eta_{02R}]} + \\
&& \frac{|k_{1}+ \bar{k}_{2}|}{|k_1-k_2|^2}
\cosh{[ (k_{1R}-k_{2R})\widetilde{x}-2(k_{1I} k_{1R} - k_{2I} k_{2R} )\widetilde{t}+\eta_{01R}-\eta_{02R}]} + \\
&& \frac{4|k_{1R} k_{2R}|}{|k_1-k_2|^2} \cos{[(k_{1I}-k_{2I})\widetilde{x}-(k_{1R}^2 -k_{1I}^2 + k_{2I}^2- k_{2R}^2 )\widetilde{t}+\eta_{01I}-\eta_{02I}]} .
\er

Let us write the conditions for the two solitons to be located some distance apart  at $t=0$. For $\eta_{0j}=\bar{\eta}_{0j}=0\,(j=1,2),$ one imposes the conditions, $\eta_1 + \eta_3 =0\,$\,and\, $\eta_2 + \eta_4 =0,$ evaluated at the coordinates $(x=x_{01}, t=0)$ and $(x=x_{02}, t=0)$, respectively. These conditions  provide a linear system of eqs. which can be solved for the special coordinates $(x_{\Delta}\,,\,t_{\Delta} )$, such that
\br
x_{\Delta} &=& \frac{k_{1I} x_{02} - k_{2I} x_{01}}{k_{1I}-k_{2I}}\\
t_{\Delta} &=& \frac{1}{2} (\frac{ x_{02} - x_{01}}{k_{1I}-k_{2I}}).
\er

Therefore, for $\eta_{0j}=\bar{\eta}_{0j}=0\,(j=1,2)\,$,  one has that $\widetilde{{\cal P}}(F_2)= \bar{F}_2,\,\widetilde{{\cal P}}(G_2)= \bar{G}_2$, and so, $\psi_{2sol}$ satisfies the special space-time symmetry property (\ref{par2}), i.e.  $\bar{\psi}_{2sol}(\widetilde{x}, \widetilde{t}) = \psi_{2sol}(-\widetilde{x}, -\widetilde{t})$. 

\subsubsection{3-bright soliton}

The eq. (\ref{hirota}) for the case $N=3$ becomes
\br
\label{3sol}
\psi_{3sol} &=& \sqrt{\frac{2}{\eta}} \,  \,\(\frac{f_3}{g_3} \),
\er
where $f_3$ and $g_3$ are provided in appendix \ref{ap:3sol}. Similarly as the above constructions, introducing the new functions $\eta_j\,(j=1,2,3)$ through (\ref{etas1})-(\ref{etas2}) one can rewrite $g_3$ and $f_3$ as
\br 
\label{g3}
\frac{g_3}{2} &=& e^{\frac{\sum_{j} \eta_j}{2}} G_3,\\
\label{f3}
\frac{f_3}{2} &=& e^{\frac{\sum_{j} \eta_j}{2}} F_3,
\er
where $G_3$ and $F_3$ are provided in the appendix  \ref{ap:3sol}.

So, the  3-soliton solution can be written as
\br
\label{3solsym}
\psi_{3sol} = \frac{F_3}{G_3}.
\er
Let us analyze the properties of the arguments appearing into the $\exp{(.)}$ and $\cosh{(.)}$ functions composing the expressions of $G_3$ and $F_3$ in (\ref{g31}) and (\ref{f31}), respectively. Notice that the combinations of type $\frac{(\eta_{j}+\eta_{3+j})}{2}\, (j=1,2,3)$, are real functions, whereas the type $\frac{(\eta_{j}-\eta_{3+j})}{2}\, (j=1,2,3)$, are purely imaginary functions; namely
\br
\frac{(\eta_{1}-\eta_{4})}{2} &=& i \Big[k_{1I}  \widetilde{x}  + (k_{1R}^2- k_{1I}^2)  \widetilde{t}\Big] + \frac{\eta_{01}-\bar{\eta}_{01}}{2},\,\,\,\,  \frac{(\eta_{1}+\eta_{4})}{2}  =  \Big[k_{1R}  \, \widetilde{x} - 2 k_{1I} k_{1R}  \, \widetilde{t}\Big] + \frac{\eta_{01}+\bar{\eta}_{01}}{2}\\
\frac{(\eta_{2}-\eta_{5})}{2} &=& i \Big[k_{2I}  \widetilde{x}  + (k_{2R}^2- k_{2I}^2)  \widetilde{t} \Big]+ \frac{\eta_{02}-\bar{\eta}_{02}}{2},\,\,\,\,  \frac{(\eta_{2}+\eta_{5})}{2}  =   \Big[k_{2R}  \, \widetilde{x} - 2 k_{2I} k_{2R}  \, \widetilde{t} \Big] + \frac{\eta_{02}+\bar{\eta}_{02}}{2}\\
\frac{(\eta_{3}-\eta_{6})}{2} &=& i \Big[k_{3I}  \widetilde{x}  + (k_{3R}^2- k_{3I}^2)  \widetilde{t} \Big] + \frac{\eta_{03}-\bar{\eta}_{03}}{2},\,\,\,\,  \frac{(\eta_{3}+\eta_{6})}{2}  =   \Big[k_{3R}  \, \widetilde{x} - 2 k_{3I} k_{3R}  \, \widetilde{t}  \Big]+ \frac{\eta_{03}+\bar{\eta}_{03}}{2}.
\er 

Therefore, taking $\eta_{0j}=\bar{\eta}_{0j}=0\,(j=1,2,3)\,$ one has that  $\widetilde{{\cal P}}(F_3)= \bar{F}_3$\, and $\widetilde{{\cal P}}(G_3)= \bar{G}_3$, as can be directly verified using  (\ref{f31}) and (\ref{g31}), respectively. So,  $\psi_{3sol}$ satisfies the special space-time symmetry property (\ref{par2}), i.e.  $\bar{\psi}_{3sol}(\widetilde{x}, \widetilde{t}) = \psi_{3sol}(-\widetilde{x}, -\widetilde{t})$. 

For the 3-soliton solution let us consider a configuration such that each soliton is located some distance apart from the others at $t=0$. So, for $\eta_{0j}=\bar{\eta}_{0j}=0\,(j=1,2,3),$ one imposes the conditions, $\eta_1 + \eta_4 =0\,,\eta_2 + \eta_5 =0$\,and\, $\eta_3 + \eta_6 =0,$ evaluated at the coordinates $(x=x_{0j}, t=0) \, (j=1,2,3)$, respectively. These conditions  provide a linear system of eqs. which can be solved for the special coordinates $(x_{\Delta}\,,\,t_{\Delta} )$, i.e.
\br
x_{\Delta} &=& \frac{k_{1I} x_{02} - k_{2I} x_{01}}{k_{1I}-k_{2I}}\\
t_{\Delta} &=& \frac{1}{2} (\frac{ x_{02} - x_{01}}{k_{1I}-k_{2I}}).
\er
The position of the third soliton at  $t=0$ becomes $x_{03}=\frac{k_{1I} x_{02} - k_{2I} x_{01}+k_{3I} (x_{01} - x_{02})}{k_{1I}-k_{2I}}$.

Then, following the above examples for the  ${\cal C}-{\cal P}_s-{\cal T}_d$ invariant $N=1,2,3$ solitons, the general expression (\ref{hirota}) for $N-$soliton can be written in terms of ${\cal C}-{\cal P}_s-{\cal T}_d$ invariants, such as hyperbolic and exponential functions (as in the expressions of $G_3$ and $F_3$, see (\ref{g31}) and (\ref{f31})); however the final expression is not very instructive and we will not present here. So, according to the above constructions, we can show that the general $N-$soliton solution (\ref{hirota}) can be conveniently rewritten in order to satisfy  
\br
\label{ChPhT}
{\cal C} \widetilde{{\cal P}}\Big[\psi_{Nsol}(\widetilde{x}, \widetilde{t})\Big] = \psi_{Nsol}(\widetilde{x}, \widetilde{t})  \Rightarrow  {\cal C}\Big[{\psi}_{Nsol}(\widetilde{x}, \widetilde{t})\Big] = \psi_{Nsol}(-\widetilde{x}, -\widetilde{t}).
\er

Notice that, for the above special $N-$bright  soliton solutions sector satisfying (\ref{ChPhT}), i.e. for the solutions invariant under the symmetry  ${\cal C}-{\cal P}_s-{\cal T}_d$  and which break the space-time translation symmetry (\ref{trxt}), the standard NLS  (\ref{nls0}) can be rewritten as  
\br
\label{nonloc}
i \frac{\partial}{\partial \widetilde{t}} \psi(\widetilde{x},\widetilde{t}) +   \frac{\partial^2}{\partial \widetilde{x} ^2} \psi(\widetilde{x},\widetilde{t}) + 2 \eta   \Big[\psi(-\widetilde{x},-\widetilde{t}) \psi(\widetilde{x},\widetilde{t})\Big]  \psi(\widetilde{x},\widetilde{t}) =  0,\,\,\,\,\psi(\widetilde{x},\widetilde{t}) \in C;\,\,\,\, \eta > 0, 
\er
where the space-time translations (\ref{trxt}),  for $x_0=-x_{\Delta}\,$ and $\,t_0 = -t_{\Delta}$, have been performed and the second relationship of (\ref{ChPhT}) has been used in the standard NLS eq (\ref{nls0}).

The equation (\ref{nonloc}) is precisely the one of the list  studied in \cite{ablowitz, reverse}, the so-called reverse space-time nonlocal NLS \cite{reverse}. So,  one can argue that the standard NLS model, in the sector described by  $N-$bright solitons possessing ${\cal C}{\cal P}_s{\cal T}_d$  symmetry invariance and broken space-time translation symmetry,  belongs to  the family of non-local generalization of the NLS model considered in the recent literature. 

New types of non-local integrable models have been introduced in \cite{alice} in the context of Alice-Bob KdV models. In fact, the equation (\ref{nonloc}) belongs to a family of the so-called Alice-Bob NLS (AB-NLS) models recently introduced in the literature \cite{lou1}. For example, among the  AB-NLS models the following family of equations have been put forward \cite{lou1, kchen}
\br
\label{abnls1}
i \pa_t A + \pa_x^2 A + \frac{1}{2} \eta (A+B)[2 A \bar{A} + \bar{B} (A-B)] &=&0 \\
 B&=& \hat{f}_k (A),\,\,\, \hat{f}_k \equiv \{ {\cal P},\, {\cal C} {\cal T}, \, {\cal C} {\cal P} {\cal T}\}.\label{abnls2}
\er
In fact, the model  (\ref{nonloc}) belongs to the above system (\ref{abnls1})-(\ref{abnls2}), in the $N-$soliton sector of the model,  provided that the operator $\hat{f}_3$ is taken into account, i.e. $B =  {\cal C} {\cal P} {\cal T} (A)$.

Regarding the quasi-conservation laws of the standard NLS described in section \ref{sec:NLSqc} one must conclude that all the anomalies will vanish upon integration in space-time for the N-soliton satisfying  (\ref{ChPhT}), since the relevant anomaly densities possess odd parities for soliton configurations satisfying the parity symmetry (\ref{par2}). Consequently, their associated  charge densities will be  asymptotically conserved charges of the standard NLS model. So, the above results show an example of an analytical, and not only numerical, demonstration of the vanishing of the space-time integrals of the anomalies associated to the infinite towers of infinitely many quasi-conservation laws in soliton theory.

We believe that, for deformed NLS models, the existence of asymptotically conserved charges associated to infinite towers of  
infinitely many  quasi-conservation laws reflects, as in the integrable soliton theories, in the special dynamics of the deformed model, in such a way that the solitary wave solutions emerge from the scattering region basically as they have entered it.

The above patterns will be qualitatively reproduced below in our numerical simulations of the relevant anomalies for the 2-bright soliton interactions of the deformed NLS model (\ref{mnls}), for a variety of soliton configurations and a wide range of values of the deformation parameter $\epsilon$.

According to the Liouville's integrability criterion one must have an infinite number of conservation laws whose conserved charges are in involution \cite{das, faddeev}. In this context, the existence of the novel towers of asymptotically conserved charges as above, even in the standard NLS model, is restricted to special soliton configurations satisfying the symmetry property (\ref{par2}).  Even though those anomalous charges are infinitely many, one can not use them  in order to match to the number of degrees of freedom of the NLS model. In fact, the true conserved charges hold for general field configurations, being solitonic or not. So far, the only explanation,  which we have put forward above, for the relationship between the anomalous charges and the set of true conserved charges of the standard NLS model is the mechanism of anomaly cancellation in order to construct an exact conservation law out of the relevant quasi-conservation laws. The consequences for the dynamics of the solitons of the deformed model and their mutual interactions remain to be investigated.

In the context of harmonic analysis it has been introduced the method of {\sl almost conservation laws} ($I$-method) for integrable systems \cite{tao}. In particular, it has been considered the KdV model and provided a proof on how the so-called {\sl almost conservation laws} can be used to recover infinitely many conserved charges that make the model an integrable system. We hope our results to be useful for some analysts in order to establish more definitive statements about the role played by the above asymptotically conserved charges for general field configurations and, then, provide some clarifications on the properties of the deformations of integrable systems.  

\section{Numerical simulations}

\label{simul}
 
In this section we will compute the space and space-time integrals of the anomaly densities $\hat{\alpha}_1$,  $\hat{\beta}_2$,  $\hat{\gamma}_2$ and  $\hat{\delta}_1$, appearing in (\ref{an1d}), (\ref{beta2d}), (\ref{gam33d}) and (\ref{ano4}), respectively, for three type of two-soliton collisions in the deformed NLS model (\ref{mnls}). The deformed NLS model (\ref{nlsd}) with potential (\ref{pot10}) possesses a solitary wave solution (\ref{solitary}). Then, we will take two one-bright solitary waves located some distance apart as the initial condition for our numerical simulations.  So,  the collision of bright solitons in the deformed  NLS equation (\ref{nlsd}) will be simulated  numerically by considering the initial condition $\psi_0(x,0)$ defined as
\br
\label{nls2}
\psi_0(x,0) =  \Big[\frac{2 + \epsilon}{2 \eta} \frac{\rho^2_1}{\cosh^2{[(1+ \epsilon) \rho_1 (x -x_{01})]}}\Big]^{\frac{1}{2(1+ \epsilon)}} \,\, e^{i  \frac{v_1}{2} x}+ \Big[\frac{2 + \epsilon}{2 \eta} \frac{\rho^2_2}{\cosh^2{[(1+ \epsilon) \rho_2 (x  -x_{02})]}}\Big]^{\frac{1}{2(1+ \epsilon)}} \,\, e^{i  \frac{v_2}{2} x},
\er  
where two one-bright soliton solutions of the deformed  NLS  model have been located at $x_{01}$ and $x_{02}$, respectively. Notice that for equal amplitudes $\rho_1=\rho_2$ and equal and opposite velocities $v_1 = - v_2$ one has a space-reflection symmetric complex function, $\psi_0(-x) = \psi_0(x)$, provided that $x_{01} = - x_{02} = x_{\Delta}$.  

The domain of simulation in the $x-$coordinate is considered to be ${\cal D} =[-L,L] $ with $L=15$, mesh size $h = 0.017$ and time step $\tau = 0.00011$. The length  $L$ is chosen such that the effect of the extreme regions near the points $x = \pm L$ do not interfere the dynamics of the solitons, i.e. the boundary conditions (\ref{bcs}) are satisfied for each time step. In our numerical simulations we have used the so-called time-splitting cosine pseudo-spectral finite difference (TSCP) method  \cite{bao1, bao2}.
 
 The two solitons are initially centered at $x_{01}$ and $x_{02}$, respectively. Notice that the direction of motion of each soliton is related to the sign of its phase slope. In addition, we will consider initially well-separated solitons, such that the parameters difference  $(x_{01}-x_{02})$ is chosen to be  several times the width of the solitons ($\sim \frac{1}{(1+ \epsilon)\rho_i},\,i=1,2$)  and $(x_{01}-x_{02}) <  2 L$. So, the initial condition considers  two deformed NLS bright solitary waves  which are stitched together at the middle point, and then we allow  the scattering of them, absorbing the radiation at the edges of the grid. It amounts to maintain the vanishing boundary conditions  (\ref{bcs}) at the edges of the grid for each time step of the numerical simulation.

\subsection{Two-bright solitons: different amplitudes and opposite velocities}

Let us consider two solitons initially centered at $\pm x_0$ ($x_0>0$), the soliton centered initially at $-x_0$ ($t=0$) moves to the right with velocity $v_2 > 0$, whereas the soliton initially ($t=0$) centered  at $x_0$ travels to the left with velocity ($v_1 <0$).  In the Fig.1 one presents the numerical simulation for the collision of two solitons  with amplitudes $|\psi_1| = 6.83$ and $|\psi_2| = 4.44$ and velocities $v_2=-v_1 = 10$ for $\epsilon= -0.06$ and $x_0 = 6$.

In the Fig. 2 we present the simulations for the anomaly density $\hat{\alpha}_1$ (\ref{an1d}) for the two soliton collision of Fig. 1.  In the top figures we plot the density anomaly as follows: in the top-left side one plots this density for three successive times, $t_i$, before collision (green), $t_c$, collision (blue ), and, $t_f$, after collision (red) times, respectively; whereas, in the top-right side we plot the density for the collision time $t_c$. In the bottom figures we present the anomaly integrals versus $t$. In the bottom left side it is plotted the $x-$integration of the anomaly versus  $t$; whereas in the bottom right side it is plotted the $(x,t)-$integration of the anomaly versus $t$.   Notice that the $(x,t)-$integration of the anomaly  $\hat{\alpha}_1$ vanishes, within numerical accuracy. In fact, the bottom right side of Fig. 2 shows a vanishing function of time with an error of order $10^{-9}$.  

In the Fig. 3 we present the simulations for the anomaly density $\hat{\beta}_2$ (\ref{beta2d}) for the two soliton collision of Fig. 1.  In the top-left  one plots the anomaly density for three successive times, $t_i$, before collision (green), $t_c$, collision (blue ), and, $t_f$, after collision (red) times, respectively. In the top right figure we present the $x-$integration of the anomaly versus  $t$; whereas in the bottom right side it is plotted the $(x,t)-$integration of the anomaly versus $t$.   Notice that the $(x,t)-$integration of the anomaly  $\hat{\beta}_2$ vanishes, within numerical accuracy. In fact, the bottom right side of Fig. 3 shows a vanishing function of time with an error of order $10^{-14}$.  

In the Fig. 4 we present the simulations for the anomaly density $\hat{\gamma}_2$ (\ref{gam33d}) for the two soliton collision of Fig. 1.  In the top-left we present the anomaly density for three successive times, $t_i$, before collision (green), $t_c$, collision (blue ), and, $t_f$, after collision (red) times, respectively. In the top right figure we present the $x-$integration of the anomaly versus  $t$; whereas in the bottom right side it is plotted the $(x,t)-$integration of the anomaly versus $t$.   Notice that the $(x,t)-$integration of the anomaly  $\hat{\gamma}_2$ vanishes, within numerical accuracy. In fact, the bottom right side of Fig. 4 shows a vanishing function of time with an error of order $10^{-7}$. 

Similarly, in the Fig. 5 we present the simulations for the anomaly density $\hat{\delta}_1$ (\ref{ano4}) for the two soliton collision of Fig. 1.  In the top-left we present the anomaly density for three successive times, $t_i$, before collision (green), $t_c$, collision (blue ), and, $t_f$, after collision (red) times, respectively. In the top right figure we present the $x-$integration of the anomaly versus  $t$; whereas in the bottom right side it is plotted the $(x,t)-$integration of the anomaly versus $t$.   Notice that the $(x,t)-$integration of the anomaly  $\hat{\delta}_1$ vanishes, within numerical accuracy. In fact, the bottom right side of Fig. 5 shows a vanishing function of time with an error of order $10^{-9}$. 

\begin{figure}
\centering
\label{fig1}
\includegraphics[width=4cm,scale=6, angle=0,height=6cm]{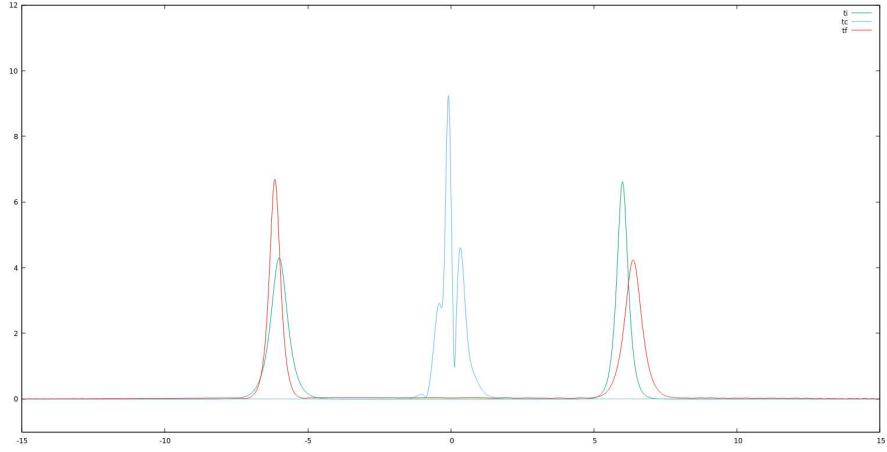} 
\parbox{6in}{\caption{(color online) 2-bright solitons with amplitudes $|\psi_1| = 6.83$ and $|\psi_2| = 4.44$ and velocities $v_2 = - v_1 = 10$ for $\epsilon = - 0.06$, for initial (green), collision (blue) and final (red) successive times. }}
\end{figure}

\begin{figure}
\centering
\label{fig2}
\includegraphics[width=0.5cm,scale=4, angle=0,height=3.5cm]{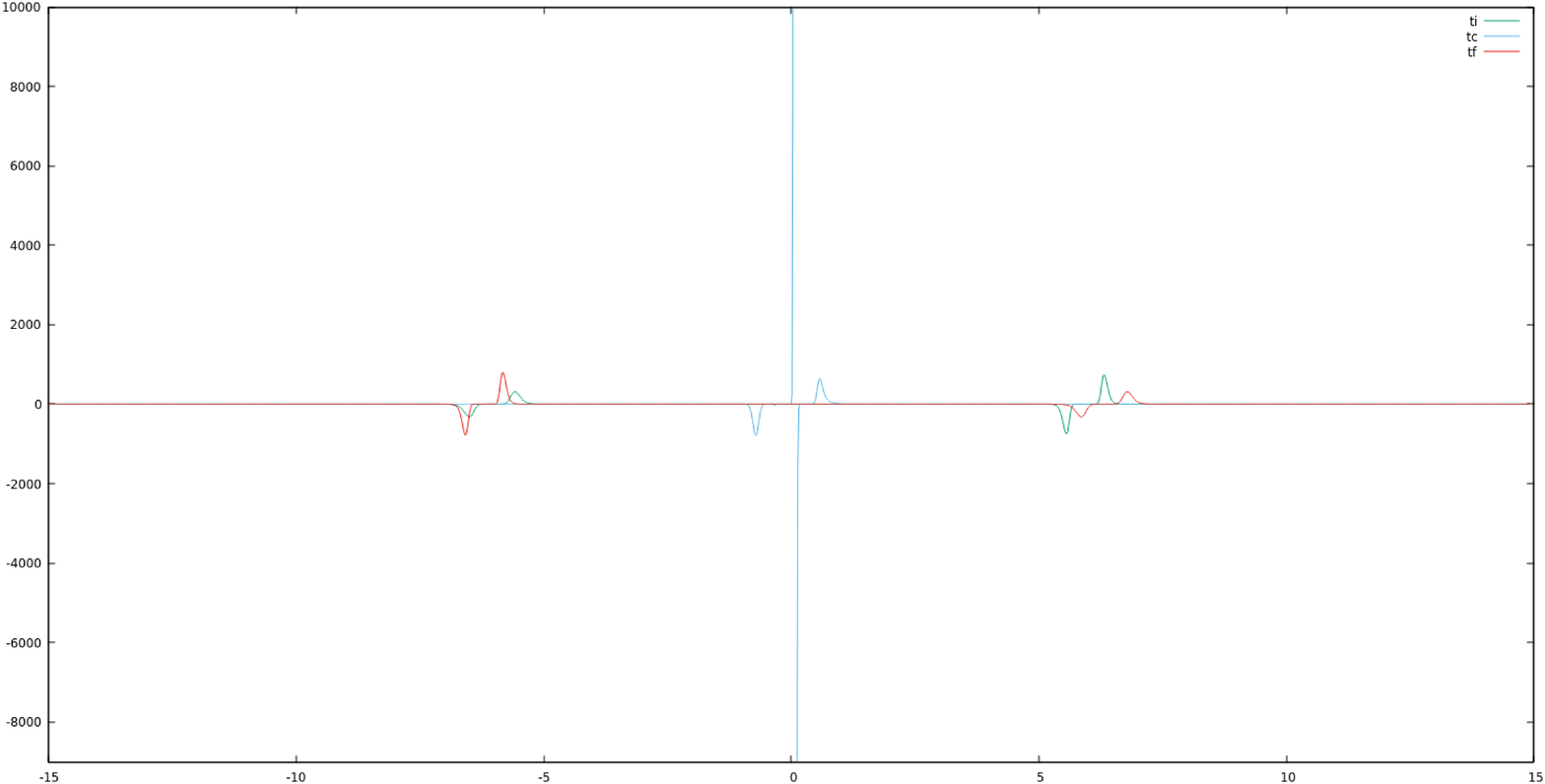} 
\includegraphics[width=0.5cm,scale=4, angle=0,height=3.5cm]{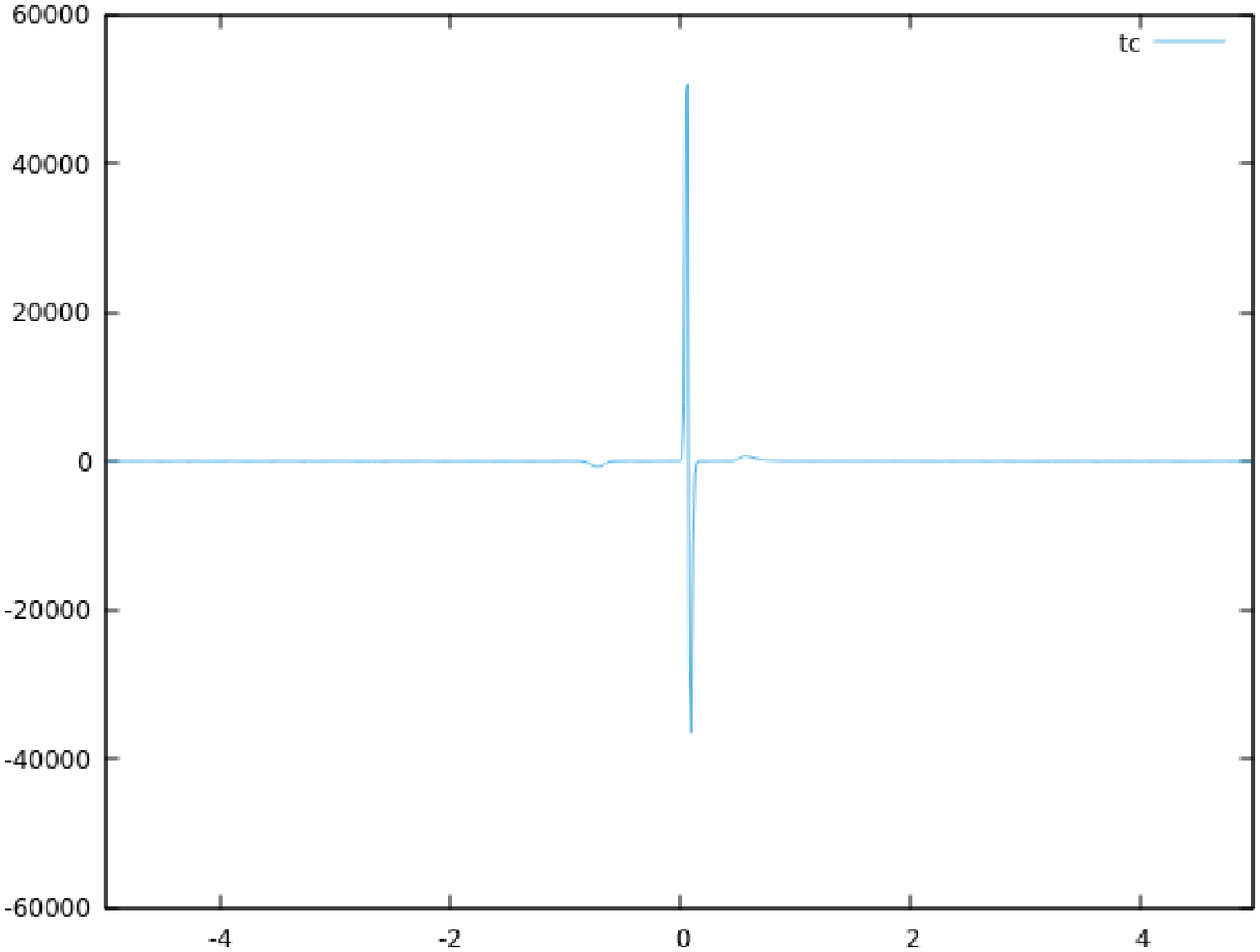} 
\includegraphics[width=0.5cm,scale=4, angle=0,height=3.5cm]{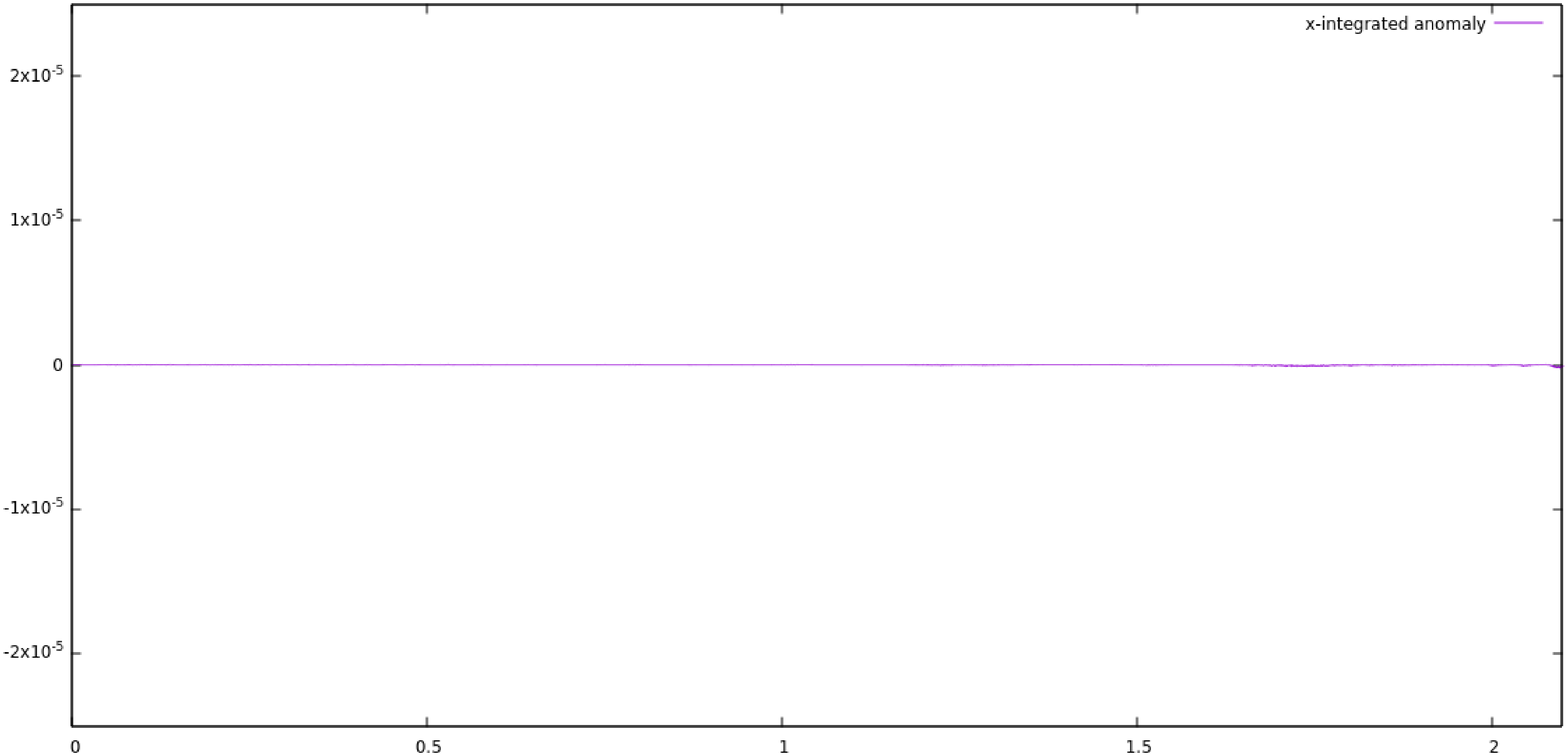}
\includegraphics[width=0.5cm,scale=4, angle=0,height=3.5cm]{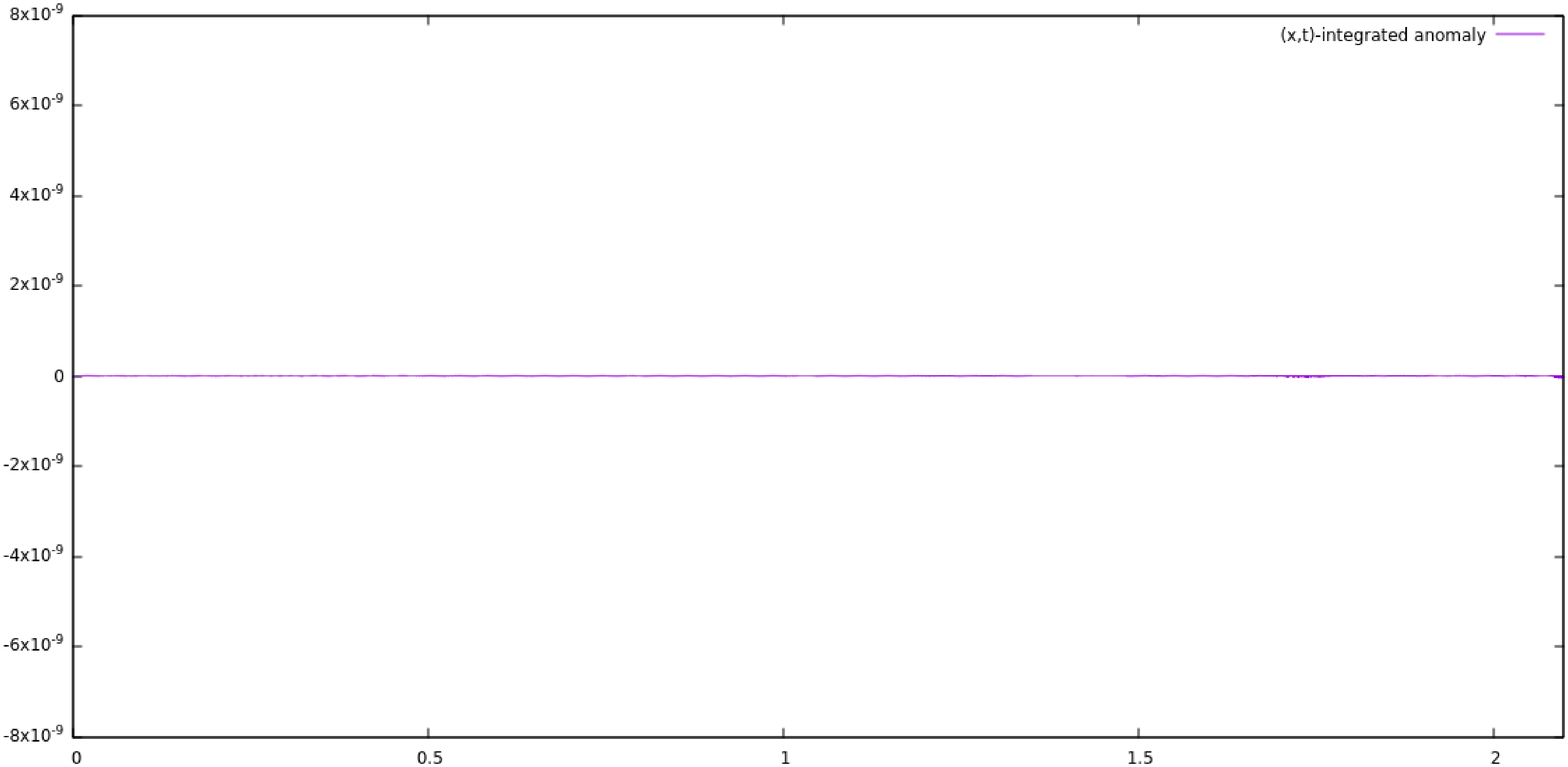}  
\parbox{6in}{\caption{(color online) Top figures show the anomaly $\hat{\alpha}_1\,\, vs\, \,x$ for the 2-soliton collision of Fig. 1. Top left shows the  initial (green), collision (blue) and final (red) times of the density profile. The top right shows the anomaly density profile for collision time. Bottom left shows the plot $\int \hat{\alpha}_1 dx\,\, vs\,\, t$ and the bottom right shows the plot $\int dt \int dx\, \hat{\alpha}_1\,\, vs\,\, t$.}}
\end{figure}

\begin{figure}
\centering
\label{fig3}
\includegraphics[width=1.5cm,scale=4, angle=0,height=3.5cm]{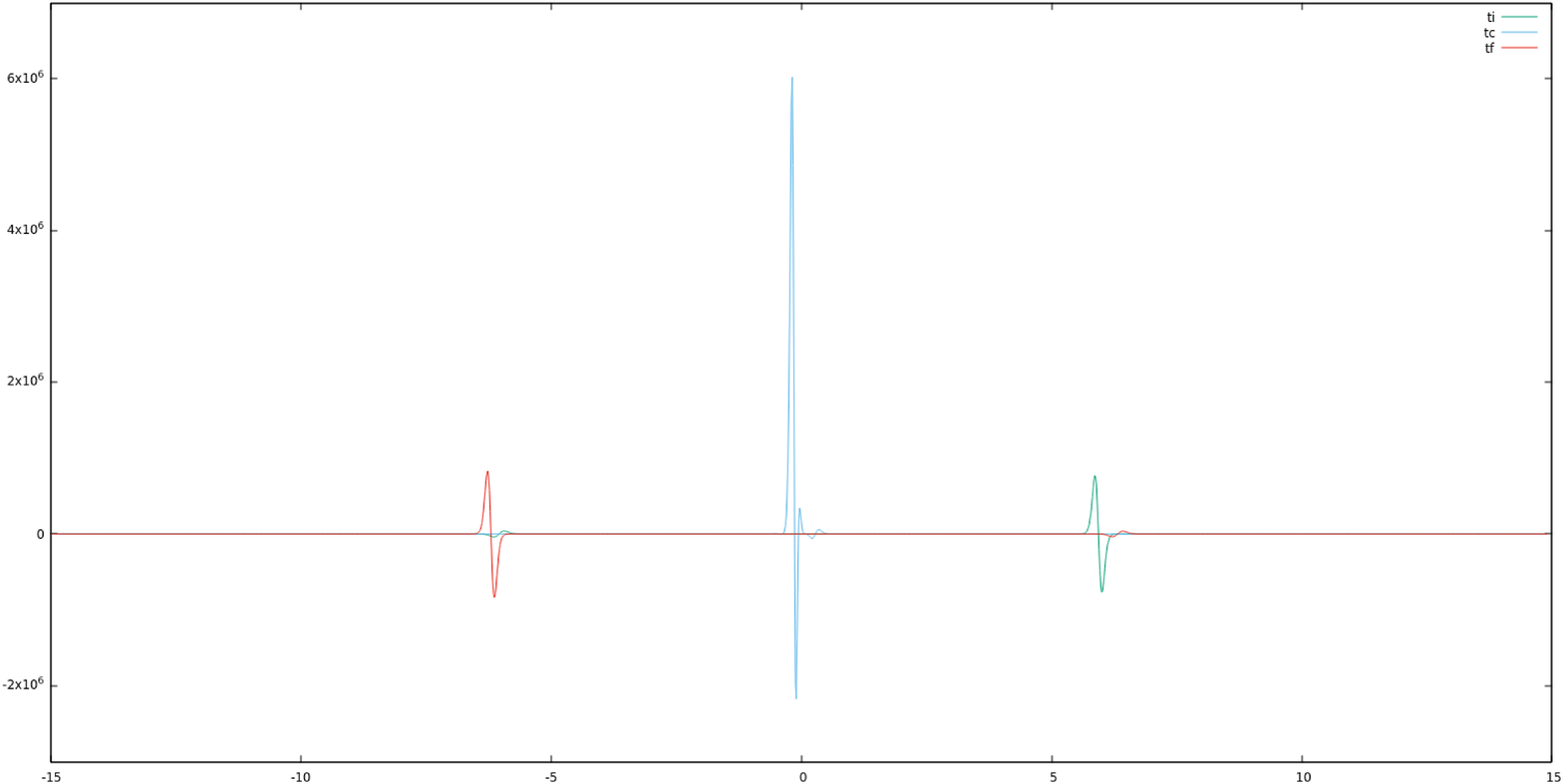} 
\includegraphics[width=1.5cm,scale=4, angle=0,height=3.5cm]{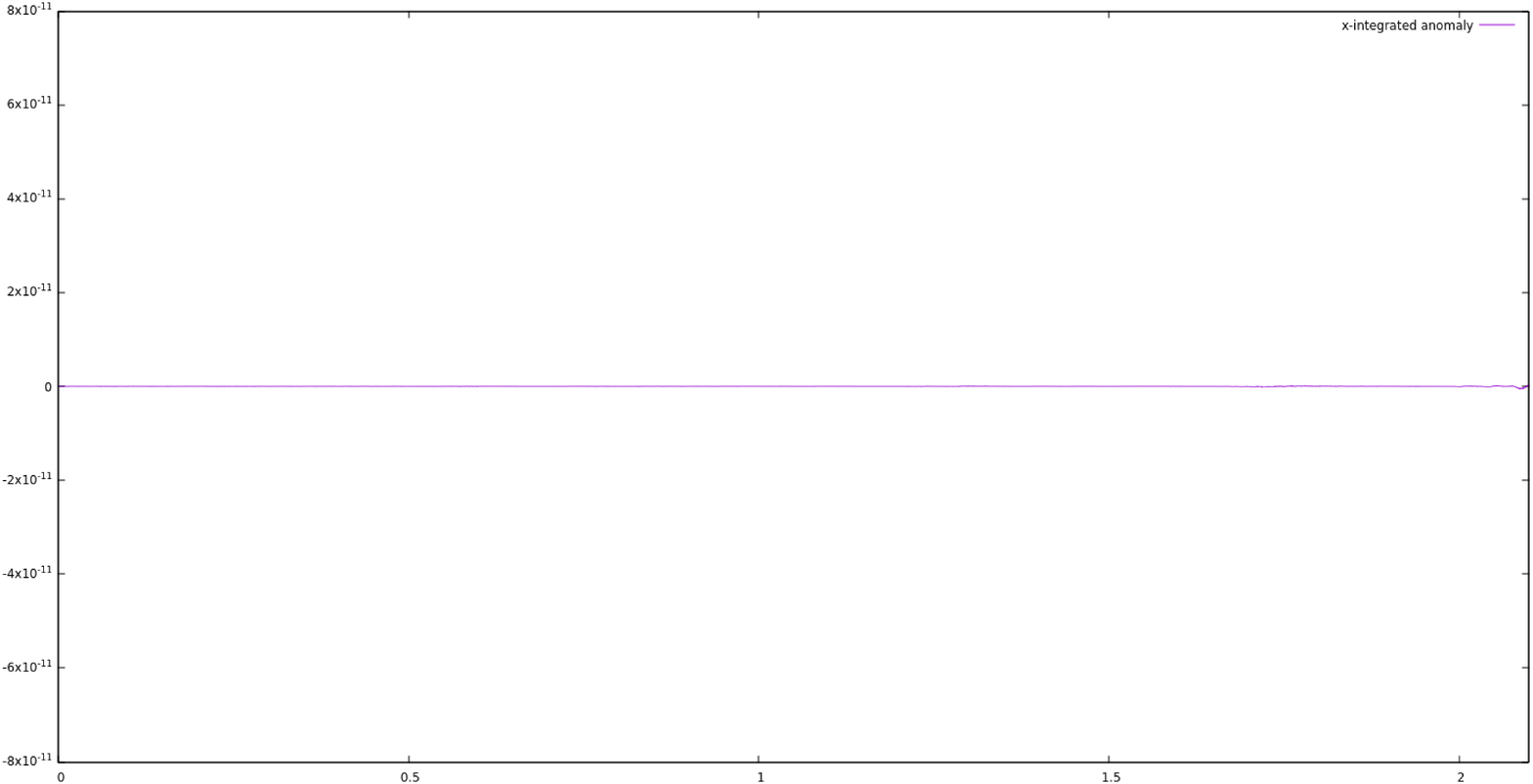}
\includegraphics[width=1.5cm,scale=4, angle=0,height=3.5cm]{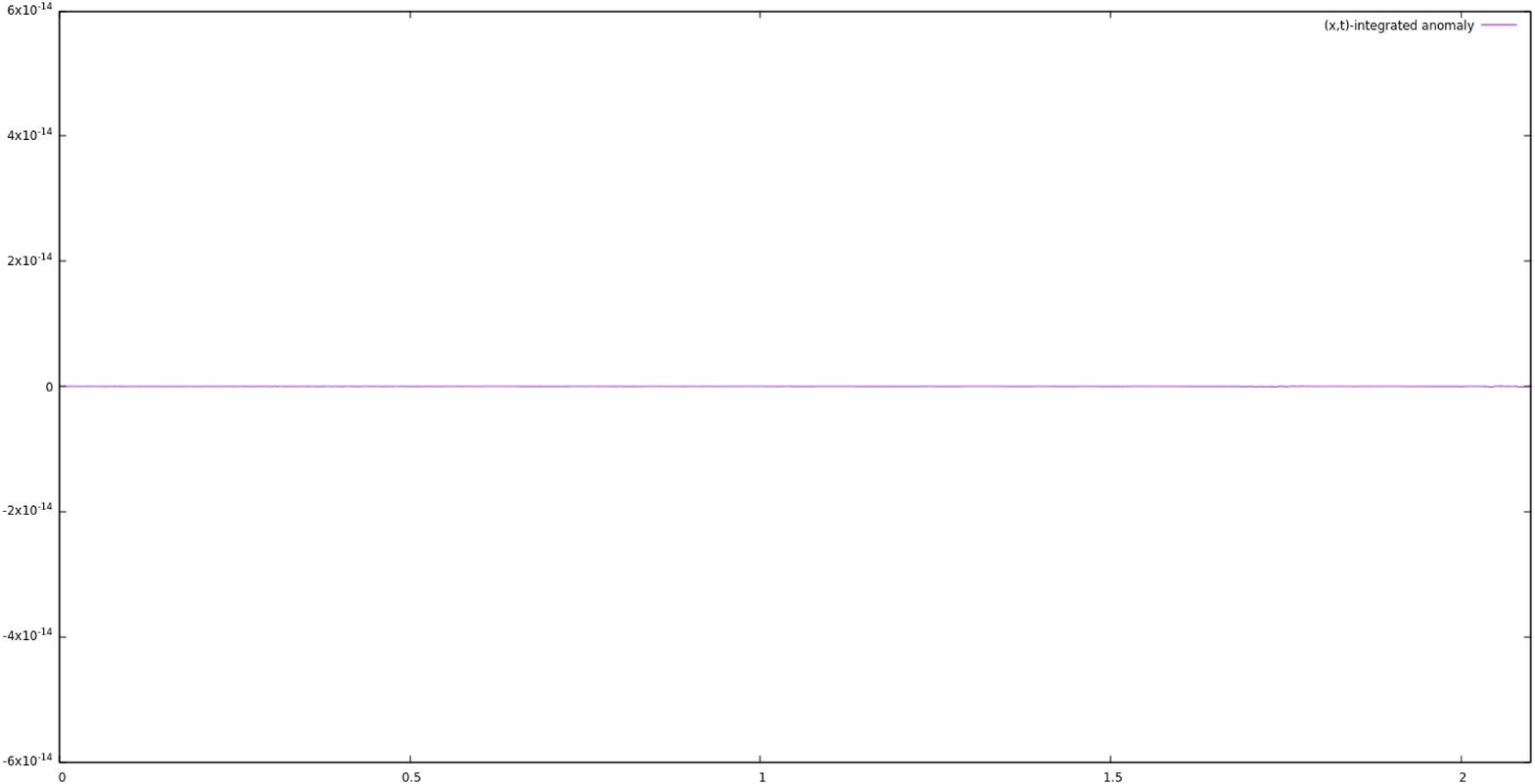}  
\parbox{6in}{\caption{(color online) Top left shows the  profile at initial (green), collision (blue) and final (red) times of the anomaly density $\hat{\beta}_2 $ for the 2-soliton collision of Fig. 1. The top right shows the plot $\int \hat{\beta}_2 dx\,\, vs\,\, t$ and the bottom right shows the plot $\int dt \int dx\, \hat{\beta}_2\,\, vs\,\, t$.}}
\end{figure} 

\begin{figure}
\centering
\label{fig4}
\includegraphics[width=1.5cm,scale=4, angle=0,height=3.5cm]{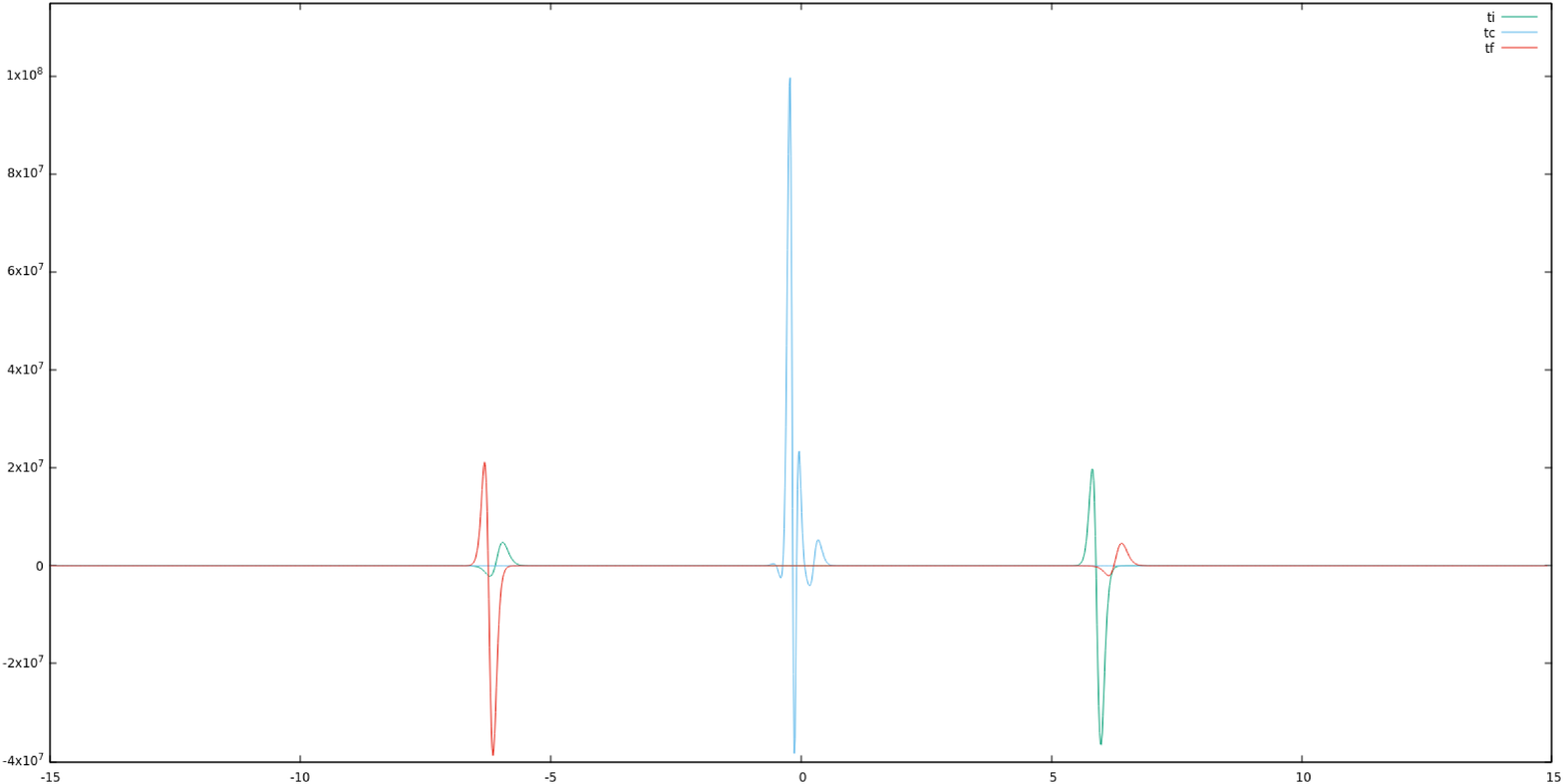} 
\includegraphics[width=1.5cm,scale=4, angle=0,height=3.5cm]{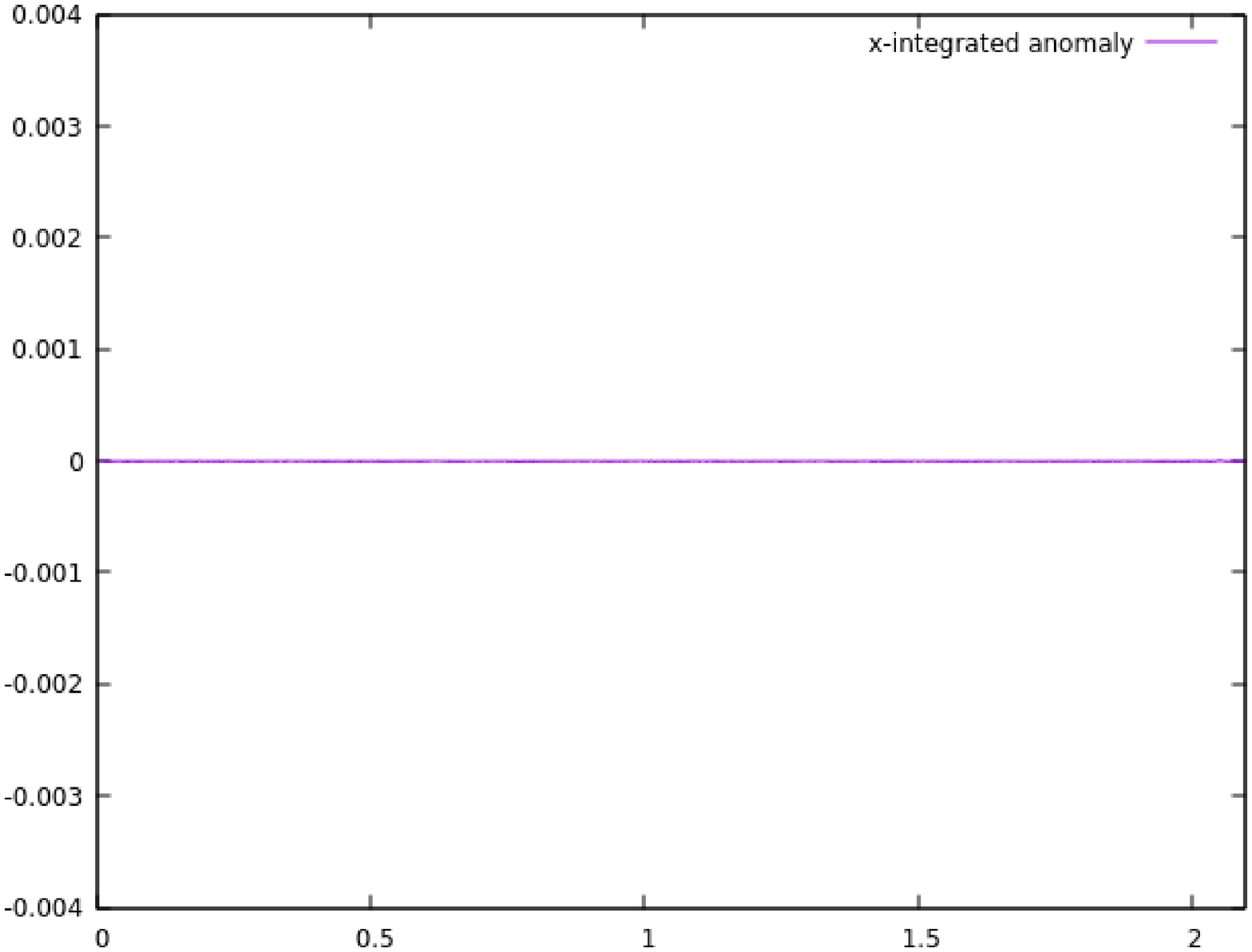}
\includegraphics[width=1.5cm,scale=4, angle=0,height=3.5cm]{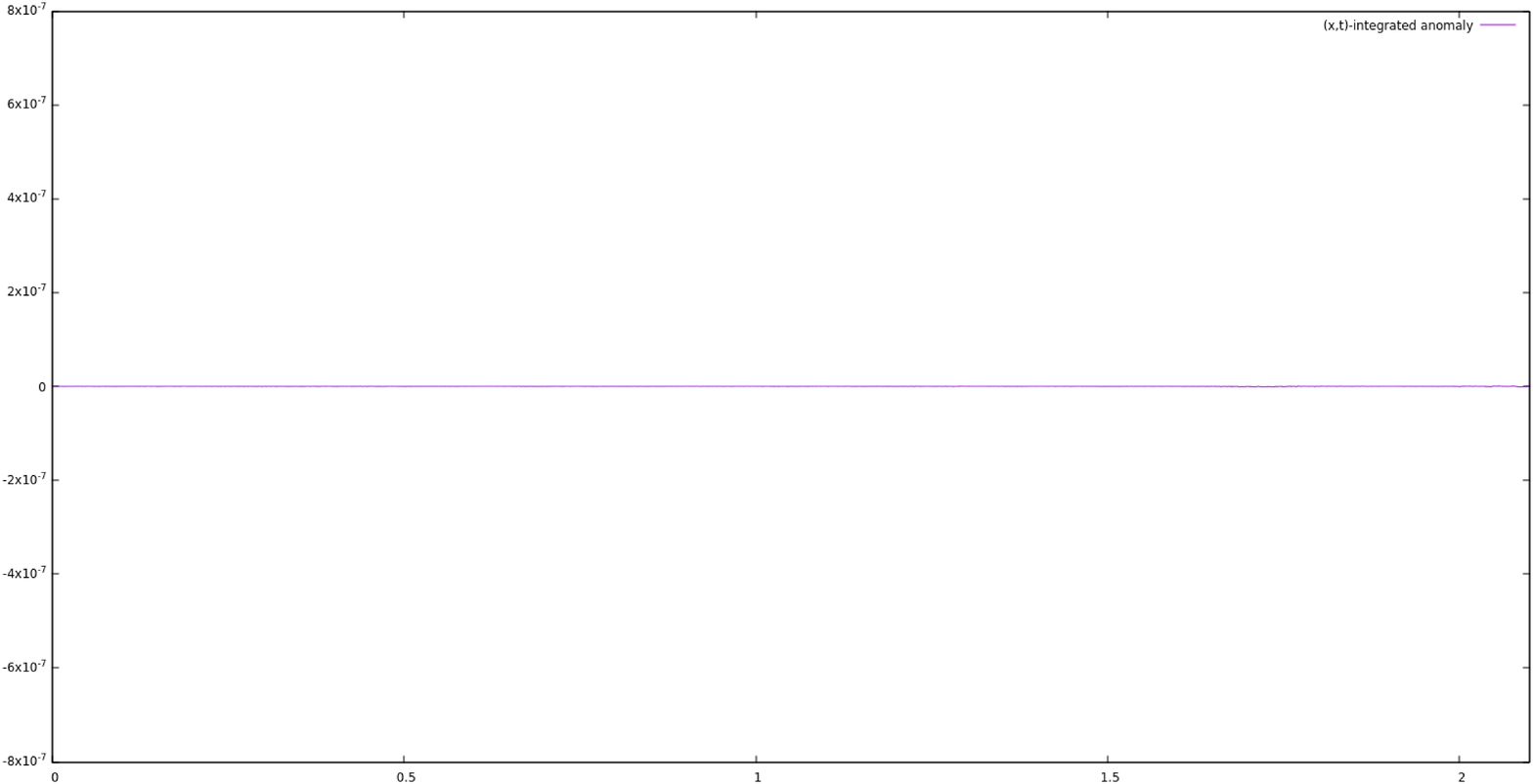}  
\parbox{6in}{\caption{(color online)Top left shows the  profile at initial (green), collision (blue) and final (red) times of the anomaly density $\hat{\gamma}_2$ for the 2-soliton collision of Fig. 1 . The top right shows the plot $\int \hat{\gamma}_2 dx\,\, vs\,\, t$ and the bottom right shows the plot $\int dt \int dx\, \hat{\gamma}_2\,\, vs\,\, t$.}}
\end{figure} 

\begin{figure}
\centering
\label{fig5}
\includegraphics[width=1.5cm,scale=4, angle=0,height=3.5cm]{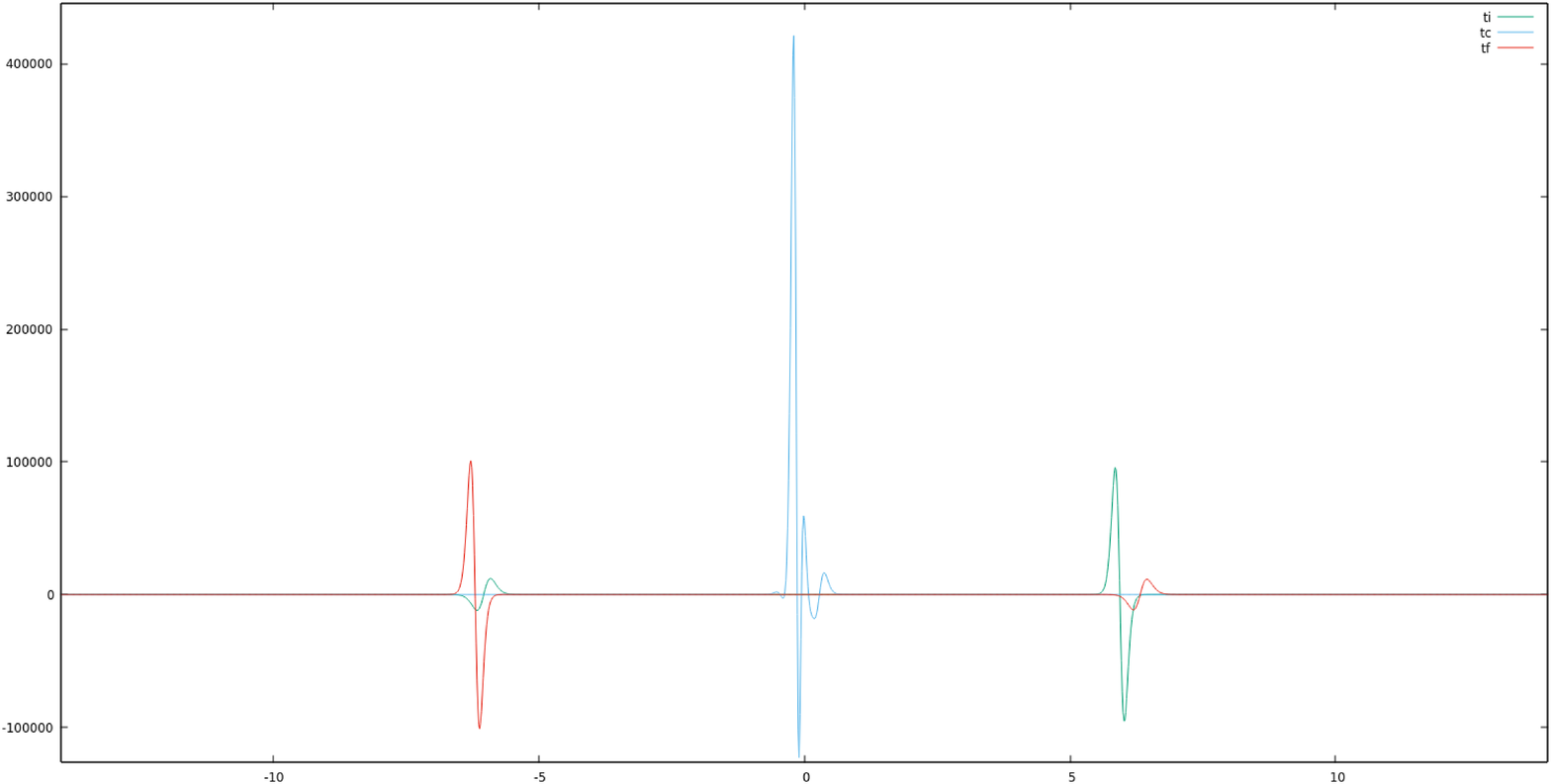} 
\includegraphics[width=1.5cm,scale=4, angle=0,height=3.5cm]{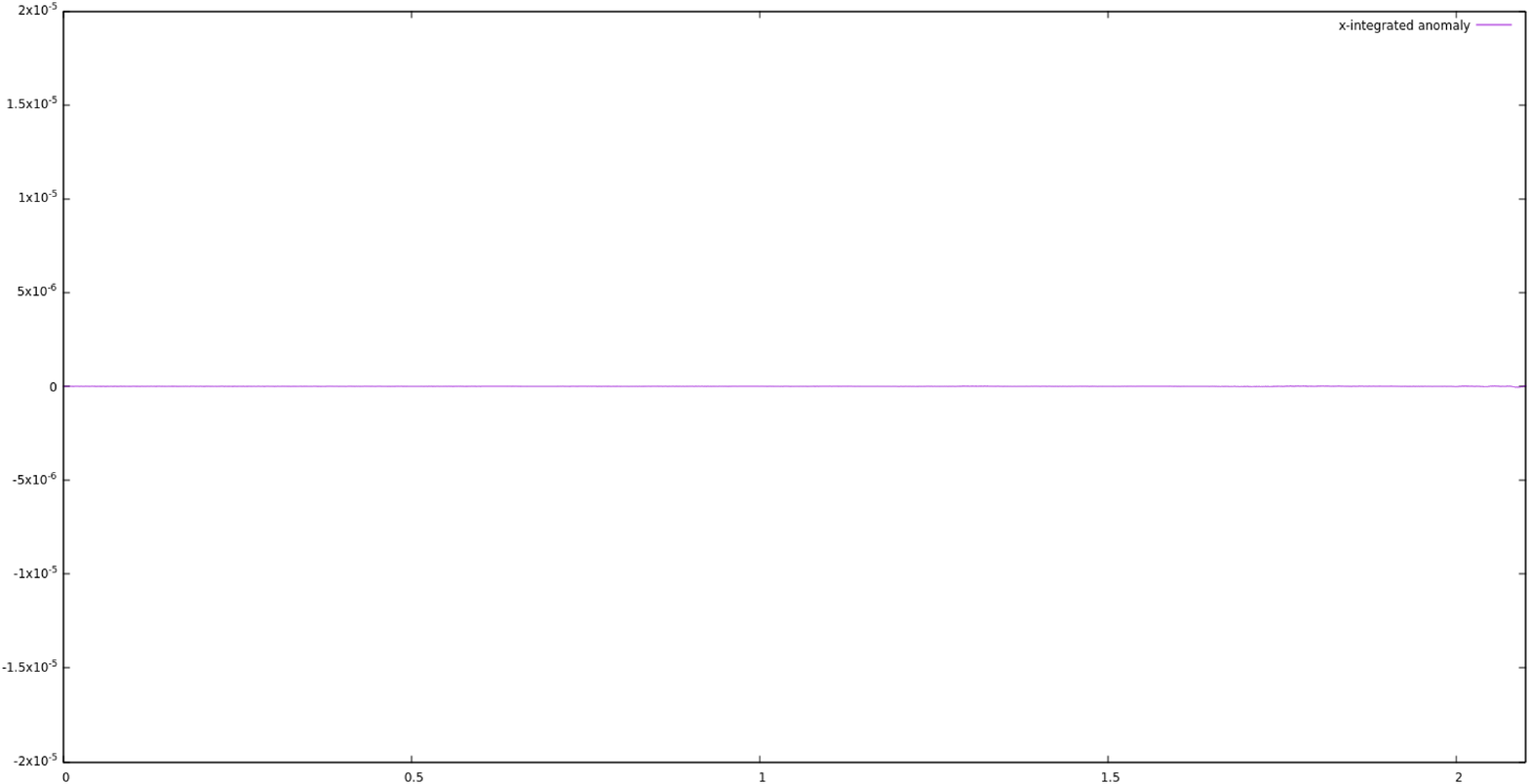}
\includegraphics[width=1.5cm,scale=4, angle=0,height=3.5cm]{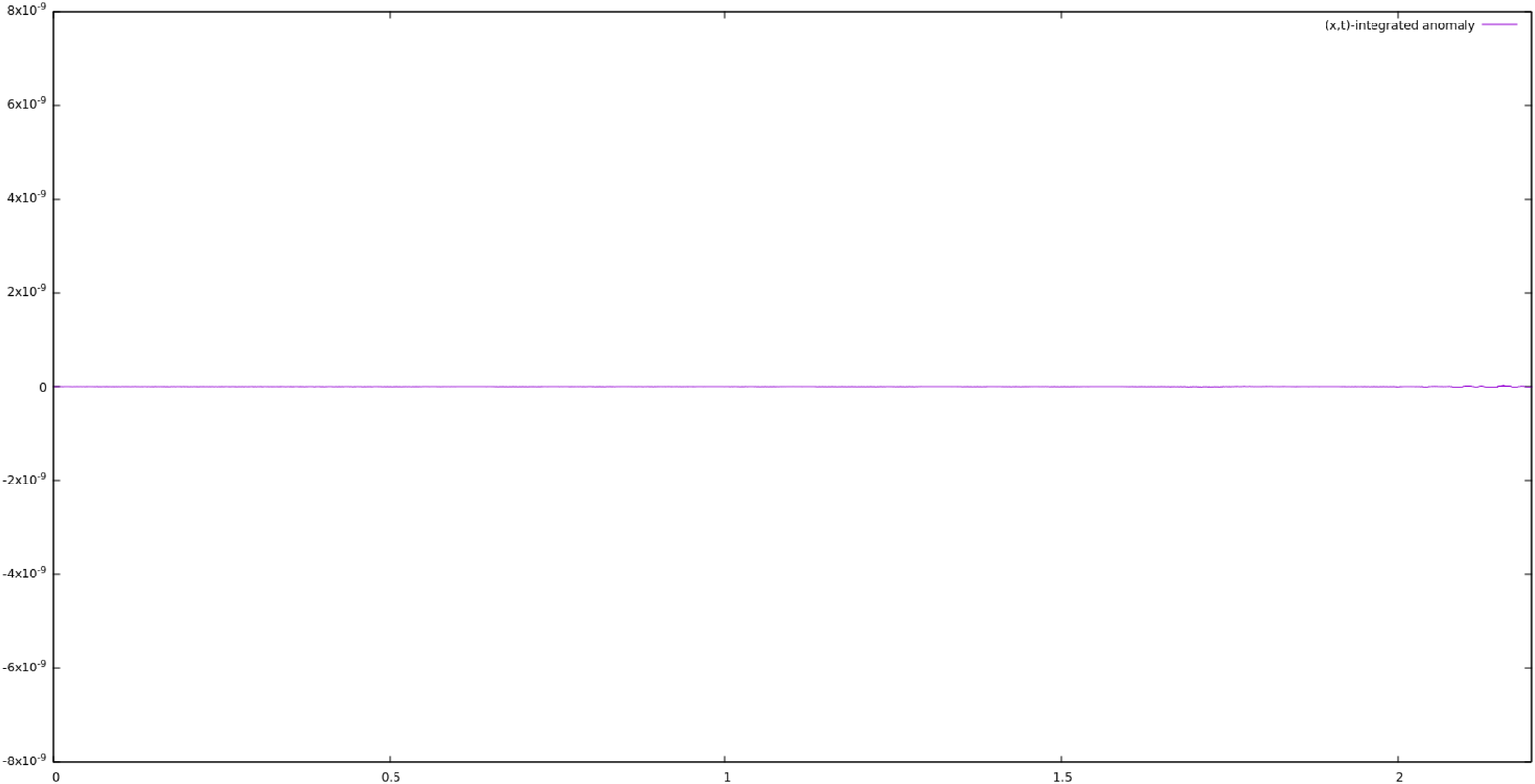}  
\parbox{6in}{\caption{(color online) Top left shows the  profile at initial (green), collision (blue) and final (red) times of the anomaly density $\hat{\delta}_1$ for the 2-soliton collision of Fig. 1. The top right shows the plot $\int \hat{\delta}_1 dx\,\, vs\,\, t$ and the bottom right shows the plot $\int dt \int dx\, \hat{\delta}_1\,\, vs\,\, t$.}}
\end{figure}

\subsection{Two-bright solitons: equal amplitudes and opposite velocities}

Let us consider two solitons with equal amplitudes initially centered at $\pm x_0$ ($x_0>0$), the soliton centered initially at $-x_0$ ($t=0$) moves to the right with velocity $v > 0$, whereas the soliton initially ($t=0$) centered  at $x_0$ travels to the left with velocity ($-v$).  In the Fig. 6 one presents the numerical simulation for the collision of two solitons  with amplitudes $|\psi_1| = |\psi_2| = 6.83$  and velocities $v_2=-v_1 = 10$ for $\epsilon= -0.06$ and $x_0 = 6$.

In the Fig. 7 we present the simulations for the anomaly density $\hat{\alpha}_1$ (\ref{an1d}) for the two soliton collision of Fig. 6.  In the left figure we plot the density anomaly for three successive times, $t_i$, before collision (green), $t_c$, collision (blue ), and, $t_f$, after collision (red) times, respectively. In the middle figure it is plotted the $x-$integration of the anomaly versus  $t$; whereas in the right figure it is plotted the $(x,t)-$integration of the anomaly versus $t$.  Notice that the $(x,t)-$integration of the anomaly  $\hat{\alpha}_1$ vanishes, within numerical accuracy, with an error of order $10^{-8}$.  

In the Fig. 8 we present the simulations for the anomaly density $\hat{\beta}_2$ (\ref{beta2d}) for the two soliton collision of Fig. 6.  In the left it is plotted the anomaly density for three successive times, $t_i$, before collision (green), $t_c$, collision (blue ), and, $t_f$, after collision (red) times, respectively. In the middle figure it is plotted the $x-$integration of the anomaly versus $t$; whereas in the right figure it is plotted the $(x,t)-$integration of the anomaly versus $t$. Notice that the $(x,t)-$integration of the anomaly  $\hat{\beta}_2$ vanishes, within numerical accuracy,  with an error of order $10^{-14}$.  

In the Fig. 9 we present the simulations for the anomaly density $\hat{\gamma}_2$ (\ref{gam33d}) for the two soliton collision of Fig. 6. In the left it is plotted the anomaly density for three successive times, $t_i$, before collision (green), $t_c$, collision (blue ), and, $t_f$, after collision (red) times, respectively. In the middle figure it is plotted the $x-$integration of the anomaly versus $t$; whereas in the right figure it is plotted the $(x,t)-$integration of the anomaly versus $t$. Notice that the $(x,t)-$integration of the anomaly  $\hat{\gamma}_2$ vanishes, within numerical accuracy,  with an error of order $10^{-7}$. 

Similarly, in the Fig. 10 we present the simulations for the anomaly density $\hat{\delta}_1$ (\ref{ano4}) for the two soliton collision of Fig. 6.  In the left it is plotted the anomaly density for three successive times, $t_i$, before collision (green), $t_c$, collision (blue ), and, $t_f$, after collision (red) times, respectively. In the middle figure it is plotted the $x-$integration of the anomaly versus $t$; whereas in the right figure it is plotted the $(x,t)-$integration of the anomaly versus $t$.  Notice that this space-time integration  vanishes, within numerical accuracy,  with an error of order $10^{-9}$. 

\begin{figure}
\centering
\label{fig6}
\includegraphics[width=1cm,scale=4, angle=0,height=5cm]{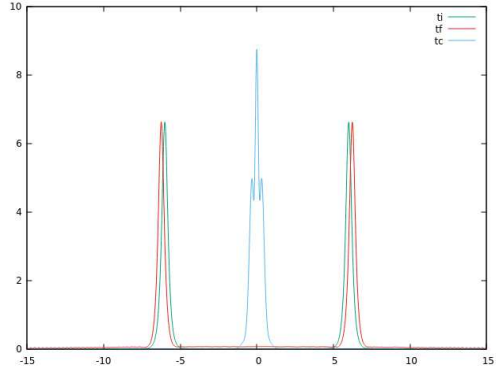} 
\parbox{6in}{\caption{(color online) Two-bright solitons with equal amplitudes and (opposite) velocites for  initial (green), collision (blue) and final (red) succesive times. }}
\end{figure}

\begin{figure}
\centering
\label{fig7}
\includegraphics[width=1.5cm,scale=4, angle=0,height=3.2cm]{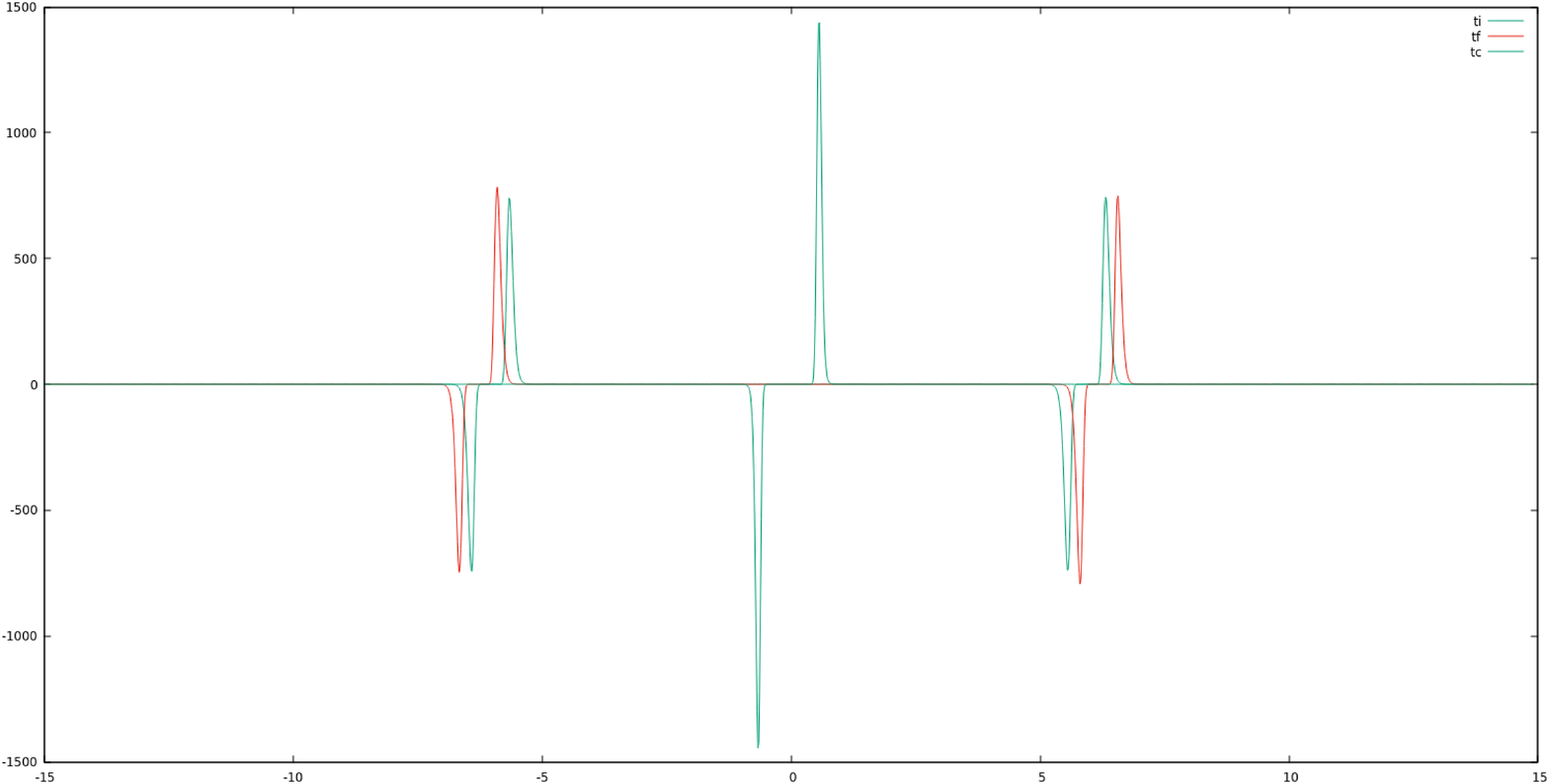} 
\includegraphics[width=1.5cm,scale=4, angle=0,height=3.2cm]{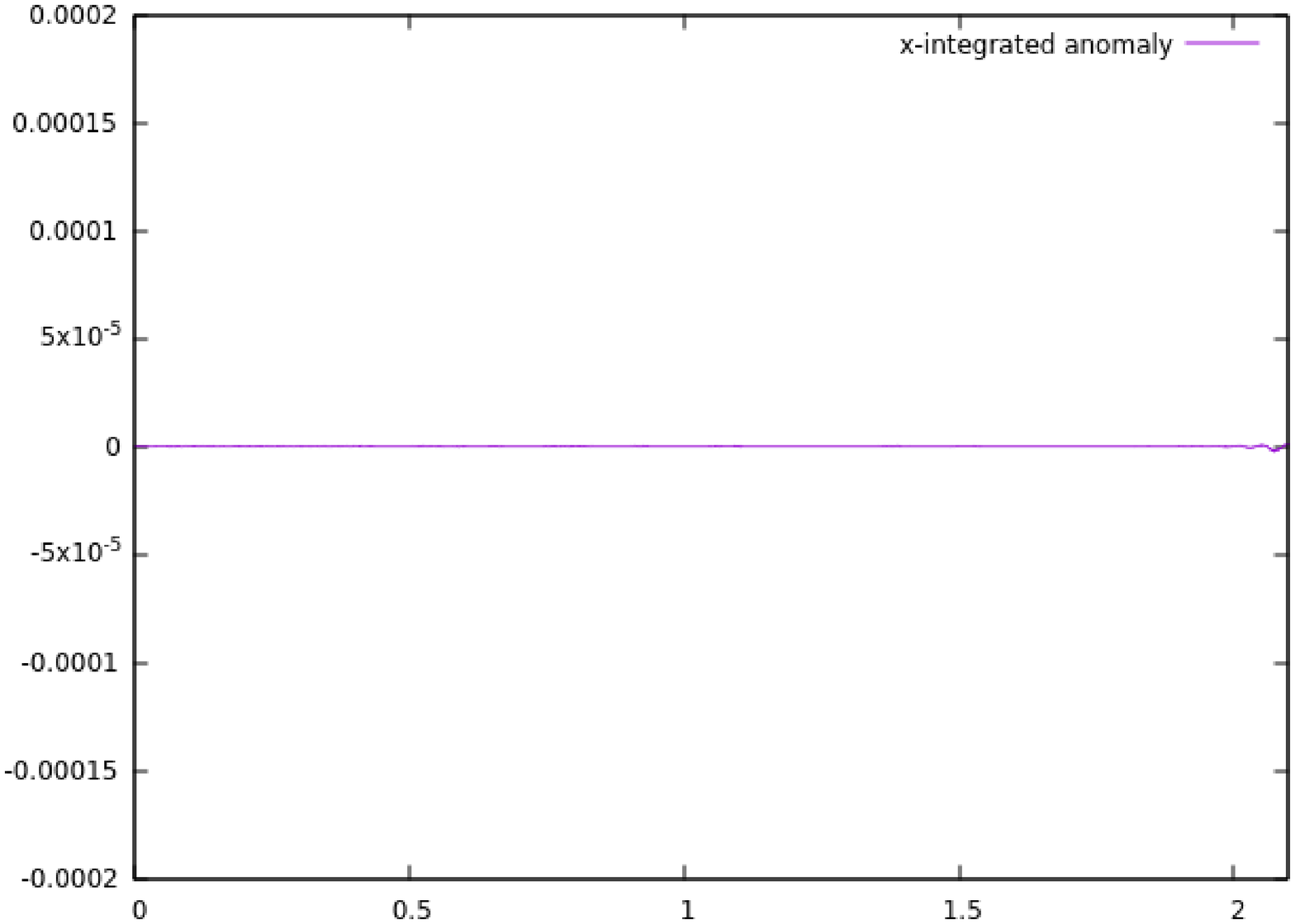}
\includegraphics[width=1.5cm,scale=4, angle=0,height=3.2cm]{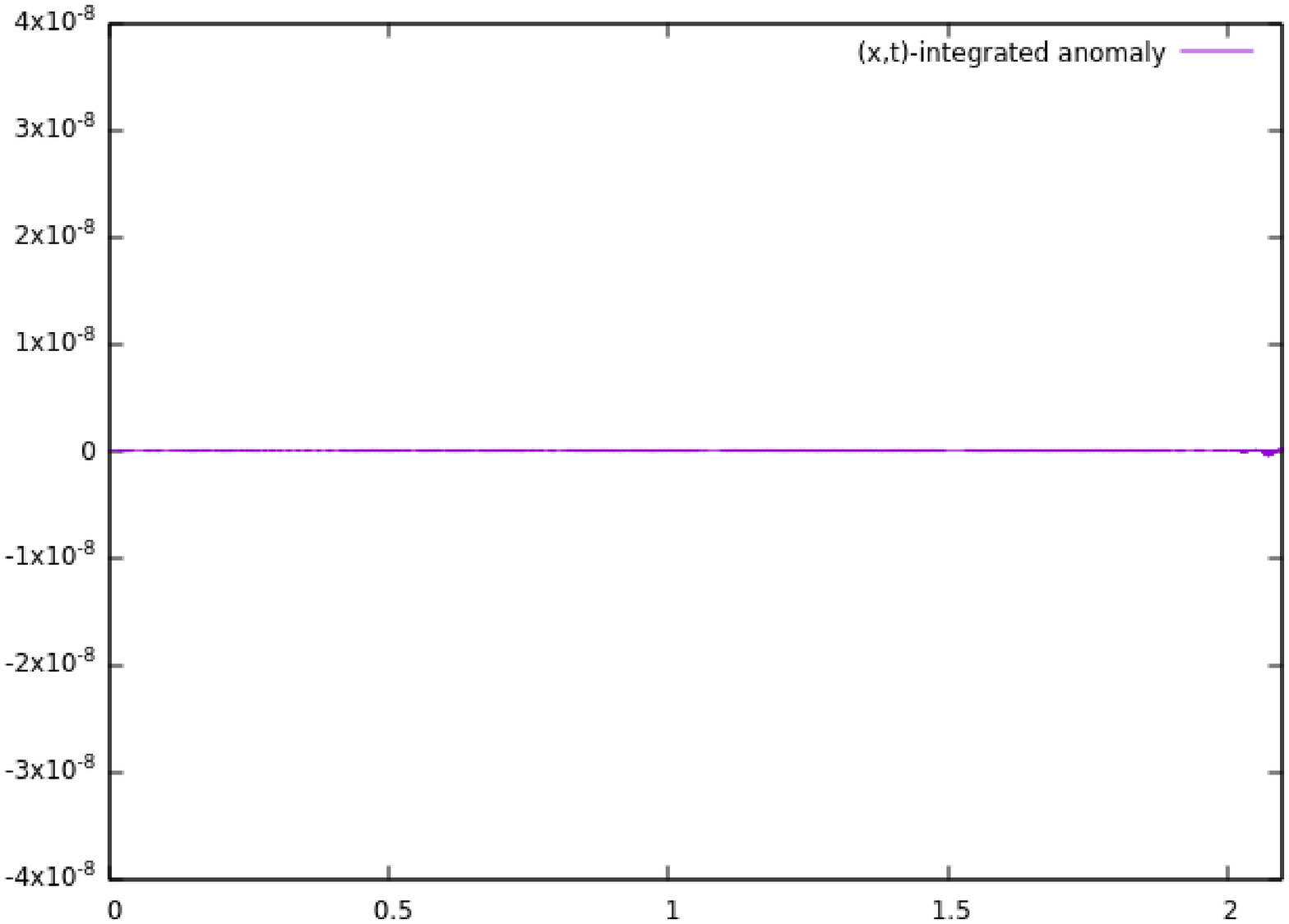}  
\parbox{6in}{\caption{(color online) Left figure shows the density profile of anomaly $\hat{\alpha}_1\,\, vs\, \,x$ for the 2-soliton collision of Fig. 6 for initial (green), collision (blue) and final (red) times. Middle figure shows the plot $\int \hat{\alpha}_1 dx\,\, vs\,\, t$ and the right shows the plot $\int dt \int dx\, \hat{\alpha}_1\,\, vs\,\, t$.}}
\end{figure} 

\begin{figure}
\centering
\label{fig8}
\includegraphics[width=1.5cm,scale=4, angle=0,height=3.2cm]{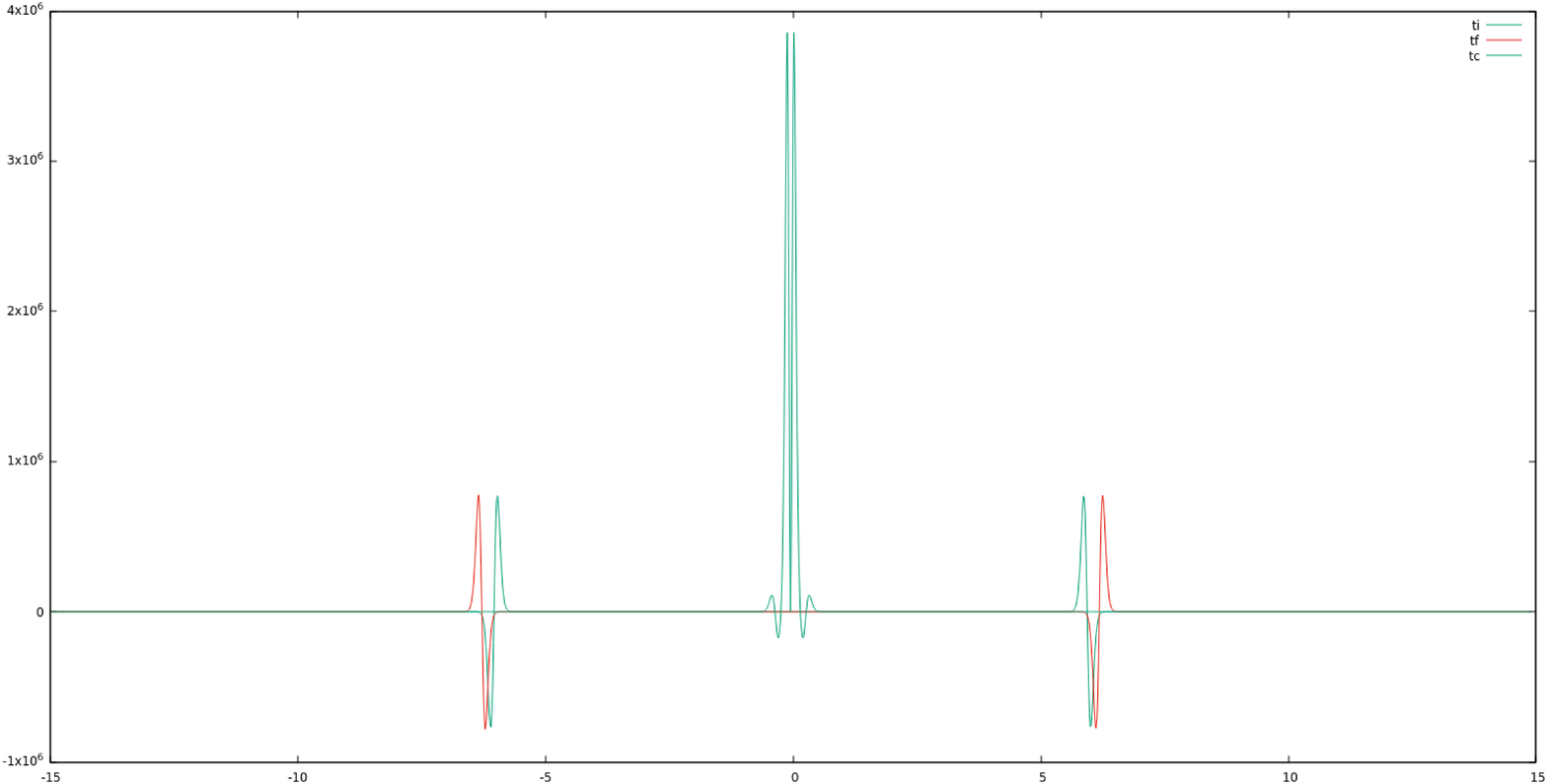} 
\includegraphics[width=1.5cm,scale=4, angle=0,height=3.2cm]{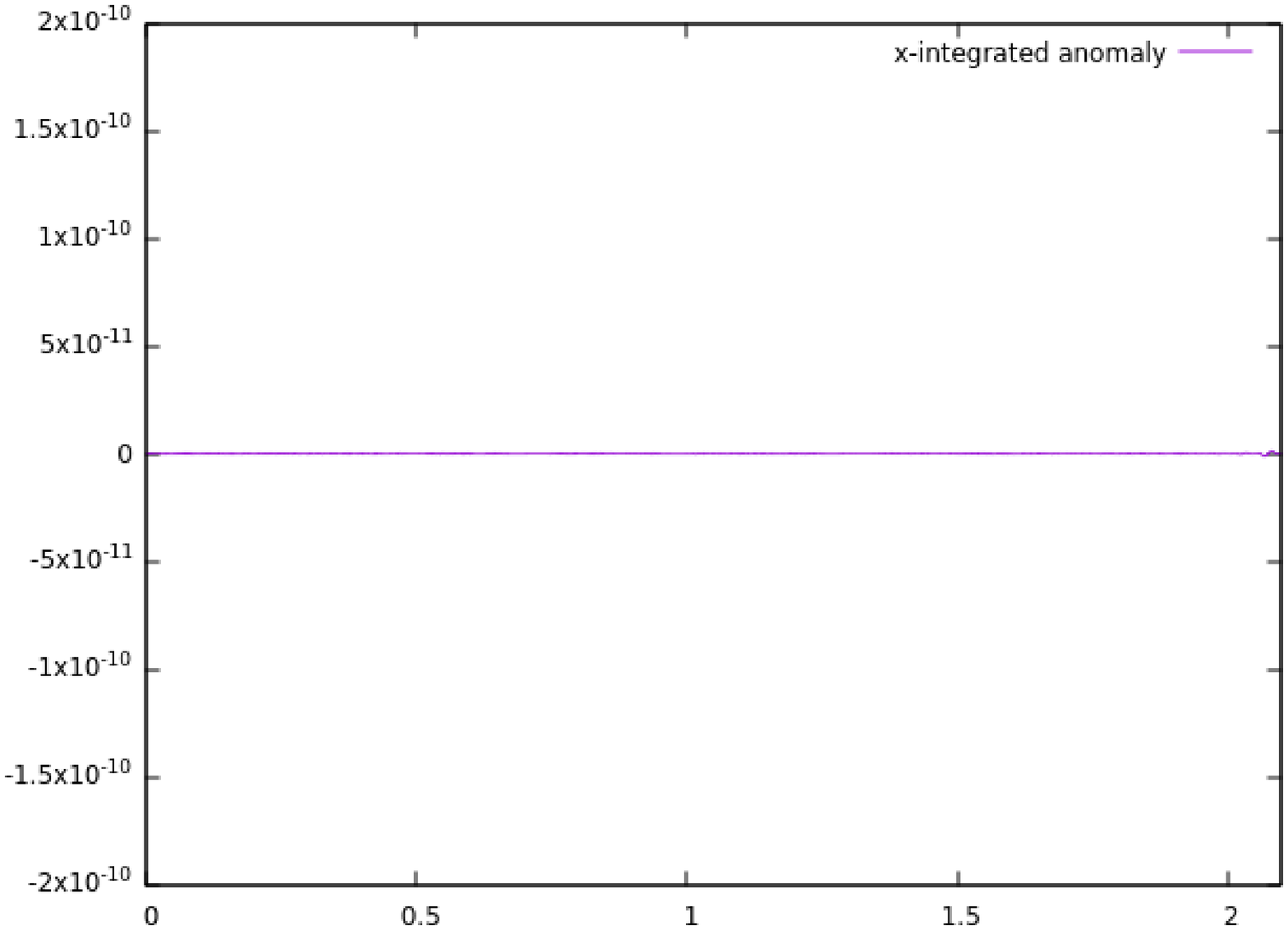}
\includegraphics[width=1.5cm,scale=4, angle=0,height=3.2cm]{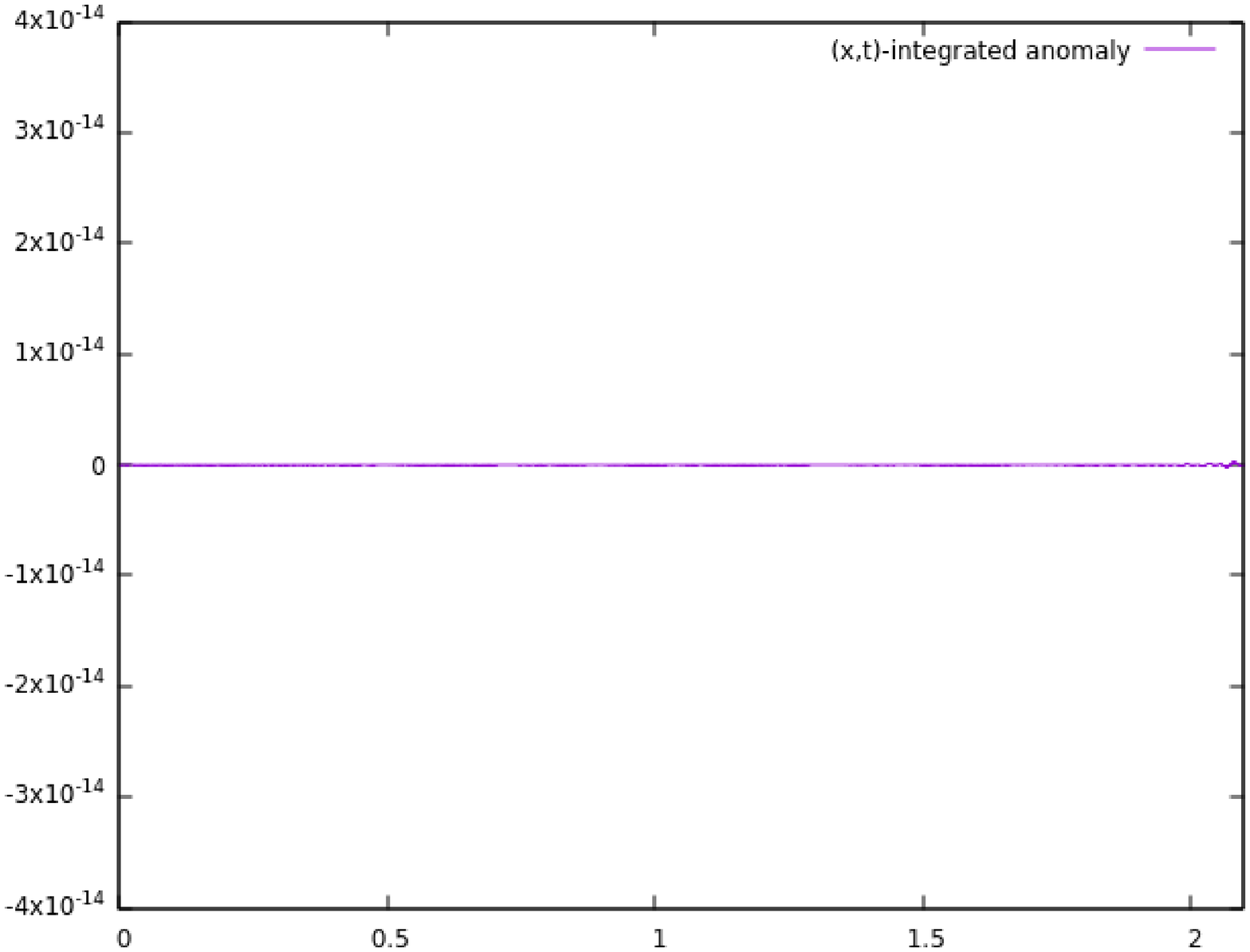}  
\parbox{6in}{\caption{(color online)Left figure shows the profile of anomaly $\hat{\beta}_2\,\, vs\, \,x$ for the 2-soliton collision of Fig. 6 for initial (green), collision (blue) and final (red) times. Middle figure shows the plot $\int \hat{\beta}_2 dx\,\, vs\,\, t$ and the right shows the plot $\int dt \int dx\, \hat{\beta}_2\,\, vs\,\, t$.}}
\end{figure} 

\begin{figure}
\centering
\label{fig9}
\includegraphics[width=1.5cm,scale=4, angle=0,height=3.2cm]{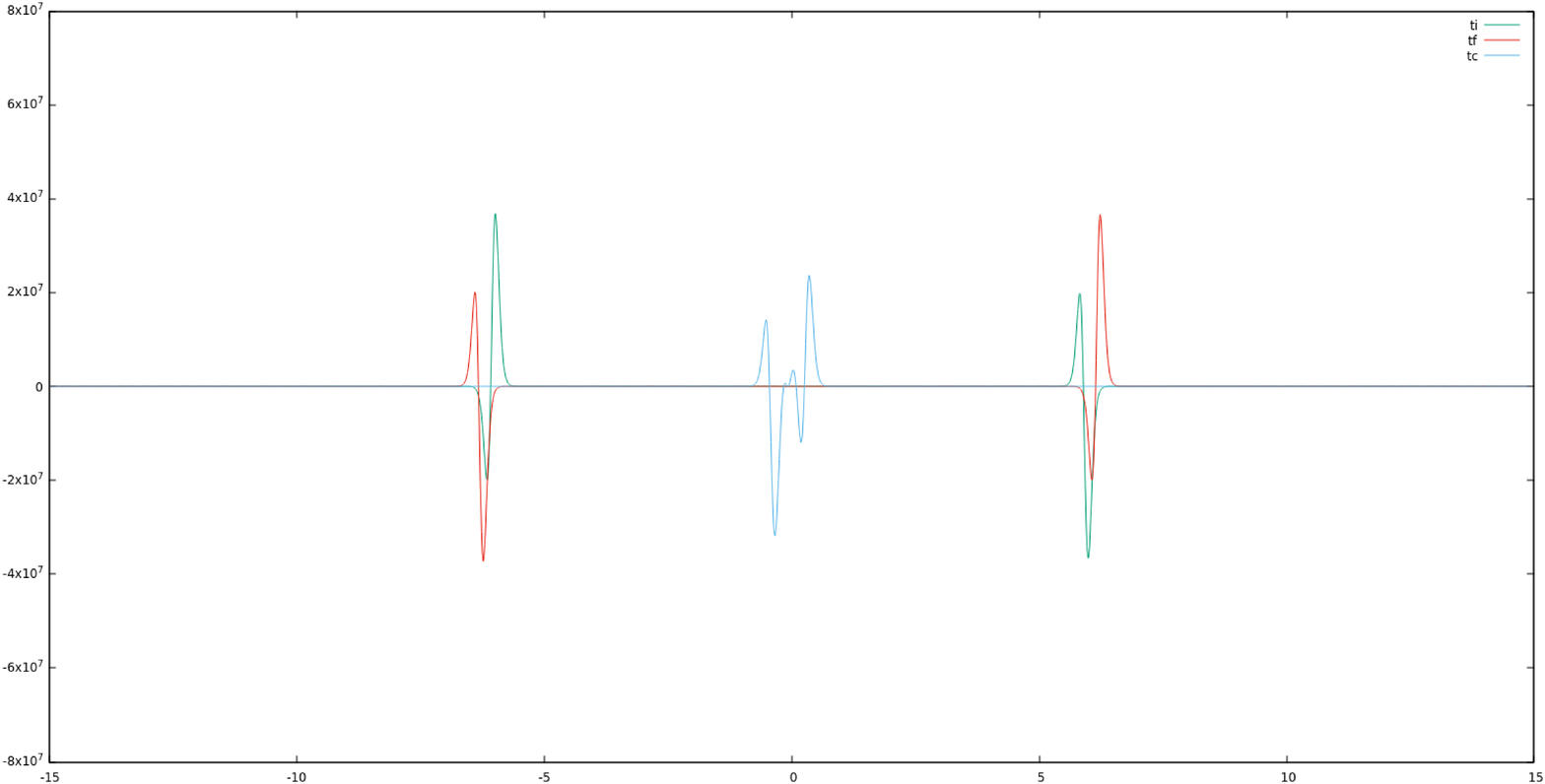} 
\includegraphics[width=1.5cm,scale=4, angle=0,height=3.2cm]{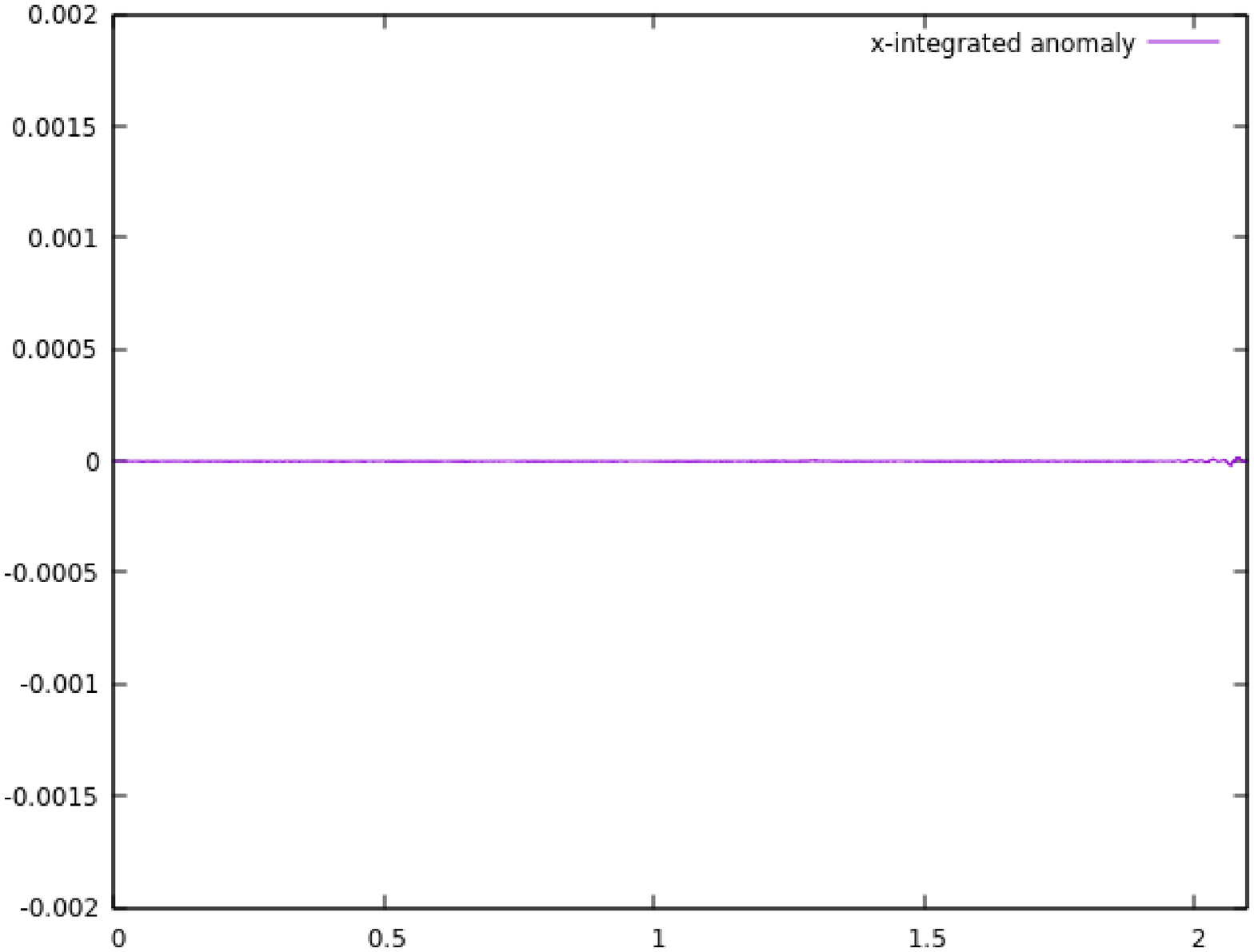}
\includegraphics[width=1.5cm,scale=4, angle=0,height=3.2cm]{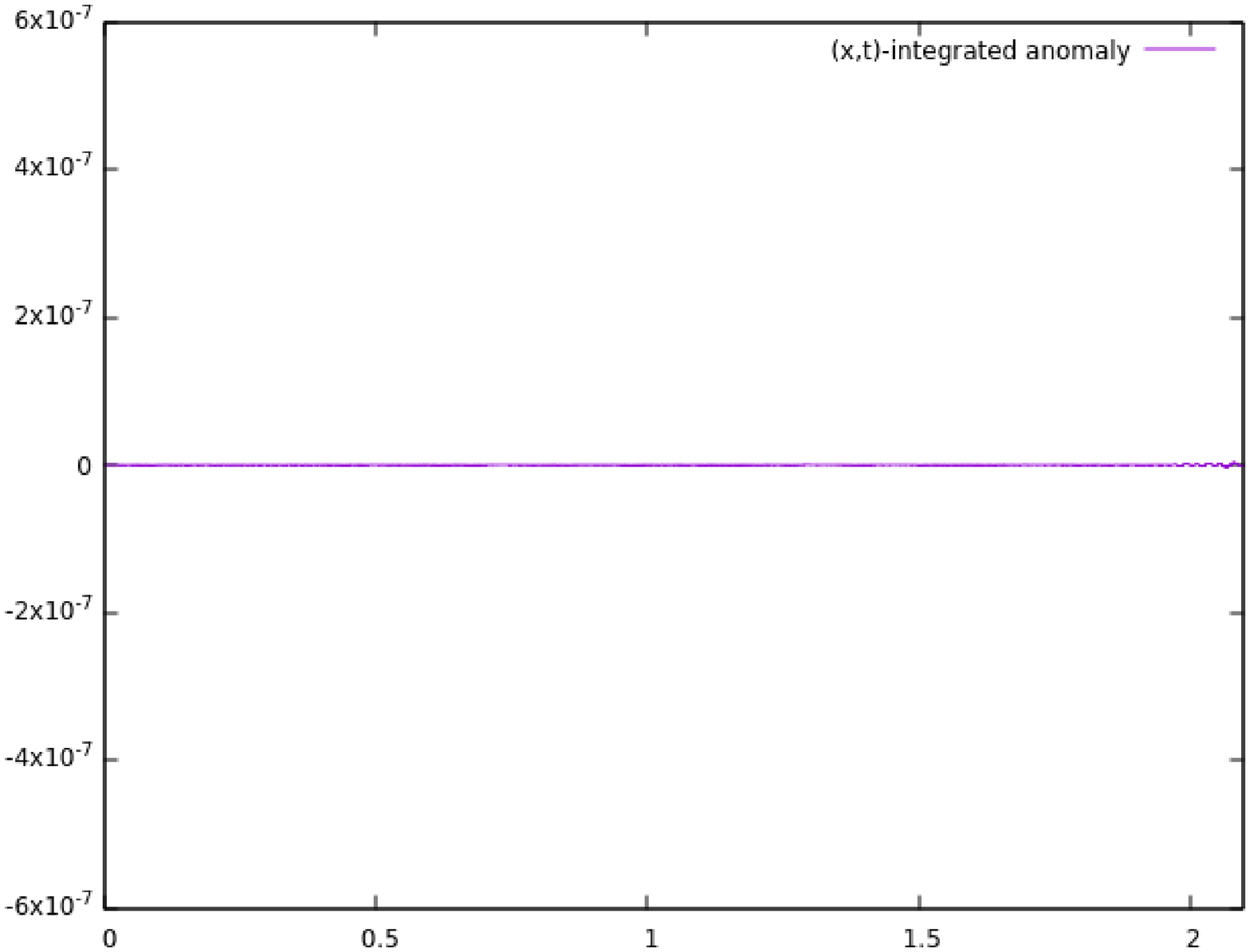}  
\parbox{6in}{\caption{(color online)Left figure shows the profile of the anomaly $\hat{\gamma}_2\,\, vs\, \,x$ for the 2-soliton collision of Fig. 6 for initial (green), collision (blue) and final (red) times. Middle figure shows the plot $\int \hat{\gamma}_2 dx\,\, vs\,\, t$ and the right shows the plot $\int dt \int dx\, \hat{\gamma}_2\,\, vs\,\, t$.}}
\end{figure} 

\begin{figure}
\centering
\label{fig10}
\includegraphics[width=1.5cm,scale=4, angle=0,height=3.7cm]{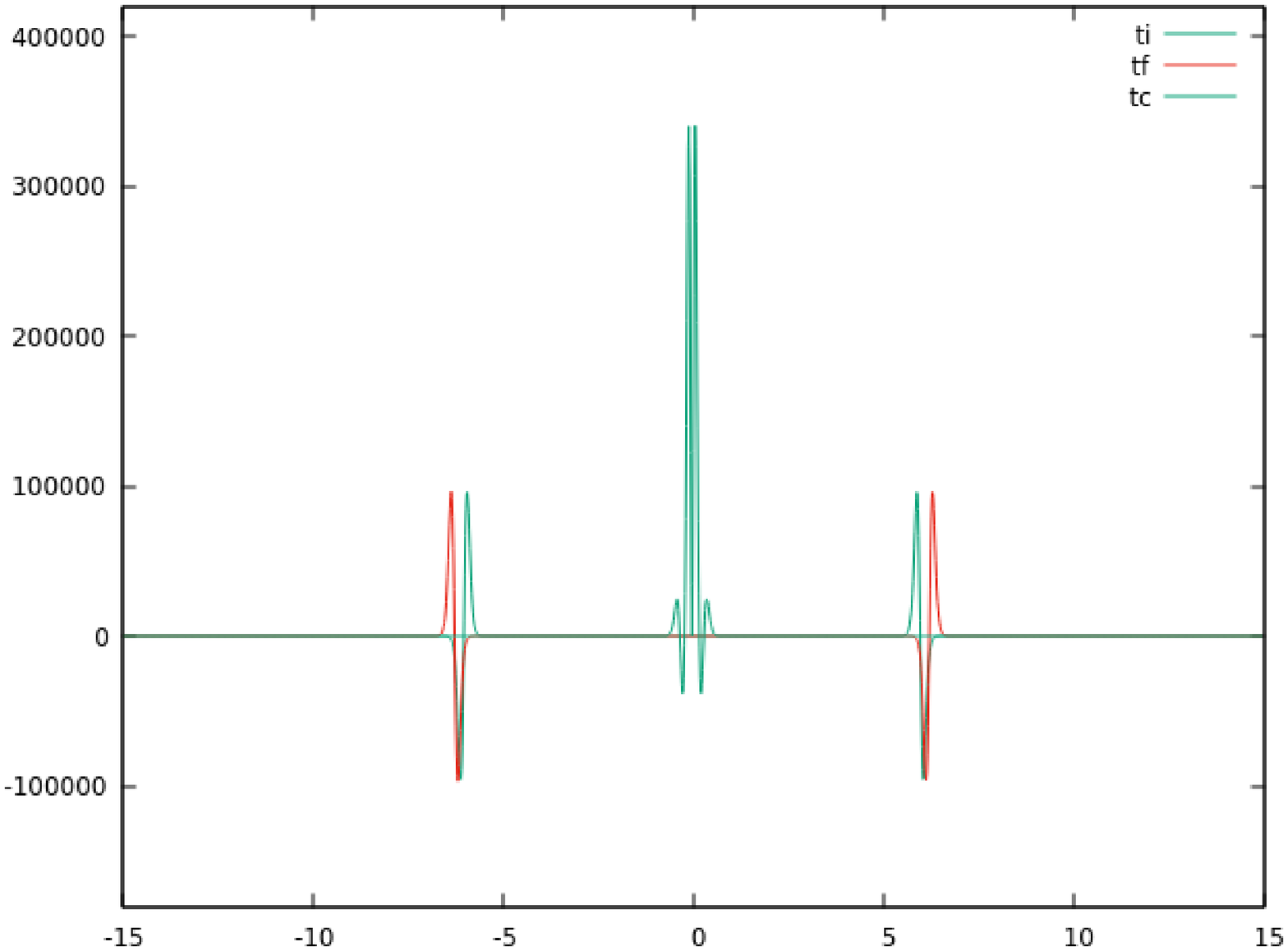} 
\includegraphics[width=1.5cm,scale=4, angle=0,height=3.7cm]{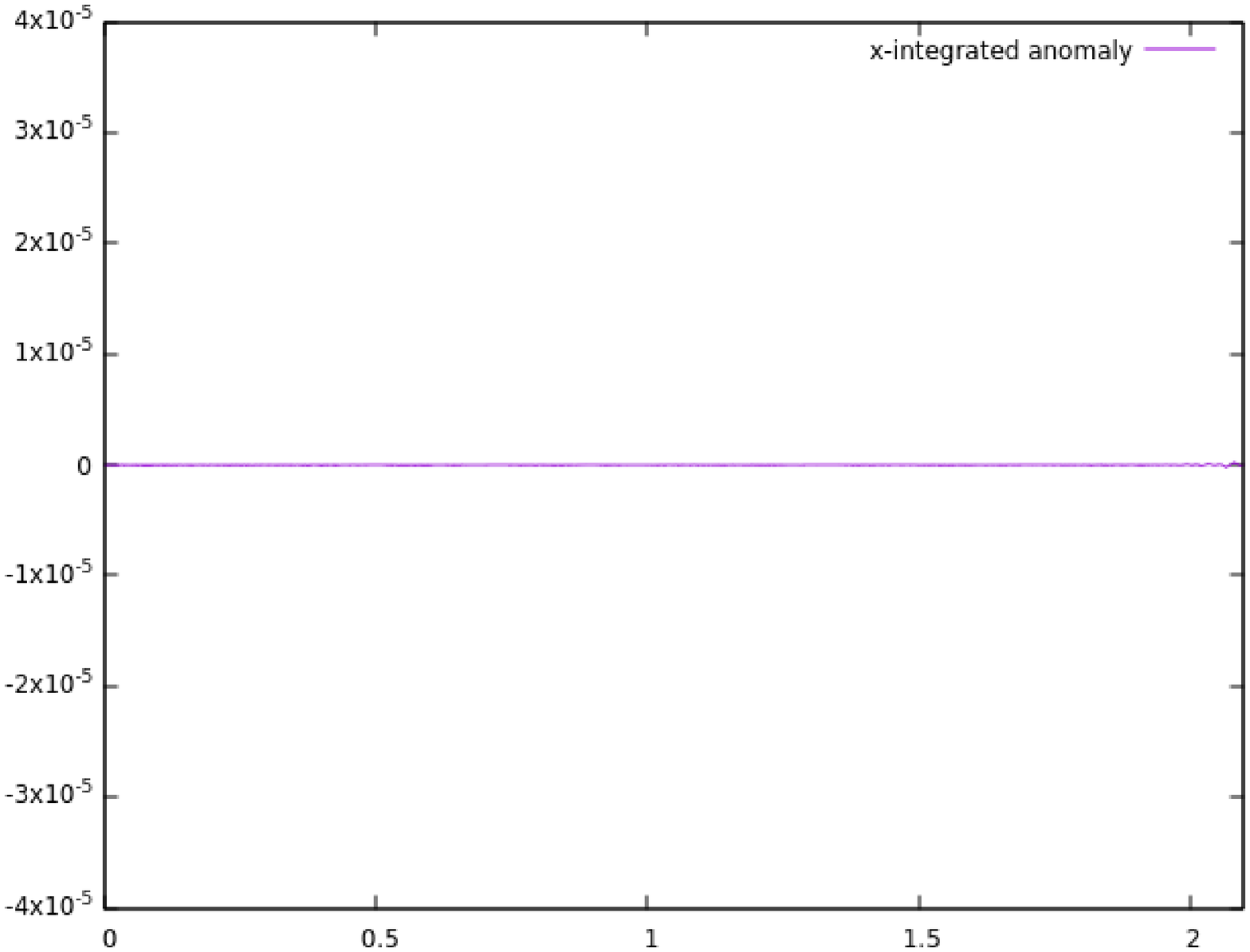}
\includegraphics[width=1.5cm,scale=4, angle=0,height=3.7cm]{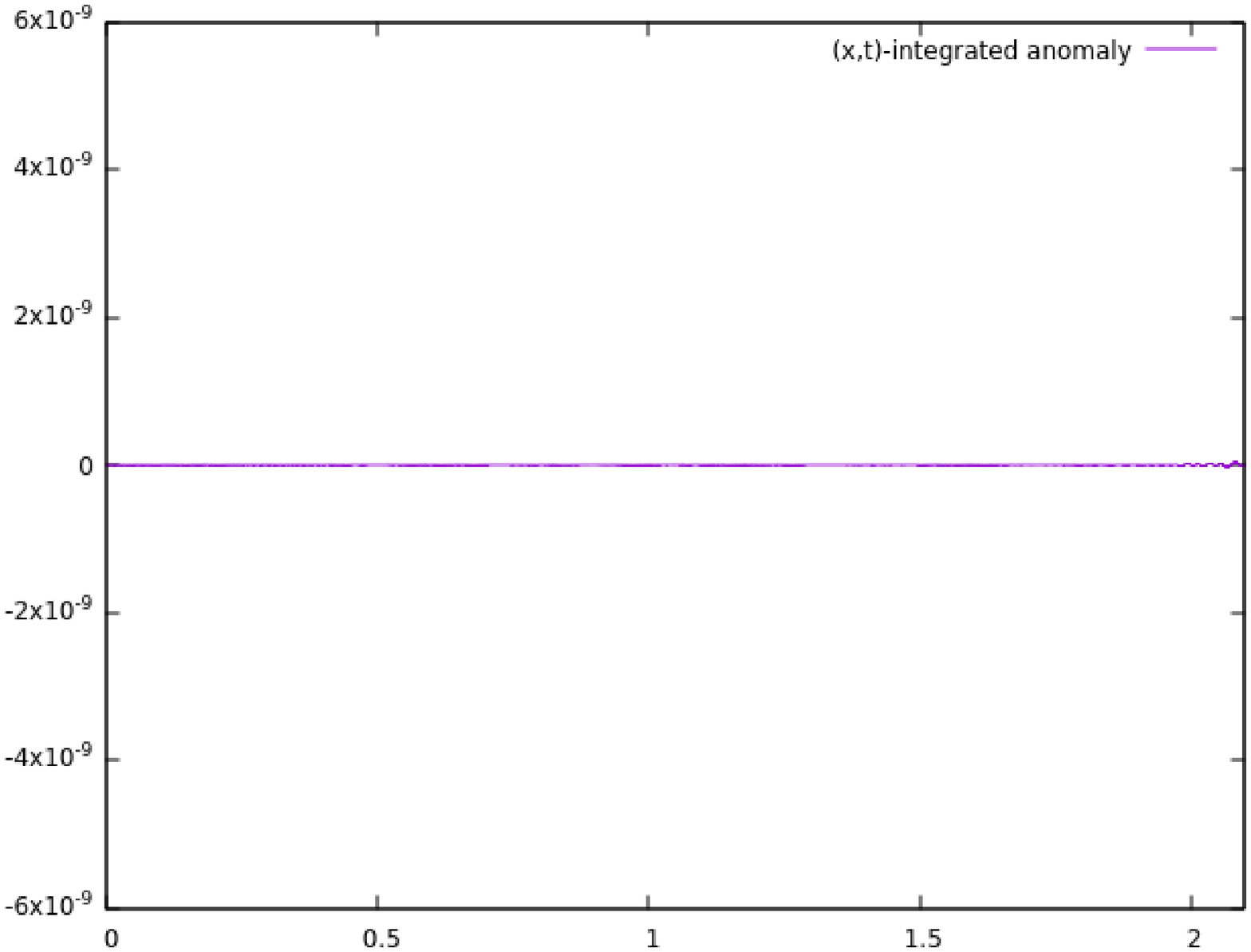}  
\parbox{6in}{\caption{(color online)Left figure shows the profile of anomaly $\hat{\delta}_1\,\, vs\, \,x$ for the 2-soliton collision of Fig. 6 for initial (green), collision (blue) and final (red) times. Middle figure shows the plot $\int \hat{\delta}_1 dx\,\, vs\,\, t$ and the right shows the plot $\int dt \int dx\, \hat{\delta}_1\,\, vs\,\, t$.}}
\end{figure}

 \subsection{Two-bright solitons: different amplitudes and positive velocities}

Let us consider the collision of two solitons with different  amplitudes initially centered at $x_{01}= -10$ and $x_{02}=-5$), traveling with positive velocities $v_1=12, v_2=6$. The Fig. 11 shows the numerical simulation for the collision of two solitons  with amplitudes $|\psi_1| =6.83,  |\psi_2| = 4.446$ for  $\epsilon= -0.06$.

In the Fig. 12 we present the simulations for the anomaly density $\hat{\alpha}_1$ (\ref{an1d}) for the two soliton collision of Fig. 11.  In the left figure we plot the density anomaly for three successive times, $t_i$, before collision (green), $t_c$, collision (blue ), and, $t_f$, after collision (red) times, respectively. In the middle figure it is plotted the $x-$integration of the anomaly versus  $t$; whereas in the right figure it is plotted the $(x,t)-$integration of the anomaly versus $t$.  The $(x,t)-$integration of this anomaly  vanishes, within numerical accuracy, with an error of order $10^{-10}$.  

In the Fig. 13 we present the simulations for the anomaly density $\hat{\beta}_2$ (\ref{beta2d})  for the two soliton collision of Fig. 11.  In the left it is plotted the anomaly density for three successive times, $t_i$, before collision (green), $t_c$, collision (blue ), and, $t_f$, after collision (red) times, respectively. In the middle figure it is plotted the $x-$integration of the anomaly versus $t$; whereas in the right figure it is plotted the $(x,t)-$integration of the anomaly versus $t$. Notice that the $(x,t)-$integration of this anomaly  vanishes, within numerical accuracy,  with an error of order $10^{-16}$.  

In the Fig. 14 we present the simulations for the anomaly density $\hat{\gamma}_2$ (\ref{gam33d}) for the two soliton collision of Fig. 11. In the left it is plotted the anomaly density for three successive times, $t_i$, before collision (green), $t_c$, collision (blue ), and, $t_f$, after collision (red) times, respectively. In the middle figure it is plotted the $x-$integration of the anomaly versus $t$; whereas in the right figure it is plotted the $(x,t)-$integration of the anomaly versus $t$. The $(x,t)-$integration of this anomaly  vanishes, within numerical accuracy,  with an error of order $10^{-8}$. 

Finally, in the Fig. 15 we present the simulations for the anomaly density $\hat{\delta}_1$ (\ref{ano4}) for the two soliton collision of Fig. 11.  In the left it is plotted the anomaly density for three successive times, $t_i$, before collision (green), $t_c$, collision (blue ), and, $t_f$, after collision (red) times, respectively. In the middle figure it is plotted the $x-$integration of the anomaly versus $t$; whereas in the right figure it is plotted the $(x,t)-$integration of the anomaly versus $t$.  The $(x,t)-$integration of this anomaly vanishes, within numerical accuracy,  with an error of order $10^{-10}$. 

Some comments are in order here.  First, in our numerical simulations of the 2-soliton collisions  of the deformed model (\ref{mnls}) we have not observed appreciable emission of radiation during the collisions; so, it can be argued that the linear superposition of two solitary waves (\ref{solitary}) of the deformed NLS model  is an adequate initial condition, as compared to the explicit  2-soliton, or alternatively, to two one-solitons stitched together of the standard NLS model,  which could have been taken as initial conditions. Second, we have shown  the vanishing of the space-time integrals of the anomaly densities $\hat{\alpha}_1$,  $\hat{\beta}_2$,  $\hat{\gamma}_2$ and  $\hat{\delta}_1$, appearing in (\ref{an1d}), (\ref{beta2d}), (\ref{gam33d}) and (\ref{ano4}), respectively, within numerical accuracy, for three type of two-soliton collisions in the deformed NLS model (\ref{mnls}). In fact, the space and space-time integrals of the anomaly densities vanish within the errors less than $10^{-4}$ and $10^{-6}$, respectively; and sometimes,  within the approximations $10^{-11}$ and $10^{-15}$, respectively. Third, we have performed extensive numerical simulations for a wide range of values in the parameter space; i.e. the deformation parameter $ |\epsilon| < 1$ and coupling constant $\eta \approx 1$, several amplitudes and relative velocities for 2-soliton collisions, obtaining the vanishing of those anomalies, within numerical accuracy.  Fourth, qualitatively similar results have recently been obtained for the space and space-time integrals of the anomaly densities $\hat{\alpha}_1$,  $\hat{\beta}_2$,  $\hat{\gamma}_2$ and  $\hat{\delta}_1$ corresponding to the collision of 2-dark solitons in the modified NLS model with defocusing nonlinearity \cite{nodycon21}. 

\begin{figure}
\centering
\label{fig11}
\includegraphics[width=1cm,scale=4, angle=0,height=5cm]{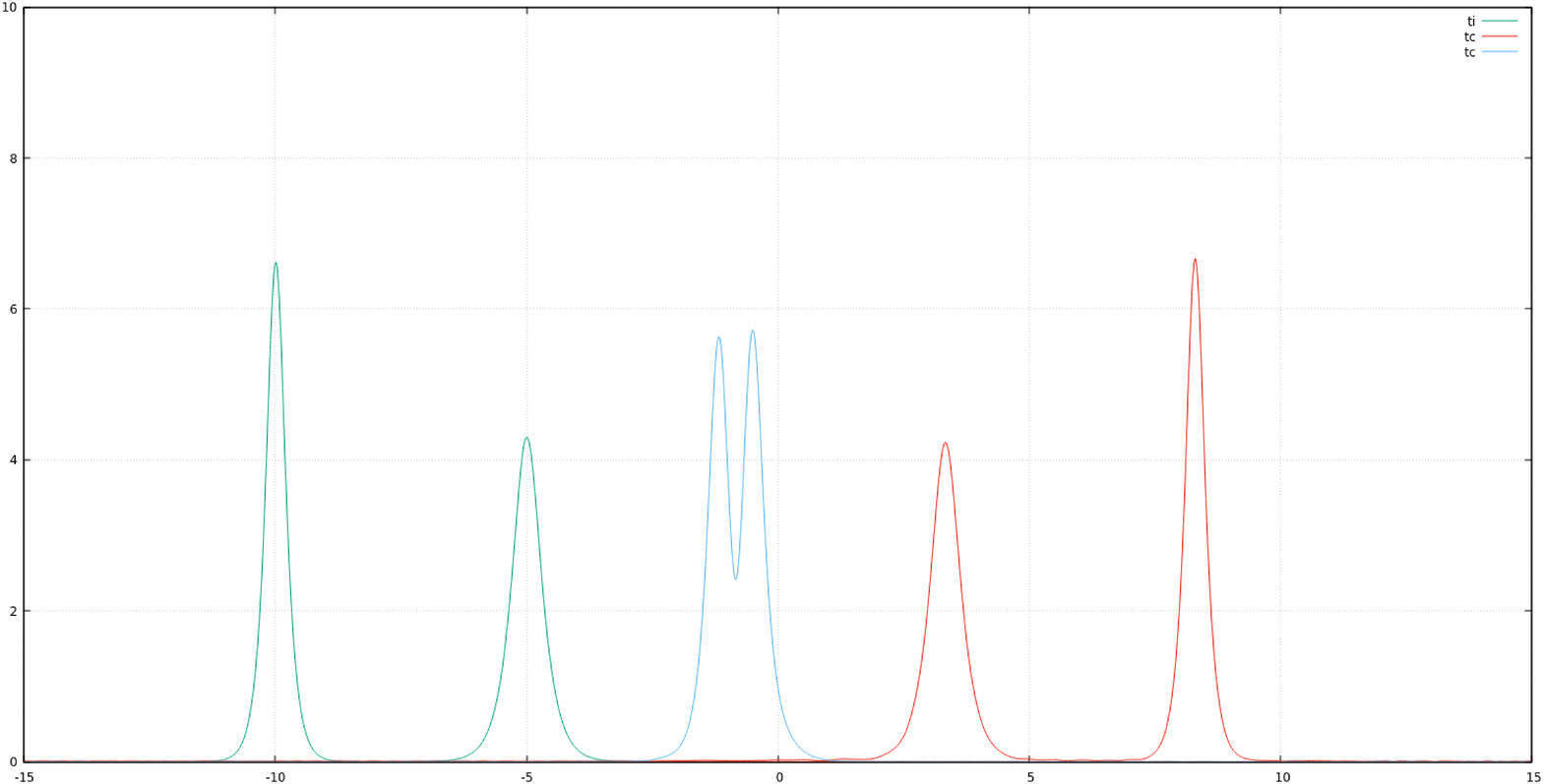} 
\parbox{6in}{\caption{(color online) Two-bright solitons with amplitudes $|\psi_1|=6.83,\,|\psi_2| = 4.446$ and velocities $v_1=12, \, v_2 = 6,$ and $\epsilon= -0.06$, for initial (green), collision (blue) and final (red) succesive times.}}
\end{figure}

\begin{figure}
\centering
\label{fig12}
\includegraphics[width=1.5cm,scale=3, angle=0,height=3.2cm]{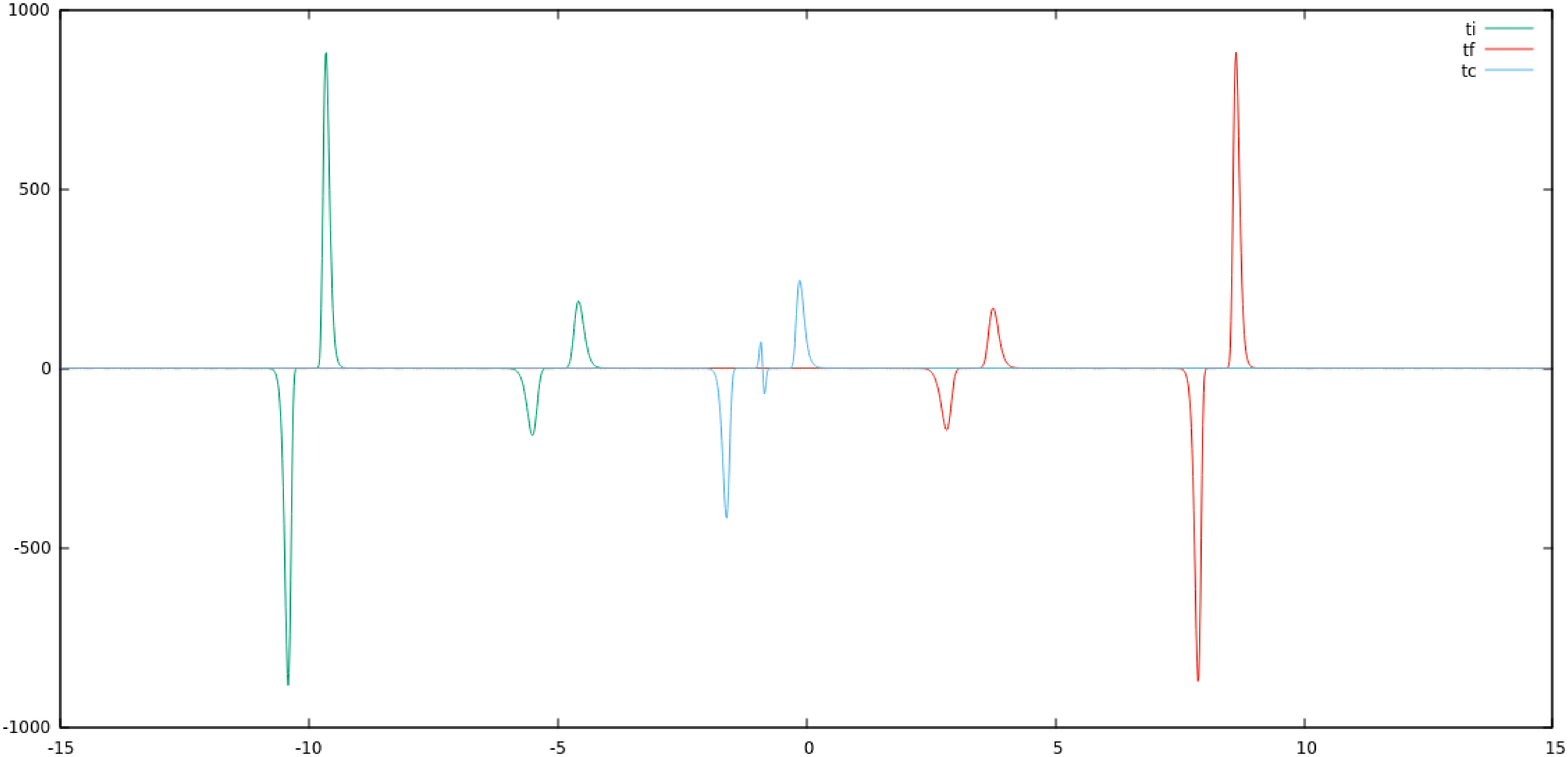} 
\includegraphics[width=1.5cm,scale=3, angle=0,height=3.2cm]{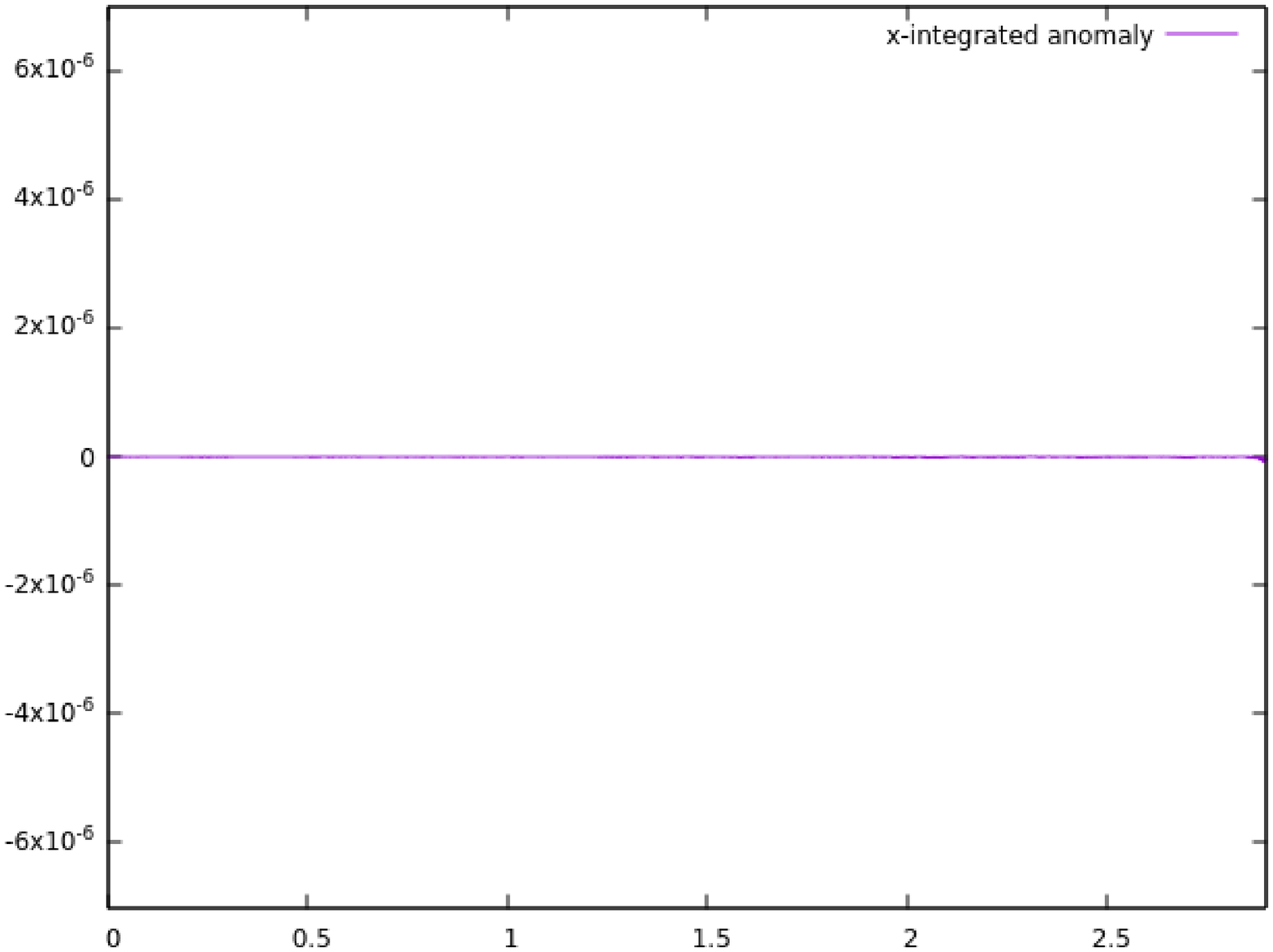}
\includegraphics[width=1.5cm,scale=3, angle=0,height=3.2cm]{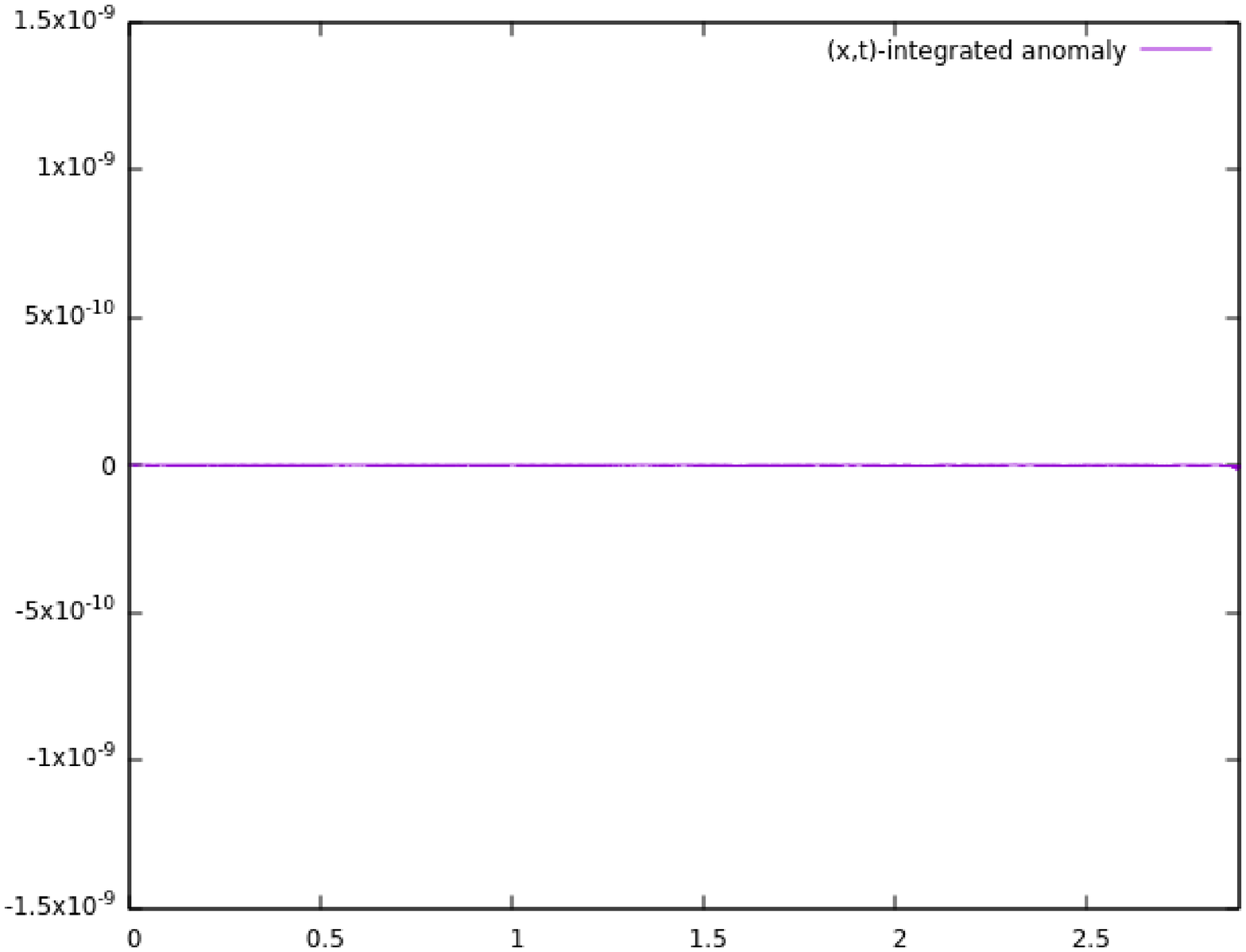}  
\parbox{6in}{\caption{(color online)Left figure shows the profile of anomaly $\hat{\alpha}_1\,\, vs\, \,x$ for the 2-soliton collision of Fig. 11 for initial (green), collision (blue) and final (red) times. Middle figure shows the plot $\int \hat{\alpha}_1 dx\,\, vs\,\, t$ and the right shows the plot $\int dt \int dx\, \hat{\alpha}_1\,\, vs\,\, t$.}}
\end{figure} 
\begin{figure}
\centering
\label{fig13}
\includegraphics[width=1.5cm,scale=3, angle=0,height=3.2cm]{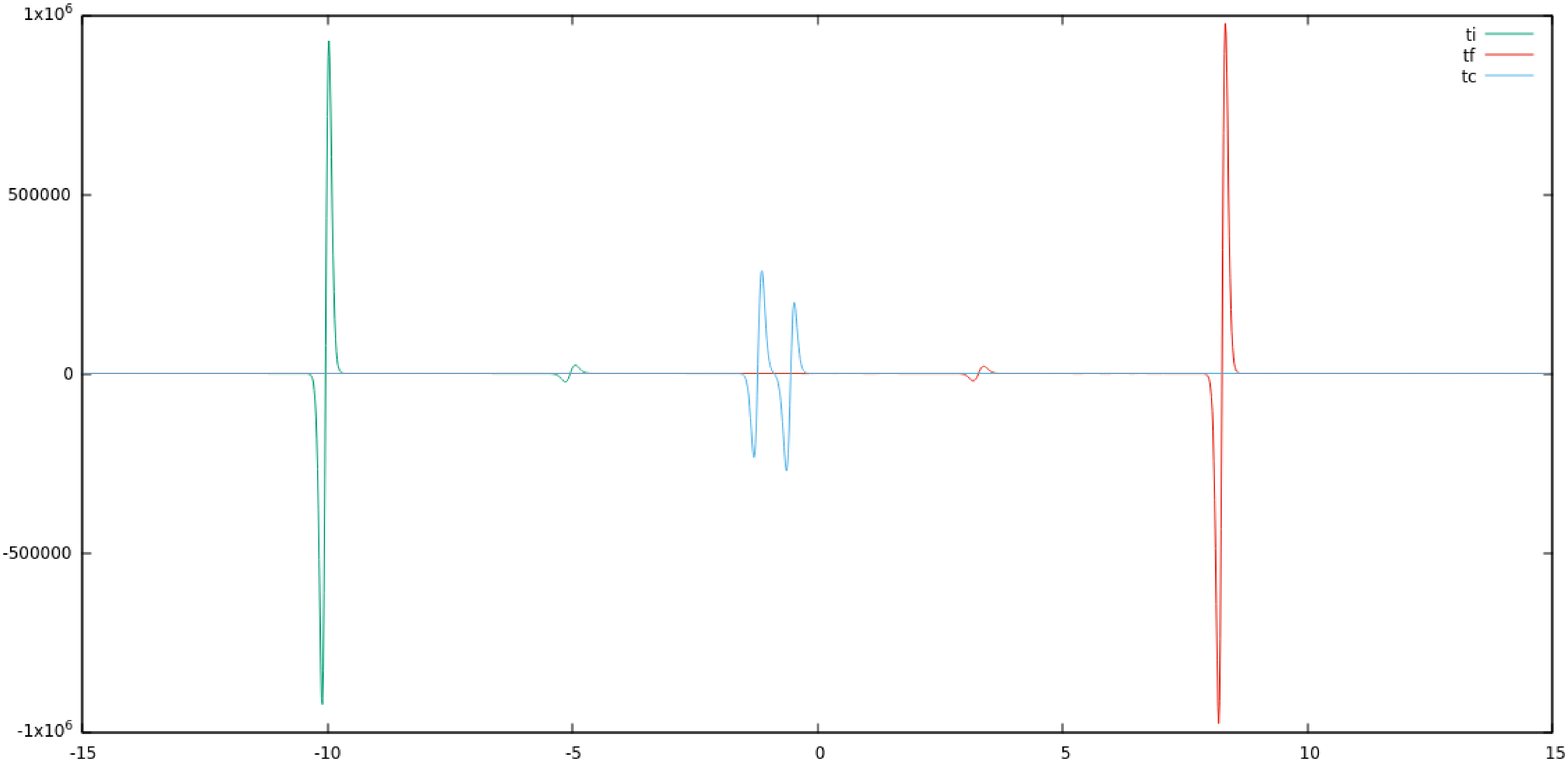} 
\includegraphics[width=1.5cm,scale=3, angle=0,height=3.2cm]{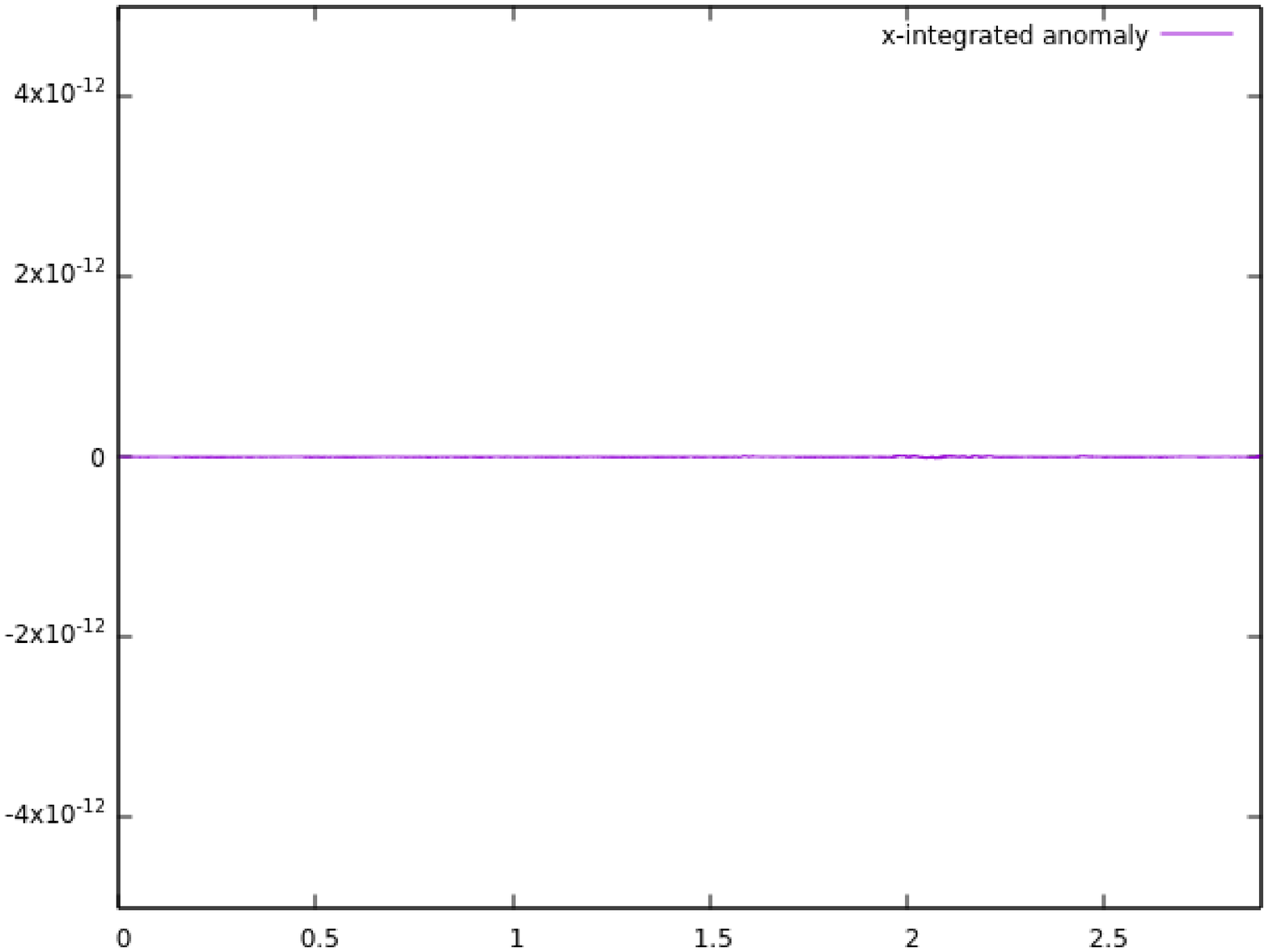}
\includegraphics[width=1.5cm,scale=3, angle=0,height=3.2cm]{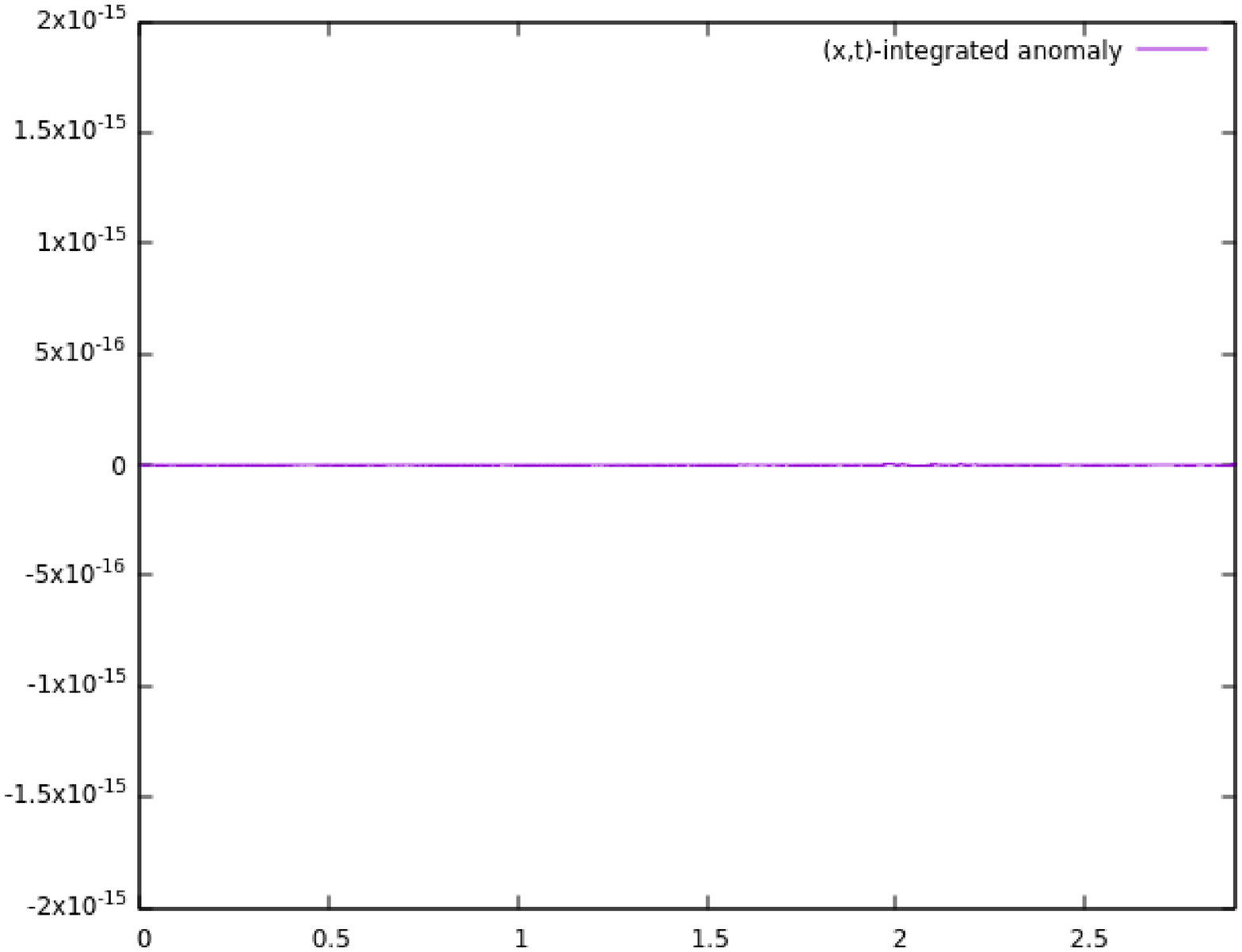}  
\parbox{6in}{\caption{(color online)Left figure shows the profile of anomaly density $\hat{\beta}_2\,\, vs\, \,x$ for the 2-soliton collision of Fig. 11 for initial (green), collision (blue) and final (red) times. Middle figure shows the plot $\int \hat{\beta}_2 dx\,\, vs\,\, t$ and the right shows the plot $\int dt \int dx\, \hat{\beta}_2\,\, vs\,\, t$.}}
\end{figure} 

\begin{figure}
\centering
\label{fig14}
\includegraphics[width=1.5cm,scale=4, angle=0,height=3.2cm]{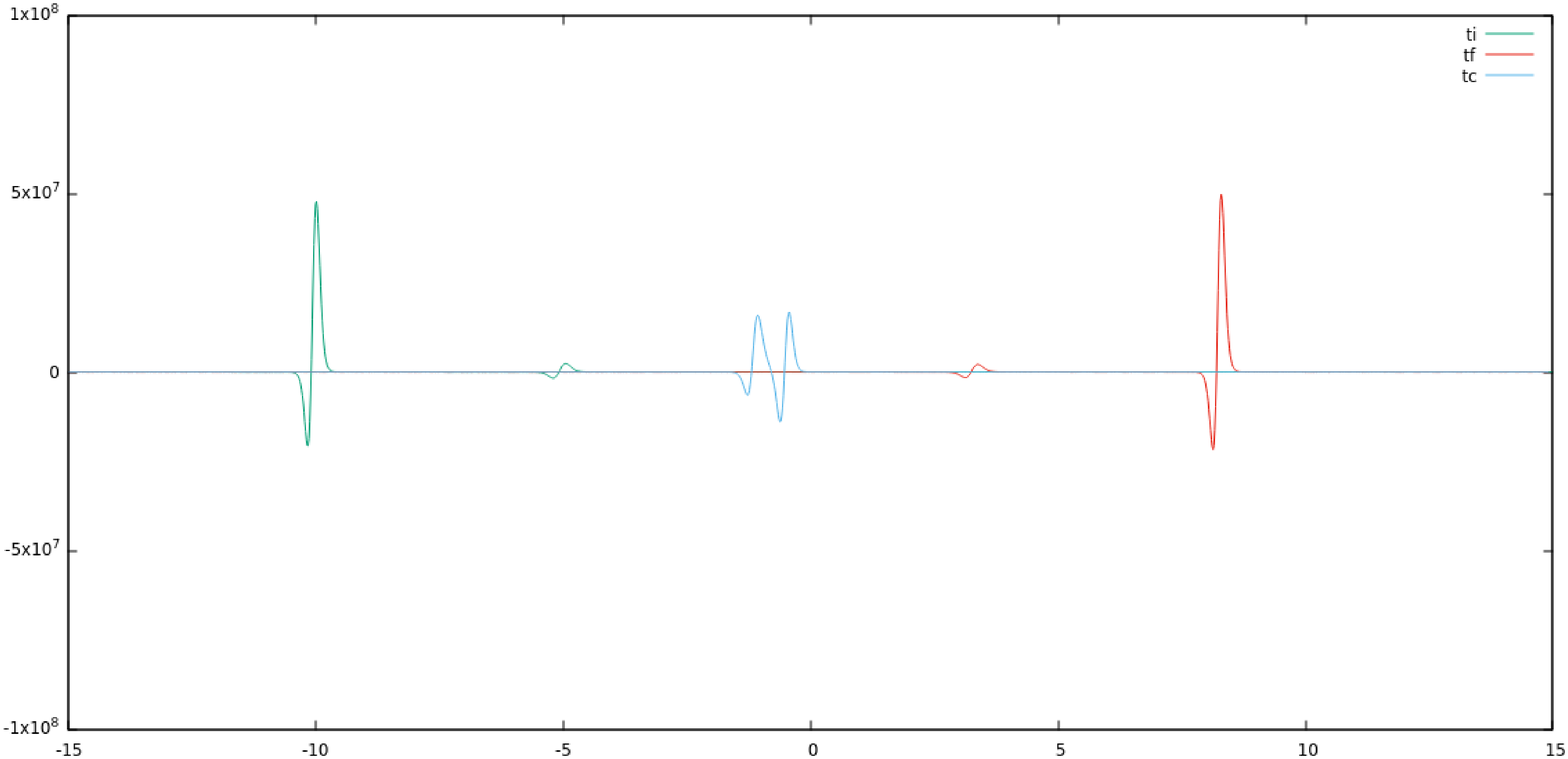} 
\includegraphics[width=1.5cm,scale=4, angle=0,height=3.2cm]{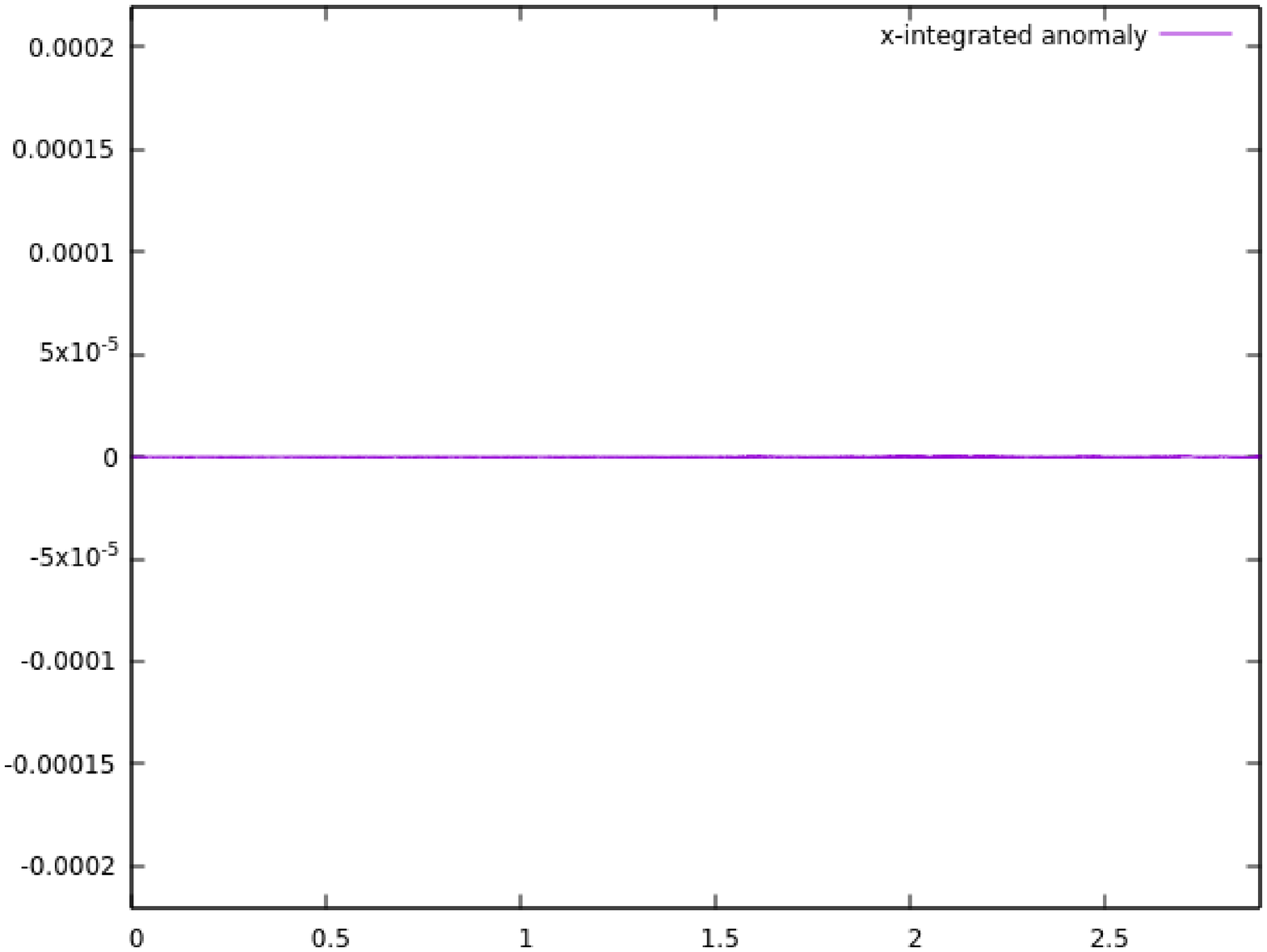}
\includegraphics[width=1.5cm,scale=4, angle=0,height=3.2cm]{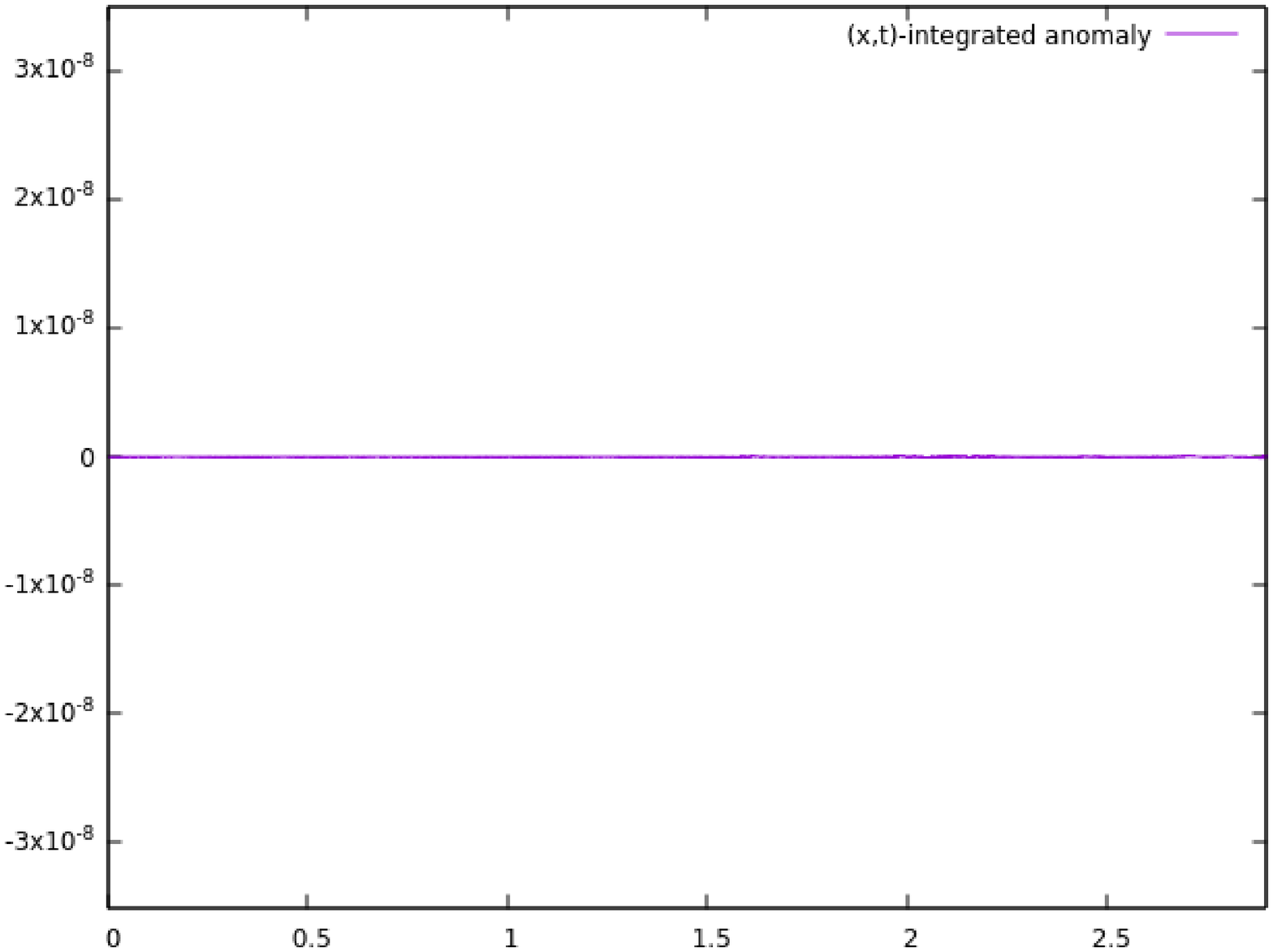}  
\parbox{6in}{\caption{(color online)Left figure shows the profile of anomaly $\hat{\gamma}_2\,\, vs\, \,x$ for the 2-soliton collision of Fig. 11 for initial (green), collision (blue) and final (red) times. Middle figure shows the plot $\int \hat{\gamma}_2 dx\,\, vs\,\, t$ and the right shows the plot $\int dt \int dx\, \hat{\gamma}_2\,\, vs\,\, t$.}}
\end{figure} 

\begin{figure}
\centering
\label{fig15}
\includegraphics[width=1.5cm,scale=4, angle=0,height=3.5cm]{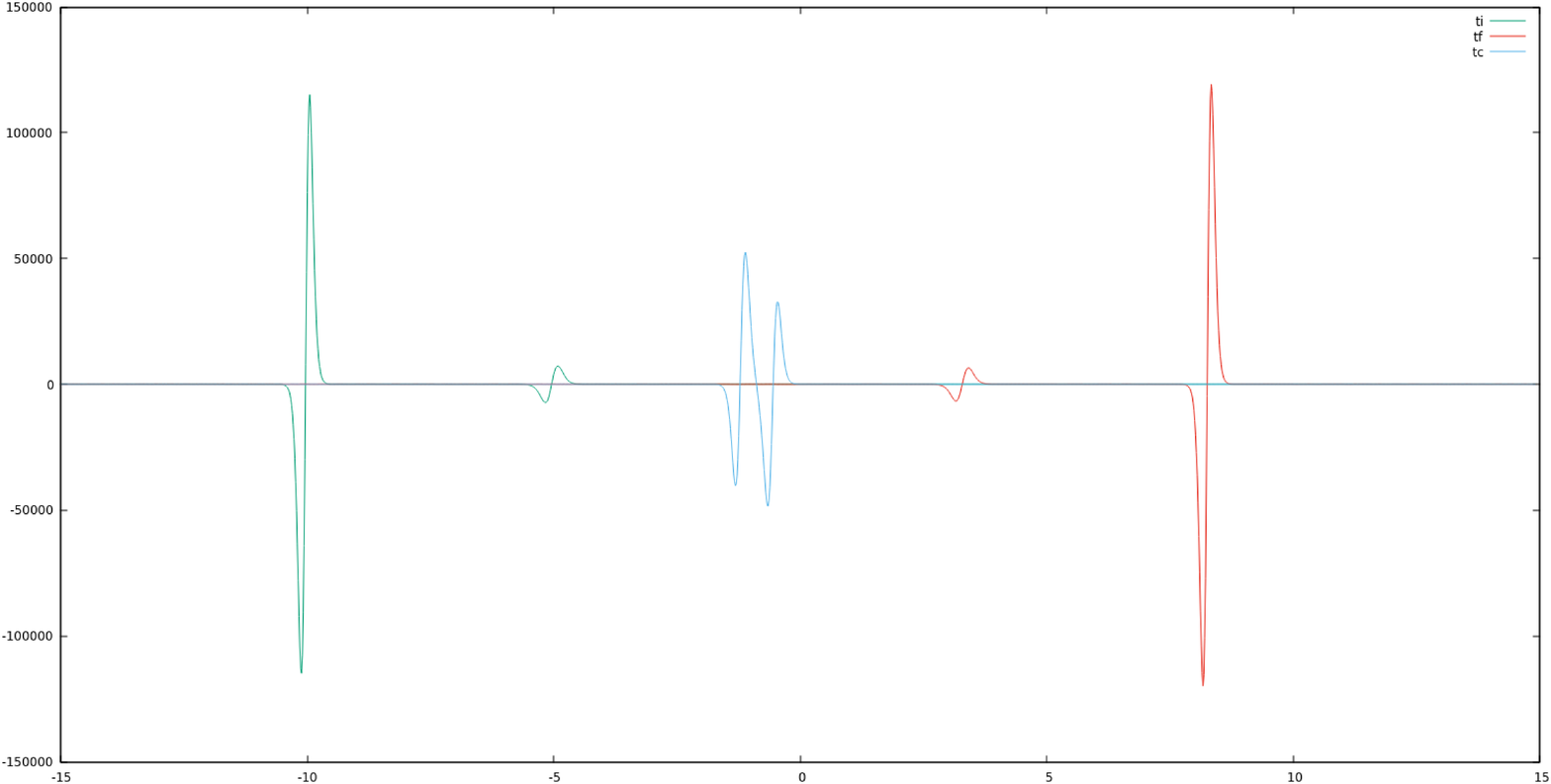} 
\includegraphics[width=1.5cm,scale=4, angle=0,height=3.5cm]{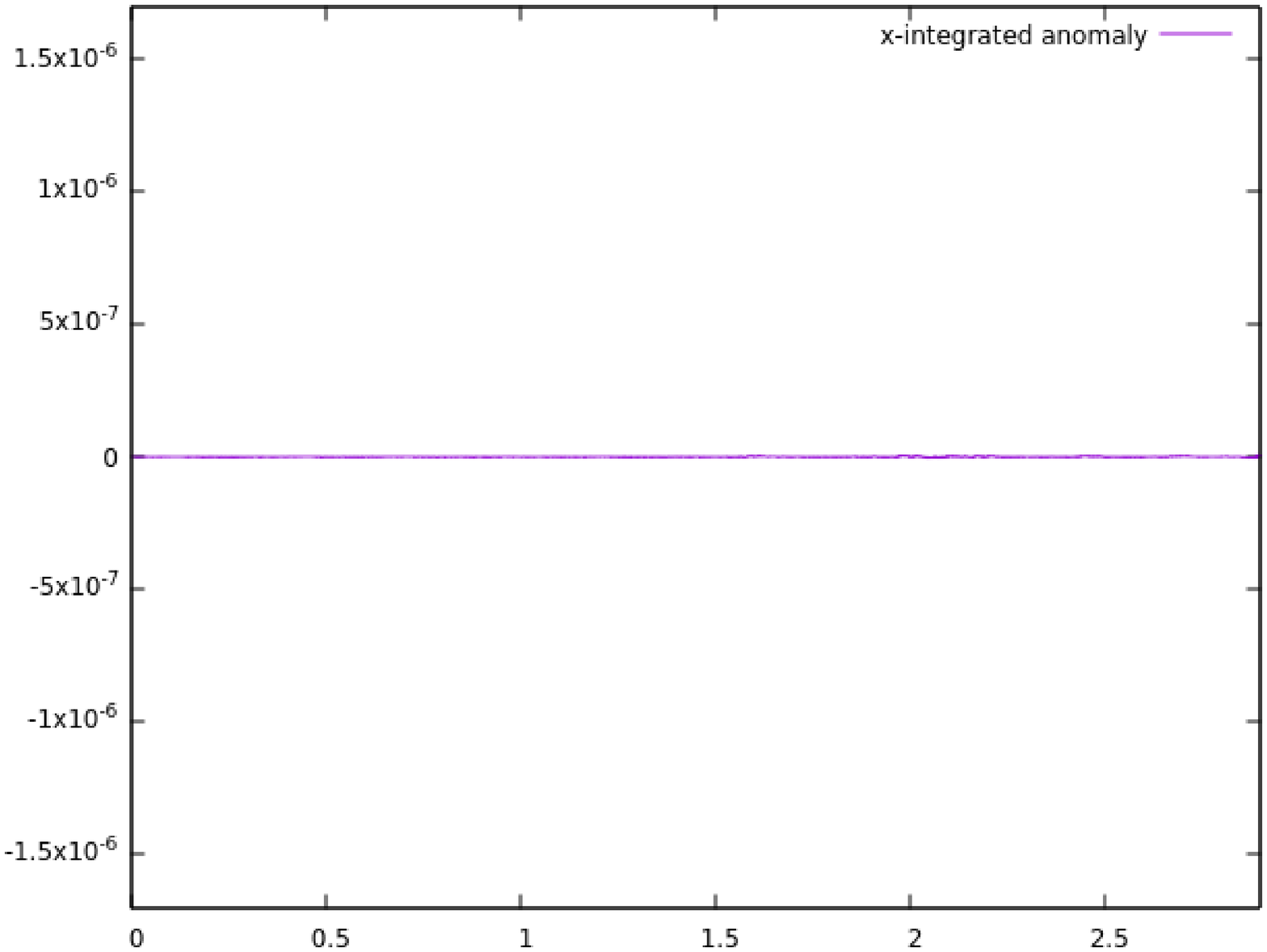}
\includegraphics[width=1.5cm,scale=4, angle=0,height=3.5cm]{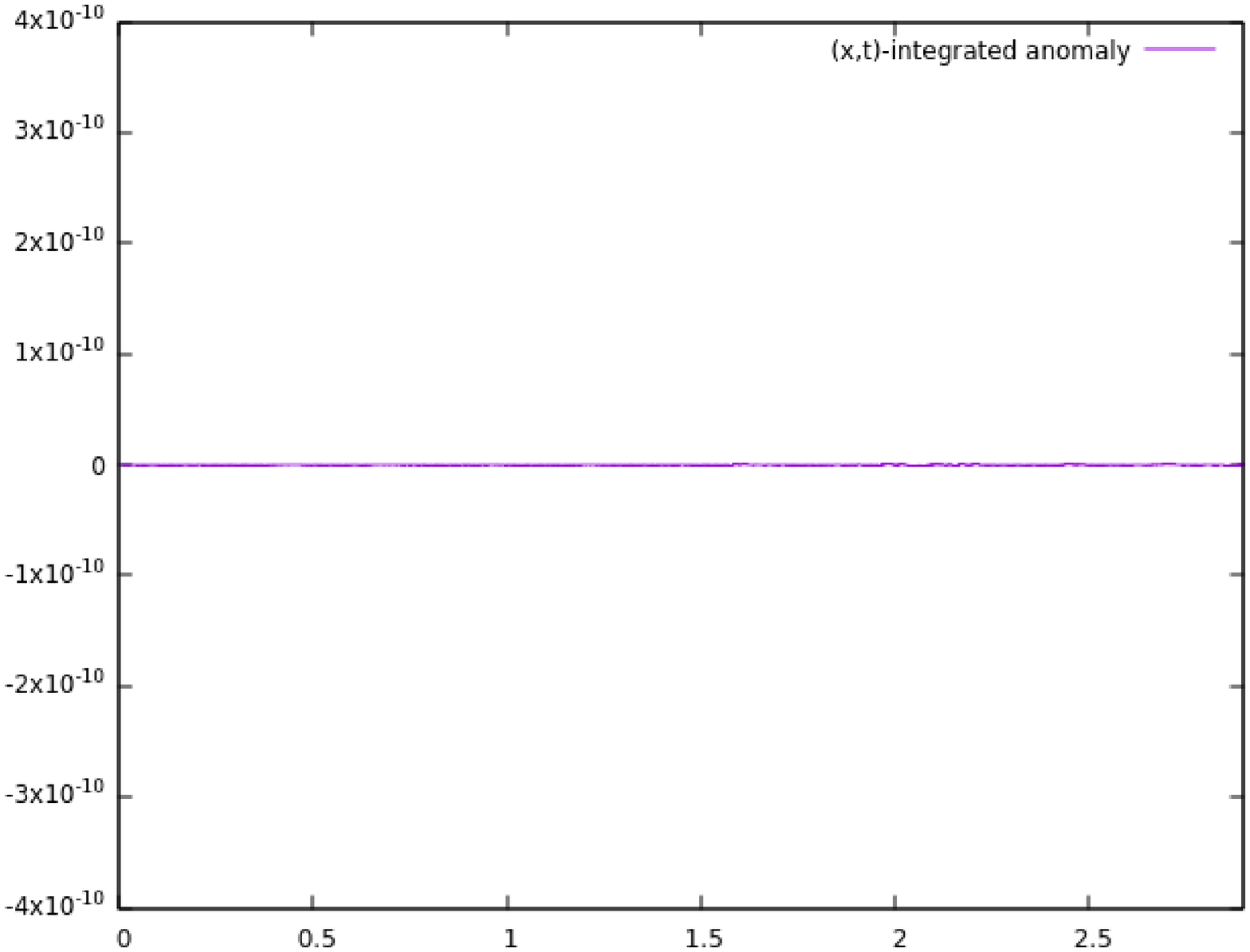}  
\parbox{6in}{\caption{(color online)Left figure shows the profile of anomaly $\hat{\delta}_1\,\, vs\, \,x$ for the 2-soliton collision of Fig. 11 for initial (green), collision (blue) and final (red) times. Middle figure shows the plot $\int \hat{\delta}_1 dx\,\, vs\,\, t$ and the right shows the plot $\int dt \int dx\, \hat{\delta}_1\,\, vs\,\, t$.}}
\end{figure} 

These developments strongly suggest that the quasi-integrable models set forward in the literature \cite{jhep1, jhep2,jhep3,jhep4, jhep5, jhep6, npb, jhep33}, and in particular the deformed NLS model (\ref{nlsd}), would possess more specific integrability structures, such as an infinite set of exactly conserved charges, and some type of Lax pairs for certain deformed potentials. In this context, the Riccati-type representations have recently been presented for the deformed KdV and sine-Gordon models \cite{jhep33, npb1}. So, in the following we will tackle the problem of extending the Riccati-type pseudo-potential formalism, which has been used for a variety of well known integrable systems, to the deformed NLS model (\ref{nlsd}).  

\section{Riccati-type pseudo-potential and modified $sl(2)$ AKNS model}
 
\label{sec:riccati}

The standard NLS model can be  obtained as a special reduction of the AKNS system; so, in the next sections we consider a convenient deformation of the usual pseudo-potential approach to the AKNS  integrable field theory. Subsequently, we will discuss its reduction process leading to the modified NLS model. In \cite{nucci} it has been generated the both Lax equations and Backlund transformations 
for well-known non-linear evolution equations using the concept of pseudo-potentials and the related  
properties of the Riccati equation. These applications have been done in the context of a variety of  integrable systems (sine-Gordon, KdV, NLS, etc), and allow the Lax pair formulation, the construction of  conservation laws and the Backlund transformations for them \cite{nucci, prl1}. 

So, let us consider the system of Riccati-type equations 
\br
\label{ricc1}
\pa_x u &=& -2 i \zeta \, u + q + \, \bar{q} \,\, u^2,\\
\label{ricc2}
\pa_t u &=&  2 A\, u - C\, u^2 + B + r - u\, s,
\er
where $u$ is the Riccati-type pseudo-potential, $r$ and  $s$ are auxiliary fields, and $q$ and $\bar{q}$ are the fields of the model. Let us assume
\br
\label{A11}
A & \equiv & - 2 i \zeta^2 -  \frac{1}{2} i\, V^{(1)},\\
\label{A22}
B & \equiv & 2 \zeta q + i \pa_x q,\\
\label{A33}
C & \equiv & -2 \zeta \bar{q} + i \pa_x \bar{q},\,\,\,\,\,V^{(1)} \equiv  \frac{d V[\rho]}{d \rho},\,\,\,\rho \equiv \bar{q} q, 
\er
where $V(\bar{q} q)$ is the potential of the modified AKNS equation (MAKNS) and $\zeta$ is the spectral parameter. We consider the following equations for the auxiliary fields $r(x,t)$ and  $s(x,t)$
\br
\label{ricc1r}
\pa_x r &=& q \, s +  (-2 i \zeta + Q \, u)\, r,\\
\label{ricc2s}
\pa_x s &=& Q \,r - 2  \, \bar{q}\, r + u \bar{q}\, s + 2 i X,\\
X & \equiv & - \pa_x  \( \frac{1}{2} V^{(1)} + \bar{q} q \),  \label{Xanom}
\er
where $Q$ is an arbitrary field. So, one has a set of two deformed Riccati-type equations for the pseudo-potential $u$ (\ref{ricc1})-(\ref{ricc2}) and a system of equations (\ref{ricc1r})-(\ref{ricc2s}) for the auxiliary fields $r$ and  $s$. 

Notice that, for the integrable AKNS model one has the potential 
\br
\label{nlspot}
V_{NLS}(\bar{q} q) = - \( \bar{q} q \)^2\, \rightarrow \, V_{NLS}^{(1)}(\bar{q} q) = - 2(\bar{q} q),
\er
and, therefore,  $X=0$ in (\ref{Xanom}), and so the auxiliary system of  eqs. (\ref{ricc1r})-(\ref{ricc2s}) possesses the trivial solution $r=s=0$. Inserting this trivial solution into the system  (\ref{ricc1})-(\ref{ricc2}) and considering  the potential (\ref{nlspot}), one has a set of two Riccati equations for the standard AKNS model and they  play an important role in order to study its properties, such as the derivation of the infinite number of conserved charges and the Backlund transformations, relating the fields $(q ,\bar{q})$ with another set of solutions $(q',\bar{q}')$ \cite{prl1}.

Note that only the $t-$component $\pa_t u$ of the Riccati equation associated to the ordinary AKNS equation has been  deformed away from the AKNS potential (\ref{nlspot}), and it carries all the information regarding the deformation of the model which are encoded in the potential $V(\bar{q} q )$ and the auxiliary fields $r(x,t)$ and  $s(x,t)$. The form of the $x-$component $\pa_x u $ remains the same as the usual Riccati equation associated to the AKNS model. 
 
We have computed the compatibility condition   $[\pa_{t}\pa_x  u - \pa_x \pa_{t} u ]=0$ for the Riccati-type equations (\ref{ricc1})-(\ref{ricc2}), taking into account  the auxiliary system of equations (\ref{ricc1r})-(\ref{ricc2s}) and then derived  the eqs. of motion for the fields $q $ and $\bar{q}$
\br
\label{mnls11}
i \pa_t q + \pa^2_x q - V^{(1)} q &=& 0,\\
-i \pa_t \bar{q} + \pa^2_x \bar{q} - V^{(1)} \bar{q} &=& 0.\label{mnls22}
\er
This is a modified $sl(2)$ AKNS system (MAKNS) for arbitrary potential of type $V(\bar{q} q)$. An important observation in the constructions above is that $\frac{\pa}{\pa t} \zeta =0$, as it can be checked by direct computation using 
 the system (\ref{ricc1})-(\ref{ricc2}) and (\ref{ricc1r})-(\ref{ricc2s}), provided that the system of eqs. (\ref{mnls11})-(\ref{mnls22}) is satisfied. So, the modified system MAKNS possesses an isospectral parameter $ \zeta $.

The standard NLS model (\ref{nls0}) can be obtained  provided that the identifications  
\br
\label{nlspsi}
q \equiv  i (-\eta)^{1/2}\, \psi,\,\,\, \bar{q} \equiv  -  i (-\eta)^{1/2} \, \bar{\psi},\,\,\,\,\eta \in \IR
\er 
are performed, where $\bar{\psi}$ stands for complex conjugation of the field $\psi$ and the potential and its derivatives are taken as in (\ref{potder}). This is a process, we have just mentioned above, through which the standard NLS model is obtained as a special reduction of the AKNS system.

Let us emphasize that for the standard NLS model we have the  trivial solution of the system  (\ref{ricc1r})-(\ref{ricc2s}), i.e. $X=0 \rightarrow r=s=0$, and the existence of the Lax pair of de ordinary NLS model reflects in its equivalent Riccati-type representation, provided by the system (\ref{ricc1})-(\ref{ricc2}) with the well known potential (\ref{potder}) \cite{nucci, prl1}. 

We define the quasi-integrable MAKNS model for field configurations $q$ and $\bar{q}$ satisfying (\ref{mnls11})-(\ref{mnls22}) such that the fields and the deformed potential transform under the space-time transformation (\ref{par1}) as
\br 
\label{aknsparity}
\widetilde{{\cal P}}(q)  = \bar{q},\,\,\,\,\,\, \widetilde{{\cal P}}(\bar{q})  = q,\,\,\,\,\,\mbox{and } \,\,\,\,\,\widetilde{{\cal P}}[ V(\rho) ] = V(\rho).
\er
Under this transformation one has that  $X$  from (\ref{Xanom}) becomes an odd function
\br
\label{Xtr}
\widetilde{{\cal P}}( X ) =  - X.
\er

Next, let us discuss the relevant conservation laws in the context of the Riccati-type system (\ref{ricc1})-(\ref{ricc2}) and the auxiliary equations (\ref{ricc1r})-(\ref{ricc2s}). So, substituting the expression for $u^2$ from (\ref{ricc1}) into (\ref{ricc2}) and considering (\ref{ricc1r})-(\ref{ricc2s}), one gets the following relationship
\br
\label{qcons}
\pa_t [ i \bar{q} \,u  ] - \pa_x \Big[ 2 i \zeta \bar{q} \, u - \bar{q} q +  u \, \pa_x \bar{q} \Big] &=& i \bar{q} (r - s \, u). 
\er

Defining the r.h.s. of (\ref{qcons}) as
\br
\label{qcons11}
\chi \equiv i \bar{q} (r - s \, u),
\er 
and using the system  (\ref{ricc1})-(\ref{ricc2}) and the auxiliary equations (\ref{ricc1r})-(\ref{ricc2s}) one can write a first order differential equation for the auxiliary field $\chi$ 
\br
\label{chi0}
\pa_x \chi &=& \( -2 i \zeta + 2 u \bar{q} + \frac{\pa_ x \bar{q}}{\bar{q}}\) \chi + 2 \bar{q}\, u X.
\er

The eqs. (\ref{qcons}) and (\ref{chi0}) will be used below in order to uncover an infinite tower of quasi-conservation laws associated to the modified AKNS model (\ref{mnls11})-(\ref{mnls22}).  We will construct  the relevant charges order by order in powers of the parameter $\zeta$. So,  let us consider the expansions
\br
\label{expan}
 u = \sum_{n=1}^{\infty} u_n \, \zeta^{-n},\,\,\,\,  \chi= \sum_{n=1}^{\infty} \chi_n  \zeta^{-n-1}.\er
 
The coefficients $u_n$ of the expansion above  can be determined order by order in powers of $\zeta$ from  the Riccati equation (\ref{ricc1}).  In appendix \ref{fsca1} we provide the recursion relation for the $u_n\, 's$ and the expressions for the first $u_n$. Likewise, using the results for the $u_n\, 's$ we get the relevant expressions for the  $\chi_n\, 's$ from (\ref{chi0}). The first components $\chi_n$ are provided in appendix \ref{ap:chi}.

Then, making use of the $u_n$ and $\chi_n$ components of the expansions of $u$ and $\chi$, respectively, provided in (\ref{expan}), one can find the conservation laws, order by order in powers of $\zeta$. So, by inserting those expansions into  the eq. (\ref{qcons}) one has that  the coefficient of the $n'$th order term becomes
\br
\label{anocons}
\pa_{t} a_{x}^{(n)} &+& \pa_x a_{t}^{(n)} = \chi_{n-1},\,\,\,\,\,n=0,1,2,3,....;\, \chi_{0}\equiv 0\\
a_{x}^{(n)} &\equiv & i \bar{q}\, u_n ,\,\,\,\,a_{t}^{(n)} \equiv   -\( 2 i \bar{q} \, u_{n+1} - \bar{q} q \, \delta_{0, n}  + \pa_x \bar{q} \, u_{n}  \),\,\,\,\,\,u_0 \equiv 0.
\er
Notice that making the substitution $\chi_{n-1} \equiv 0$ into the eq. (\ref{anocons}) one can get the tower of exact conservation laws of the usual AKNS system. A truly conservation law character of this equation, at each order $n$, remains to be clarified, since the field components $\chi_{n-1}$ in the r.h.s. of (\ref{anocons}), as they can be seen in the appendix \ref{ap:chi}, do not present the adequate forms to be directly incorporated into  the l.h.s. of the conservation laws. We will tackle this construction order by order for each $\chi_{n-1}$ component.  Notice that analogous quasi-conservation laws have been obtained in the context of the anomalous zero-curvature formulation of the modified NLS model and its associated anomalous Lax pair in \cite{jhep3}.  

We will show below that the r.h.s. of (\ref{anocons}) for  $\chi_1, \chi_2$ and $\chi_3$ can be written in the form $ \chi_j \equiv \pa_{x} \chi^x_j + \pa_{t} \chi^t_j$, with $\chi^x_j$ and $\chi^t_j$ being certain local functions of $\{\bar{q}, q, V\}$ and their $x$ and  $t-$derivatives; i.e. there exist local expressions for some $\chi_j\, (j=1,2,3)$, such that the eq. (\ref{anocons}) provides a proper local conservation law. 

Let us compute the charges order by order in $n$ using the eq. (\ref{anocons}) and the relevant expressions presented in the appendices \ref{fsca1} and \ref{ap:chi}.  

The  {\bf zero'th} order provides a trivial identity.
 
{\bf The order $n=1$ and the field normalization}

In this case the anomaly is trivial $\chi_0 =0$. So, one has
\br
\label{n1}
\pa_t \( \frac{1}{2} \bar{q} q \) - \pa_x \( \frac{1}{2} i \bar{q} \pa_x q -  \frac{1}{2} i q \pa_x \bar{q}   \) =0.
\er
It provides the conserved charge
\br
\label{n1nor}
N = \int dx \, \bar{q} q
\er

{\bf The order $n=2$ and momentum conservation}

At this order one has 
\br
\label{n2}
\pa_t \( \frac{1}{4} i \bar{q} \pa_x q\) + \pa_x \( \frac{1}{4} [ (\bar{q} q)^2 + \bar{q} \pa_x^2 q - \pa_x \bar{q}\pa_x q]\) &=&\chi_1.\er
The function  $\chi_1$ can be rewritten as
\br
\label{chi1}
\chi_1 &=& \frac{1}{2} \pa_x [ F(\rho) ],\,\,\,\,\rho \equiv \bar{q} q . \\
\label{fv}
F(\rho) & \equiv &\frac{1}{2} \rho \frac{d}{d\rho}V(\rho)-\frac{1}{2} V(\rho)+\frac{1}{2} \rho^2. 
\er
So, from (\ref{n2}), taking into account  (\ref{chi1}), one can write the conserved charge
\br
\label{n2mo}
P =   i \int dx \, \( \bar{q} \pa_x q - q \pa_x \bar{q} \)
\er

{\bf The order $n=3$ and energy conservation}

One has the conservation law
\br
\label{n3}
\pa_t [ -\frac{1}{8} (\bar{q} q)^2 - \frac{1}{8} \bar{q} \pa^2_x q ] -\pa_x \(2 i \bar{q} u_4 + \pa_x \bar{q} u_3 \) &=& \chi_2.       
\er
The function $\chi_2$ can be rewritten as
\br
\label{chixt}
 \chi_2 \equiv  -\frac{1}{8}  \pa_t V -\frac{1}{8} \pa_t (\bar{q} q)^2 - \frac{1}{4} i \pa_x\Big[ X \bar{q} q - X (q  \pa_x \bar{q} - \bar{q}  \pa_x q)\Big].
\er
So, (\ref{n3}) provides the conserved charge
\br
\label{energy}
H_{MNLS} =   \int dx \, [\, \pa_x \bar{q} \pa_x q + V( \bar{q} q)  \,].
\er
Notice that in order to get the identity (\ref{chixt}) we have used the eqs. of motion (\ref{mnls11})-(\ref{mnls22}).
Since we have considered $\chi_2 \neq 0$ in the r.h.s. of (\ref{n3}), which carries the effect of the modified potential,  the expression of the energy  (\ref{energy})  is valid for the general MNLS model.  In particular, for the ordinary AKNS the energy follows directly  from the l.h.s. of (\ref{n3}) (provided that $\chi_2 =0 $ in the r.h.s.  of that eq.), i.e.  $H_{NLS} =   \int dx \, [\, \pa_x \bar{q} \pa_x q + V_{NLS}( \bar{q} q)  \,]$, where $V_{NLS} = - (\bar{q} q )^2$ as in (\ref{nlspot}).

{\bf The order $n=4$:  A first trivial charge and its associated anomalous charge}

One has 
\br
\label{charge4}
\pa_t \(- \frac{3}{16} i \bar{q} q \bar{q} \pa_x q -\frac{1}{32} i \pa_x (\bar{q} q)^2  - \frac{1}{16} i \bar{q} \pa^3_x q \)- \pa_x [ 2 i \bar{q} \, u_5 + \pa_x \bar{q} \, u_4 ] &=& \chi_3.
\er
Remarkably, the expression for $\chi_3$ can be written as  
\br
\label{chi3xt}
\chi_3 & \equiv & \pa_x [ \chi^{(3)}_{x} ] + \pa_t [ \chi^{(3)}_{t} ],\\
   \chi^{(3)}_{t} & = & - \frac{3}{16} i \bar{q} q \bar{q} \pa_x q  - \frac{1}{16} i \bar{q} \pa^3_x q, \label{chi3t}\\
   \label{chi3x}
    \chi^{(3)}_{x} & = & \frac{3}{8} X \bar{q} \pa_x q + \frac{3}{8}  H_1(\rho) + \frac{1}{8} \bar{q} q  \pa_x X - \frac{1}{8} X \pa_{x} (\bar{q} q) - \frac{1}{16} [\pa_{x} (\bar{q} q) ]^2+ \frac{3}{8} \bar{q} q \pa_{x}\bar{q} \pa_x q- \frac{3}{8} H_2(\rho) + \nonumber\\ 
&& \frac{3}{32} i \bar{q} \pa_t q^2 - \frac{1}{16} V^{(1)} \( q \pa^2_x \bar{q} + \bar{q} \pa^2_x q  \) + \frac{1}{16}   V^{(1)}  \pa_x q \pa_x \bar{q} +\nonumber\\ 
&& \frac{1}{16} i [\pa_t q \pa^2_x \bar{q} - \pa_x \pa_t q \pa_x \bar{q} + \bar{q} \pa_t \pa^2_x q].  
    \\   
    \frac{d}{d\rho}H_1(\rho) & \equiv & -(V^{(2)}/2+1) \rho^2,\,\,\,\,\, \frac{d}{d\rho}H_2(\rho) \equiv \rho V^{(1)},\,\,\,\,\,\rho \equiv \bar{q}  q,
\er

Therefore, at this order, the eq. (\ref{charge4}) can be written  as an exact conservation law. However, taking into account the  term $\pa_t \chi^{(3)}_{t}$ of $\chi_3$ in (\ref{chi3xt})-(\ref{chi3t}) and the relevant terms in  the l.h.s. of (\ref{charge4}) with partial $t-$derivatives one gets a fourth order trivial charge $Q^{(4)}=0$, provided that the surface term $\sim \pa_x (\bar{q} q)^2$ is dropped, since upon integration in $x$ in order to define the charge it vanishes for suitable boundary conditions. So, at this order of the above formulation, the charge $Q^{(4)}$ trivially vanishes.

However,  at this order and in the higher order ones, one can define an asymptotically conserved charge for the MAKNS model 
\br
\label{charge4a}
Q^{(4)}_a =  \frac{i}{2} \int dx \, \Big[ 3 \bar{q}  q \( \bar{q} \pa_x q - q \pa_x \bar{q} \) + \bar{q} \pa^3_x q - q \pa^3_x \bar{q}  \Big],
\er
such that 
\br
\label{qa44}
\frac{d}{dt}  Q^{(4)}_a     &=& \hat{\tau}\\
\hat{\tau} &=& - 8 \int\, dx \, \chi_3,\label{qa441}\\
\label{qa442}
                 &=&  -  \int\, dx \, [ 3(\bar{q} q)^2 X + \bar{q} q \pa^2_x X - 3 \pa_x \bar{q} \pa_x q X],
\er
where in (\ref{qa441}) the expression of $ \chi_3$ from (\ref{chis}) must be inserted and the final form of the  anomaly density in (\ref{qa442} ) is obtained by dropping a surface term. Notice that the anomaly density in (\ref{qa442}) possesses an odd parity under (\ref{par1}) and (\ref{aknsparity}) taking into account that $X$ is an odd function according to (\ref{Xtr}).  Therefore, one has $\int dt \int dx\, \chi_3 = 0$ implying the asymptotically conservation of the charge $Q^{(4)}_a$.   

The charge $Q^{(4)}_a $ in (\ref{charge4a}) takes the same form as the fourth order charge in the standard AKNS model. In fact, when the r.h.s. of (\ref{charge4}) vanishes, i.e. $\chi_3 =0$, one has a charge similar in form to the one in (\ref{charge4a}),  conveniently rewritten by discarding surface terms.  
Taking into account the reduction process (\ref{nlspsi}) one can get a similar anomalous charge for the MNLS model, as presented in  \cite{jhep3, jhep4, jhep5}. In fact, upon the reduction  (\ref{nlspsi})  the anomalous charge  $Q^{(4)}_a $ in (\ref{charge4a}) corresponds to the one for the MNLS model in (\ref{q5nlsd}).

{\bf The order $n=5$ and the quasi-conserved charge}

At this order one has
\br
\label{charge5}
 \frac{1}{32} \pa_t  \Big[ 2 (\bar{q} q)^3  + 5 \bar{q}^2 (\pa_x q)^2 + 6 \bar{q}q \( \pa_x q \pa_x \bar{q} + \bar{q}  \pa^2_x q\) + \bar{q} q^2 \pa^2_x \bar{q} + \bar{q} \pa^4_x q  \Big]- \pa_x [ 2 i \bar{q} \, u_6 + \pa_x \bar{q} \, u_5 ] = \chi_4.
\er
Likewise, the expression for $\chi_4$ can be written as  
\br
\label{chi4xt}
\chi_4 & \equiv & \pa_x [ \chi^{(4)}_{x} ] +\frac{1}{16} \pa_t [ Z(\rho)] +\beta_1 ,\\
   \label{chi4x}
    \chi^{(4)}_{x} & = & \frac{i}{16} \Big\{Z^{(1)}(\rho) ( q \pa_{x} \bar{q} - \bar{q} \pa_x q ) +   6 \bar{q} \pa^2_{x} q X + 4 \bar{q} \pa_{x} q \pa_x X  - 4 \pa_x ( \bar{q} \pa_{x} q) X - \pa_x ( \bar{q}  q) \pa_x X + \pa^2_x ( \bar{q} q) X + \nonumber \\
   &&  \bar{q} q \pa^2_{x}X - (\frac{1}{2} V^{(1)} +\bar{q} q )[-6 \pa_x(\bar{q} \pa^2_{x} q ) + 4 \pa^2_x (\bar{q} \pa_{x} q  ) + 4 \bar{q} \pa^3_{x} q - \pa^3_x (\bar{q} q)]  \Big\} \nonumber\\
\label{anomaly1}
\beta_1 &=&  \frac{i}{32} \( \frac{2 \bar{q} q}{V^{(1)} } + 1\)  \( \bar{q} \pa_{x}^4 q - q \pa^4_{x} \bar{q}  \) V^{(1)}
     \\     
 \label{zz1}   \frac{d}{d\rho} Z(\rho) & \equiv &   6 \int_{\rho_0}^{\rho} \hat{\rho} [ \frac{1}{2} V^{(2)}(\hat{\rho}) + 1 ] \, d\hat{\rho},
\er
where the function $\beta_1$ defines the anomaly associated with the quasi-conservation law (\ref{charge5}).  Let us write the next identity
\br
\( \bar{q} \pa_{x}^4 q - q \pa^4_{x} \bar{q}  \)  V^{(1)} &=&  \Big[ (-i \pa_t \bar{q} + \pa_x^2 \bar{q}) \pa_{x}^4 q - (i \pa_t q + \pa_x^2 q ) \pa^4_{x} \bar{q}  \Big] \\
&=& \Big[ \pa_{x} {\cal M} - i \pa_t (  \bar{q} \pa^4_x q )  \Big]     \label{id11}
    \\   
  {\cal M} &\equiv & \pa^2_x \bar{q} \pa^3_x q - \pa^3_x \bar{q} \pa^2_x q  - i \pa^3_x \bar{q} \pa_t q  + i \pa^2_x \bar{q} \pa_x \pa_t q -i \pa_x \bar{q} \pa^2_x \pa_t q + i \bar{q} \pa^3_x \pa_t q, 
\er
which is derived by using the eqs. of motion (\ref{mnls11})-(\ref{mnls22}). Therefore, using (\ref{id11}) the anomaly $\beta_1$ can be written as
\br
\label{anomaly11}
\beta_1 &=&  \frac{1}{32} \( \frac{2 \bar{q} q}{V^{(1)} } + 1\)  \Big[ \pa_{x} {\cal M} - i \pa_t (  \bar{q} \pa^4_x q )  \Big] .
\er

Next, taking into account the  relevant terms of $\chi_4$  in (\ref{chi4xt}) and the terms in  the l.h.s. of (\ref{charge5}) with partial $t-$derivatives and discarding the boundary terms with partial $x-$derivatives one can define the fifth order quasi-conserved charge 
\br
 \label{charge5a}
\frac{d}{dt} Q^{(5)}_a &=&  \int dx\, \beta_1,\\
Q^{(5)}_a & \equiv & \frac{1}{32} \int dx\,  \Big[ 2 (\bar{q} q)^3  - 8 \bar{q}q \pa_x q \pa_x \bar{q} - \bar{q}^2   (\pa_x q)^2 - q^2 (\pa_x \bar{q})^2 + \pa_x^2 \bar{q} \pa_x^2 q - 2 Z(\rho) \Big] \label{charge5a1},
\er
where the anomaly $\beta_1$ can take the form (\ref{anomaly1}) or, alternatively, the form (\ref{anomaly11}). Notice that the form of the anomaly in (\ref{anomaly1}) possesses an odd parity under (\ref{par1}) and  (\ref{aknsparity}). Therefore, one has $\int dt \int dx \, \beta_1 = 0$ implying the asymptotically conservation of the charge $Q^{(5)}_a$.

Therefore, the fifth order eq. (\ref{charge5}) has been written  as a quasi-conservation law. Through the reduction process (\ref{nlspsi}) one can get an anomalous charge and its relevant anomaly $\beta_1$ at this order for the MNLS model, as presented in  \cite{jhep3, jhep4, jhep5}.  In fact, upon the reduction  (\ref{nlspsi})  the anomalous charge  $Q^{(5)}_a$ in (\ref{charge5a})-(\ref{charge5a1})  can be identified, dropping surface terms, to the one for the MNLS model in (\ref{q6lc}) and (\ref{q66}).

Notice that, in the usual AKNS  limit, i.e. when $V^{(1)} = - 2 \bar{q} q$ and $V^{(2)} = - 2$ for the AKNS potential as in (\ref{nlspot}), the factor $\(\frac{2 \bar{q} q}{V^{(1)}} + 1\)$ of the anomaly $\beta_1$ in (\ref{anomaly1}) vanishes, and the term $Z(\rho)$ in the density of the charge (\ref{charge5a1}) can be set to zero (see (\ref{zz1})). Therefore, the quasi-conserved charge $Q^{(5)}_a$ in (\ref{charge5a}) becomes the fifth order charge $Q^{(5)}$  of the usual AKNS model. Actually, in this limit one has that $\chi_4 =0$ for $X=0$ (see \ref{chis}), then the r.h.s. of (\ref{charge5}) vanishes, and so, this eq. can be written as an exact conservation law.

So, we have constructed the set of (quasi-)conservation laws of type  (\ref{anocons}) using the Riccati-type approach  of the  modified AKNS model. By a suitable reduction process these charges can be identified to the ones of the MNLS model, as discussed above. In ref. \cite{jhep3} in the context of the anomalous Lax pair formulation of modified NLS models and through the abelianization procedure it has been constructed an infinite set of asymptotically conserved charges, which are similar in form to the exact conserved charges of the standard NLS model. 
 
\section{Dual Riccati-type formulation and novel anomalous charges}
\label{sec:dual1}

In this section we will derive the novel anomalous conservation laws through the dual formulation of the Riccati-type pseudo-potential approach. So, in order to discuss a dual formulation, let us rewrite the Riccati-type system (\ref{ricc1})-(\ref{ricc2}) and the auxiliary eq. (\ref{chi0}) as 
\br
\label{ricc1n1}
\pa_x u &=& -2 i \zeta \, u + q + \, \bar{q} \,\, u^2,\\
\label{ricc2n1}
\pa_t u &=&  2 A\, u - C\, u^2 + B - i \frac{\chi}{\bar{q}},\\
\pa_x \chi &=& \( -2 i \zeta + 2 u \bar{q} + \frac{\pa_ x \bar{q}}{\bar{q}}\) \chi + 2 \bar{q}\, u X \label{chi01},
\er
where the $A, B$ and $C$ functions are defined in (\ref{A11})-(\ref{A33}).  

Notice that the system of differential eqs. (\ref{mnls11})-(\ref{mnls22}) is invariant under the transformations: $q  \leftrightarrow \bar{q}$ and $i \leftrightarrow -i$. So, a dual formulation of the Ricati-type system (\ref{ricc1n1})-(\ref{chi01}) is achieved by performing the changes $q  \leftrightarrow \bar{q}$ and $i \leftrightarrow -i$, $u \rightarrow \bar{u}$ and $\chi \rightarrow \bar{\chi}$ into the system above. So, one has
\br
\label{ricc1d}
\pa_x \bar{u} &=& 2 i \zeta \, \bar{u} + \bar{q} + \, q \,\, \bar{u}^2,\\
\label{ricc2d}
\pa_t \bar{u} &=&  2 \bar{A}\, \bar{u} - \bar{C}\, \bar{u}^2 + \bar{B} +i \frac{\bar{\chi}}{q},\\
\label{chi0d}
\pa_x \bar{\chi} &=& \( 2 i \zeta + 2 \bar{u} q + \frac{\pa_ x q}{q}\) \bar{\chi} + 2 q\, \bar{u} X.
\er
where $\bar{u}$ is a new  Riccati-type pseudo-potential and $\bar{\chi}$ is a new  auxiliary field. Notice that the function $X$ defined in (\ref{Xanom}) remains the same. The functions $\bar{A},\bar{B}$ and $ \bar{C}$ become
\br
\label{A11d}
\bar{A} & \equiv &  2 i \zeta^2 + \frac{1}{2} i\, V^{(1)},\\
\label{A22d}
\bar{B} & \equiv & 2 \zeta \bar{q} - i \pa_x \bar{q},\\
\label{A33d}
\bar{C} & \equiv & -2 \zeta q - i \pa_x q.
\er
It is a simple calculation to verify that this dual Riccati-type system (\ref{ricc1d})-(\ref{chi0d}) reproduces the eqs. (\ref{mnls11})-(\ref{mnls22}).  

Considering  the expansions
\br
\label{expandual}
\bar{u} = \sum_{n=1}^{\infty} \bar{u}_n \, \zeta^{-n},\,\,\,\,  \bar{\chi}= \sum_{n=1}^{\infty} \bar{\chi}_n  \zeta^{-n-1},
\er
the coefficients $\bar{u}_n$ and $\bar{\chi}_n\, 's$ can be obtained from (\ref{ricc1d}) and (\ref{chi0d}), respectively.  The first components  are provided in appendix \ref{app:uchid}.

For the fields $q, \bar{q}$ and $X$ satisfying the transformation laws (\ref{aknsparity}) and (\ref{Xtr}), respectively, one can verify from the system of dual equations  (\ref{ricc1n1})-(\ref{chi01}) and (\ref{ricc1d})-(\ref{chi0d})  the following symmetry transformations 
\br
\label{ubu}
\widetilde{{\cal P}}(u) &=& - \bar{u},\,\,\,\,\,\,\,\widetilde{{\cal P}}(\bar{u}) = - u,\\
\widetilde{{\cal P}}(\chi) &=&- \bar{\chi},\,\,\,\,\,\,\, \widetilde{{\cal P}}(\bar{\chi})= - \chi.
\label{chibchi}
\er 
In fact, a careful inspection of the first six and five lowest order components for the expressions of $\{u,\bar{u}\}$  and  $\{\chi, \bar{\chi}\}$, respectively,  provided in the  appendices \ref{fsca1}, \ref{ap:chi} and \ref{app:uchid}, are in accordance, order by order in $n$, with the symmetries above, i.e.
\br
\label{ubun}
\widetilde{{\cal P}}(u_n) &=& - \bar{u}_n,\,\,\,\,\,\,\,\widetilde{{\cal P}}(\bar{u}_n) = - u_n,\,\,\,\,\,\, n=1,2,...,6;\\
\widetilde{{\cal P}}(\chi_n \pm  \bar{\chi}_n) &=& \mp  (\chi_n \pm \bar{\chi}_n),\,\,\,\,\,\,\,\,\,\,\,\,\,\,\,\,\,\,\,\,\,\,\,\,\,\,\,\,\,\,\,\,\,n=1,2,...,5;
\label{chibchin}
\er  
From the both dual systems of Riccati-type eqs. (\ref{ricc1n1})-(\ref{chi01}) and (\ref{ricc1d})-(\ref{chi0d}) one can write the next equations, respectively
\br
\label{qua1d}
\pa_t(i \bar{q} u) -\pa_x (2i \zeta \bar{q} u - \bar{q}q+ u \pa_x \bar{q}) = \chi
\er
and 
\br
\label{qua2d}
\pa_t(i q \bar{u}) -\pa_x (2i \zeta q \bar{u} + \bar{q}q - \bar{u} \pa_x q) = -\bar{\chi},
 \er
where (\ref{qua1d}) has already been considered in (\ref{qcons}) with $\chi$ defined in (\ref{qcons11}). Subtructing  the b.h.s. of (\ref{qua1d}) and (\ref{qua2d}) one has 
\br
\label{quasidual}
\pa_t [i \bar{q} u-i q \bar{u}] -\pa_x [2i \zeta (\bar{q} u -q \bar{u})- 2\bar{q}q+ u \pa_x \bar{q} + \bar{u} \pa_x q] = \chi + \bar{\chi}.
\er
Notice that the r.h.s. of the last equation is an odd expression under the special space-time operator; i.e. taking into account (\ref{chibchi}) one has $ \widetilde{{\cal P}}(\chi + \bar{\chi})= - (\chi+ \bar{\chi})$. So, the equation (\ref{quasidual}) defines  a quasi-conservation law. In fact,  the first five lowest order components  are indeed odd functions as written in (\ref{chibchin}). A usual computation shows that the components of the expansion in powers of $\zeta^{-n}$ of the quasi-conservation law (\ref{quasidual}) give rise to the normalization $(n=1)$, momentum $(n=2)$ and energy $(n=3)$ conserved charges; whereas, the higher order ones provide the same anomalous charges as the ones discussed in sec. \ref{sec:riccati}. An important observation is that the density charges of the (quasi-)conservation eq. (\ref{quasidual}) are even functions, since  the expression $[i \bar{q} u-i q \bar{u}]$ inside the partial time derivative in the l.h.s. of (\ref{quasidual}) is an even parity function.

However, the summation of the b.h.s. of (\ref{qua1d}) and (\ref{qua2d}) will not reproduce a quasi-conservation law,  since the anomaly $(\chi - \bar{\chi})$ is an even function according to  (\ref{chibchi}). In addition, in this case the expression $[i \bar{q} u+i q \bar{u}]$ of the charge density  will be an odd parity function, furnishing a trivial charge.

In the following we construct new towers of quasi-conservation laws with true anomalies, i.e. expressions  with odd parities under the symmetry transformation (\ref{par1}). Let us consider even parity expressions of the types:  $ \bar{u}u$, $i(\bar{u}\pa_x u- u \pa_x\bar{u})$,  $ \pa_x \bar{u} \pa_x u$, $ i(\bar{u}\pa_x^3 u- u \pa_x^3\bar{u})$, $i\bar{u} u (\bar{u}\pa_x u- u \pa_x\bar{u}),...$ So, one can write an infinite tower of quasi-conservation laws on top of every monomial or polynomial of these types. Next, we show the first examples of this novel set of infinite number of quasi-conservation laws.   

The next equation  follows from the both dual systems of eqs. (\ref{ricc1n1})-(\ref{chi01}) and (\ref{ricc1d})-(\ref{chi0d}) 
\br
\label{nor1}
\pa_t [\frac{1}{k}( \bar{u} u )^{k}] - \pa_x[i (\bar{u} u)^{k-1}(\bar{u} \pa_x u - u \pa_x\bar{u})] &=& {\cal A}^{(k)},\,\,\,\,\,\,\,\,\,\, k=1,2,3,...\\
{\cal A}^{(k)} &=& (\bar{u} u)^{k-1} {\cal A}-i(k-1)(\bar{u} u)^{k-2}[(\bar{u} \pa_x u)^2-(u \pa_x\bar{u})^2],\\
 {\cal A} &\equiv& 4 \zeta \bar{u} u [\bar{q} u + q \bar{u}  ] + 2 i \bar{u} u [\bar{u}\pa_x q - u  \pa_x\bar{q}  ] - 2 i \bar{u} u [\bar{q}^2u^2 - q^2 \bar{u}^2  ]+\nonumber\\
&& i [u \frac{\bar{\chi}}{q} -  \bar{u} \frac{\chi}{\bar{q}}].\label{and1}
\er
Notice that  ${\cal A}$ is an odd function, and so is the general function ${\cal A}^{(k)}$ for any positive integer $k$. Remarkably, the anomaly  ${\cal A}^{(k)}$ encompasses two types of anomalies. In fact, the last terms of ${\cal A}$ in (\ref{and1}) show the auxiliary potentials  $\bar{\chi}$ and $\chi$ which incorporate the information of the modification of the AKNS model. The remaining terms of ${\cal A}^{(k)}$ do not depend explicitly on those fields; and so, they will be present even in the standard AKNS model, i.e. for $\bar{\chi}=\chi=0$.  This is our first description, in the pseudo-potential approach, of the presence of these type of quasi-conservation laws even for a truly integrable system.         

Let us examine the first three lowest order equations  $(n=2,3,4)$, for the case $k=1$ in (\ref{nor1}).  The first two orders $n = 2, 3$ correspond, up to overall constant factors, to the exact conservation laws for the normalization and momentum charges, (\ref{n1nor}) and (\ref{n2mo}), respectively. The order $n=4$ can be written as 
\br
\nonumber
\pa_t \{\pa_x \bar{q} \pa_x q -\bar{q} \pa^2_x q - q \pa^2_x \bar{q} -2 (\bar{q} q)^2\}&-&\pa_x \{i [2(\pa_x\bar{q}\pa^2_x q-\pa_x q\pa^2_x \bar{q})-(\bar{q}\pa^3_x q- q\pa^3_x \bar{q})-\\
&& 2 \bar{q}q (\bar{q}\pa_x q - q \pa_x\bar{q})]\}= {\cal A}_0 + {\cal A}_{X}\label{n4dd}\\
{\cal A}_0 &\equiv&-2i \bar{q}q (\bar{q}\pa^2_x q - q \pa^2_x\bar{q}) -4i [(\bar{q}\pa_x q)^2 - (q \pa_x\bar{q})^2]\\
{\cal A}_X &\equiv&- 6i (\bar{q}\pa_x q - q \pa_x\bar{q}) X,
\er
with the odd functions ${\cal A}_0$ and ${\cal A}_X$ defining the relevant anomaly $({\cal A}_0+{\cal A}_X)$. Notice that ${\cal A}_X$ contains the deformation variable, $X$, and it will vanish for the standard AKNS model, i.e. ${\cal A}_X =0$. Whereas, the term ${\cal A}_0$ will be present as an anomaly even for the usual AKNS model. 
 
Next, let us examine the first two lowest order conservation laws for $k=2$ in (\ref{nor1}).  The first non-trivial quasi-conservation law is for the order $n = 4$
\br
\pa_t [(\bar{q} q)^2] - \pa_x [i \bar{q} q (\bar{q}\pa_x q - q \pa_x\bar{q})] = - i \bar{q} q [(\bar{q}\pa_x q)^2-(q \pa_x\bar{q})^2],
\er
where the r.h.s. is an odd function which does not depend explicitly on the deformation variable $X$. This quasi-conservation law corresponds, up to an overall factor $\frac{1}{4}$,  to the eqs. (\ref{rho1t})-(\ref{rho1tan}) (with $n=2$) for the (modified) NLS model provided that the field identifications (\ref{nlspsi}) are performed. 

The next order $n=5$ provides
\br
\nonumber
\pa_t [i \bar{q} q (\bar{q}\pa_x q - q \pa_x\bar{q})] &-& \pa_x \{\frac{1}{2} [ \pa_x q^2 \pa_x\bar{q}^2- ((\bar{q}\pa_x q)^2-(q \pa_x\bar{q})^2)- \bar{q} q  (\bar{q}\pa^2_x q + q \pa^2_x\bar{q})-\\
\nonumber
&&\frac{4}{3} (\bar{q} q)^3]\} = [\bar{q}^2 \pa_x q \pa^2_x q + q^2 \pa_x\bar{q} \pa^2_x \bar{q} ] -[ \bar{q} (\pa_x q)^2 \pa_x \bar{q} + q \pa_x q (\pa_x \bar{q})^2] -\\
&& 2 (\bar{q} q)^2 X.
\er
In this case the anomaly contains a term with $X$ and another terms which will be present in the usual AKNS model for $X=0$.
 
In the following, let us consider the quasi-conservation law
\br
\nonumber
\pa_t [\pa_x \bar{u} \pa_x u ] - \pa_x [i (\pa_x \bar{u} \pa^2_x u - \pa_x u \pa^2_x \bar{u})] &=& 2 \zeta (\pa_x\bar{u} \pa_xq_ + \pa_x u \pa_x\bar{q}) + 2\zeta (\bar{u}^2\pa_x u \pa_x q+ u^2\pa_x \bar{u} \pa_x \bar{q})+\\ \nonumber
&&
2\zeta(\bar{q} \pa_x \bar{u} \pa_x u^2+q \pa_x u \pa_x \bar{u}^2)-
i(\pa_x \bar{q} \pa_x \bar{u} \pa_x u^2-\pa_x q \pa_x u \pa_x \bar{u}^2 )- \\ 
\label{kindual}
&&i(\pa^2_x \bar{q} \pa_x u-\pa^2_x q \pa_x \bar{u} )-i(\pa^2_x \bar{q}\, u^2\pa_x \bar{u} -\pa^2_x q\, \bar{u}^2\pa_x u) - \\
&&i (u \pa_x \bar{u}  -\bar{u} \pa_x u)\pa_x V^{(1)}+
i(\pa_x \bar{q} \pa_x \bar{u} \frac{\chi}{\bar{q}^2}-\pa_x q \pa_x u \frac{\bar{\chi}}{q^2})+\nonumber \\
&&i (\pa_x u \frac{\pa_x \bar{\chi}}{q}-\pa_x \bar{u} \frac{\pa_x \chi}{\bar{q}})-i (\pa_x \bar{u} \pa^3_x u - \pa_x u \pa^3_x \bar{u}). \nonumber
\er
The lowest order of (\ref{kindual} ) is for $n=2$ and it becomes the quasi-conservation law   
\br
\pa_t [\frac{1}{4}\pa_x \bar{q} \pa_x q ] - \pa_x [\frac{i}{4} (\pa_x \bar{q} \pa^2_x q - \pa_x q \pa^2_x \bar{q})] = \frac{i}{4} [(\bar{q} \pa_x q)^2-(q \pa_x \bar{q})^2] V^{(2)}.
\er 
Similarly, this quasi-conservation law corresponds, up to an overall factor $\frac{1}{4}$,  to the eqs. (\ref{kinetic11})-(\ref{an4d}) (with $n=1$) for the modified NLS model provided that the field identifications (\ref{nlspsi}) are performed. 

Analogous construction process for the polynomial $i(\bar{u}\pa_x u- u \pa_x\bar{u})$ will provide the momentum charge, at the lowest order conservation law ($n=2$),  and in top of it an infinite set of anomalous charges with the tower of charges in (\ref{gam3d})-(\ref{gam33d}) as a subset.

We have presented rigorous constructions of the first representatives of novel type of quasi-conservation laws in the pseudo-potential approach. Let us emphasize that those quasi-conservation laws are present in the deformations of AKNS model of the type (\ref{mnls11})-(\ref{mnls22}), as well as in the standard AKNS model for potential (\ref{nlspot}). Certainly, they  deserve a more careful consideration in the context of (quasi-)integrability phenomena and, in general, the dynamics of soliton collisions, which we will address in a future work.   
 
Since the early days of integrable models the presence of an infinite number of conservation laws is among the most important features of integrability, and the presence of anomalous conservation laws, in the N-soliton sector of them, is a novelty. The anomalous charges of deformations of the SG and KdV models have also been performed in the Ricatti-type pseudo-potential approach \cite{npb1, jhep33}), reproducing the results of \cite{npb, jhep1} obtained in the zero-curvature method. In fact, this method reproduces the anomalous charges which possess the same form as the  ones of the standard integrable models. In sec. \ref{secmnlsano} of the present paper we have obtained infinite number of  anomalous charges of the modified NLS by a direct construction method, and in this section we have uncovered novel anomalous charges in the pseudo-potential approach. 

The partial differential equations have been regarded as infinite-dimensional manifolds and  the so-called differential coverings have been introduced, which have been used to construct some structures such as Lax pairs and B\"acklund transformations (see e.g.  \cite{krasil, igonin}). In particular, an auto-B\"acklund transformation is associated to an automorphism of the covering. Then, it would be interesting to study the properties of the dual system  (\ref{ricc1n1})-(\ref{chi01}) and (\ref{ricc1d})-(\ref{chi0d}) as some types of differential coverings of the MAKNS system  (\ref{mnls11})-(\ref{mnls22}).      

\section{Linear system formulation of modified AKNS system}

\label{linear}

In the next steps we will pursue a linear system of equations associated to the deformed AKNS system (\ref{mnls11})-(\ref{mnls22}). In order to tackle this problem we will resort to the Riccati-type formulation of the model presented above; so, in this context we will make use of three of the eqs. presented above,  the Riccati eq.  (\ref{ricc1}), the quasi-conservation law (\ref{qcons}) and the eq. for the auxiliary field $\chi$ in  (\ref{chi0}). Next, let us consider the transformation
\br
\label{tr12}
u = -  \pa_x \log{\phi}, 
\er
where $\phi$ represents a new pseudo-potential. 

With the substitution (\ref{tr12}) the Riccati eq. (\ref{ricc1}) becomes
\br
\label{r1phi}
\phi_{xx} = -[2 i \zeta  - \frac{\pa_x \bar{q}}{\bar{q}} ] \phi_x - \bar{q} q \phi .
\er

Next, consider the quasi-conservation law (\ref{qcons}) and integrate that eq. once in $x$. 
Then one gets 
\br
\label{ss}
s(x,t) = \bar{q} q - \frac{1}{ \phi}\Big[ i \phi_t - 2 i \zeta \phi_x -   \frac{\pa_x \bar{q}}{\bar{q}} \phi_x \Big],
\er
where 
\br
\label{ss1}
s(x,t) \equiv \int^x dx'  \chi.
\er

The auxiliary eq. (\ref{chi0}) upon substitution of (\ref{tr12}) becomes
\br
\label{sxx}
s_{xx} =  -2 X \pa_x \log{\phi} -  [2 i \zeta  - \frac{\pa_x \bar{q}}{\bar{q}} + 2 \pa_x \log{\phi}] s_x .  
\er
The compatibility condition $\pa_t[ \pa^2_{x} \phi ]- \pa^2_{x} [\pa_t \phi] =0$ gives rise to  the eq. of motion of the deformed AKNS (\ref{mnls11})-(\ref{mnls22}). 

Some comments are in order here. First, in the absence of deformations the auxiliary eq. (\ref{sxx}) becomes a trivial one, i.e. $\chi=X=0$ implies $s=0$. Second, one can get the linear system of eqs. (\ref{r1phi})-(\ref{ss}) as the linear formulation of the standard AKNS model, provided that $s\equiv 0$ in the l.h.s. of (\ref{ss}). 

So, following analogous constructions presented in \cite{npb1, jhep33} related to the deformations of the sine-Gordon and KdV models, we look for a linear system of eqs. associated to the deformed AKNS system.  Notice that the function $s$ in (\ref{ss})-(\ref{ss1}) will inherit from $\chi$ in (\ref{chi0}) a highly nonlinear dependence on $u$; then, through the transformation (\ref{tr12}), $s$ will have in general a nonlinear dependence on $\phi$. However, one can argue that the eq. (\ref{ss}) would represent a linear eq. for the pseudo-potential $\phi$ provided that the auxiliary field $s$ is written solely in terms of the fields $q,\bar{q}$ and $X$ and their derivatives. So, let us assume the next Ansatz
\br
\label{lin01}
\pa_x \phi &=& {\cal A}_x \phi,\\
\label{lin02}
\pa_t \phi &=& {\cal A}_t \phi,
\er
where the functions ${\cal A}_x$ and ${\cal A}_t$ represent the gauge connections and depend on the fields of the model. The compatibility condition of this  system will provide the eq. of motion
\br
\label{eqm}
\pa_{t} {\cal A}_x - \pa_x {\cal A}_t =0.
\er  
Using (\ref{lin01}) into (\ref{r1phi}) one gets the following Riccati eq. for ${\cal A}_x$
\br
\label{riccatiAx}
\pa_x {\cal A}_x =  - 2i \zeta  {\cal A}_x - ({\cal A}_x)^2 - \bar{q} q + \frac{\pa_ x \bar{q}}{\bar{q}} {\cal A}_x .
\er
Likewise, replacing (\ref{lin01})-(\ref{lin02}) into  (\ref{ss}) one gets a relationship for the quantity $s$
\br
\label{ss110}
s = \bar{q} q - i {\cal A}_t + ( 2 i  \zeta  +\frac{\pa_ x \bar{q}}{\bar{q}}) {\cal A}_x .
\er
Substituting this form of $s$ into the eq. (\ref{sxx}) and using the eqs.  (\ref{eqm})-(\ref{riccatiAx})  one gets the eq. of motion of the deformed AKNS (\ref{mnls11})-(\ref{mnls22}). So, the form of $s$ in (\ref{ss110}) is consistent with the dynamics of the deformed model.

Notice that the system of eqs.  (\ref{lin01})-(\ref{lin02}) are defined up to a gauge transformation of the type
\br
\label{gauge1}
\phi &\rightarrow& e^{\Lambda} \phi\\
\label{gauge2}
{\cal A}_x &\rightarrow&  {\cal A}_x + \pa_x \Lambda\\
\label{gauge3}
{\cal A}_t &\rightarrow& {\cal A}_t + \pa_t \Lambda,
\er
for an arbitrary function $\Lambda$. We will discuss this point below in more detail.

In order to find a linear system formulation of the modified AKNS it is needed a certain amount of guesswork out of the eq. (\ref{ss}). Moreover, due to the gauge symmetry (\ref{gauge1})-(\ref{gauge3}) it is possible to make a particular choice for the connections ${\cal A}_x$ and ${\cal A}_t$. Let us propose the following linear system of equations as the linear formulation of the deformed AKNS \footnote{Below we will provide a gauge transformation between the systems (\ref{sys1})-(\ref{sys2}) and (\ref{lin01})-(\ref{lin02}).}
\br
\label{sys1}
\pa_t \Phi &=& A_t \Phi\\
\label{sys2}
\pa_x \Phi &=& A_x \Phi\\
\label{AA1}
A_x &\equiv & -\zeta \pa_x\bar{q} + 2 \bar{q} \,\(\frac{a_0 \zeta +\zeta^2 a_1 }{2 \zeta \bar{q}+ i \pa_x\bar{q}}\),\\
A_t &\equiv & \zeta \, \int^{x}\, dx' \, \frac{b_0 + b_1 \zeta + b_2 \zeta^2}{(2 \zeta \bar{q}+ i \pa_{x'}\bar{q})^2},\label{AA2}
\er
such that 
\br
\nonumber
b_0 &\equiv& \frac{1}{2} i \pa_x \bar{q}^2 (2 \pa_t a_0 + \pa_x V^{(1)} \pa_x \bar{q} ) +i(\pa_x \bar{q})^2 (V^{(1)} \pa_x\bar{q} -\pa^3_x \bar{q})+2 a_0 (\bar{q}^2 \pa_x V^{(1)} + \pa_x \bar{q} \pa^2_x\bar{q} - \bar{q} \pa^3_x \bar{q})\\
b_1 &\equiv& 2 \bar{q}^2 [2 \pa_t a_0 + \pa_x V^{(1)} (a_1 + 2 \pa_x \bar{q})] + 2 a_1 \pa_x \bar{q} \pa^2_x \bar{q} + \bar{q} [2i \pa_t a_1 \pa_x \bar{q} + 4 V^{(1)} (\pa_x \bar{q})^2-2 (a_1 + 2 \pa_x \bar{q} )\pa^3_x\bar{q}],\nonumber\\
b_2 &\equiv& 8 \bar{q}^2 [\pa_t a_1 - i (\bar{q} \pa_x V^{(1)} + V^{(1)} \pa_x \bar{q} - \pa^3_x\bar{q})],\nonu
\er  
where $a_0$ and $a_1$ are some nonvanishing auxiliary functions.
The compatibility condition of the system of eqs. (\ref{sys1})-(\ref{sys2}); i.e. $\pa_t \pa_x (\Phi)-\pa_x \pa_t (\Phi)=0$, furnishes the next equation
\br
\label{eqm1}
\pa_{t} A_x - \pa_x A_t =0.
\er
Substituting the expressions of the connection components  $\{A_x\,,\, A_t\}$ defined in (\ref{AA1})-(\ref{AA2}) into (\ref{eqm1})  one gets the next expression, which is a polynomial in powers of $\zeta$ 
\br
\label{pol1}
\{ 8 i \bar{q}^2\pa_x(V^{(1)} \bar{q} + i \pa_t \bar{q} - \pa^2_x \bar{q})\} \zeta^2 -\\ 
\label{pol2}
 \{2 a_1 [\pa_x \bar{q}(\pa^2_x \bar{q}-i \pa_t \bar{q}) + \bar{q} (\bar{q} \pa_x V^{(1)} + i \pa_x \pa_t \bar{q} - \pa^3_x \bar{q}) ] +2 \pa_x\bar{q}^2 \pa_x[V^{(1)} \bar{q} + i \pa_t \bar{q} - \pa^2_x \bar{q}   ]   \} \zeta-\\
\{2 a_0 [\pa_x  \bar{q} (\pa^2_x \bar{q}  - i \pa_t \bar{q} ) + \bar{q}  (\bar{q}  \pa_x V^{(1)} + i \pa_x \pa_t \bar{q}  - \pa^3_x \bar{q} )] + i (\pa_x \bar{q} )^2\pa_x (V^{(1)} \bar{q} + i \pa_t \bar{q} - \pa^2_x \bar{q})\} \equiv 0
\label{pol3}
\er 
Therefore, equating to zero the coefficient of $\zeta^2$ in (\ref{pol1}) provides the identity
\br
\pa_x[ V^{(1)} \bar{q} + i \pa_t \bar{q} - \pa^2_x \bar{q}] = 0.
\er
Next, replacing this identity into the coefficients of $\zeta$ and $\zeta^{0}$ in the eqs. (\ref{pol2}) and (\ref{pol3}), respectively,  one can get 
\br
\label{dakns21}
a_i\, [-i \pa_t \bar{q} + \pa^2_x \bar{q} - V^{(1)} \bar{q}] =0,\,\,\,\,\,i=0,1.
\er
Since the auxiliary fields $a_i$ are non-vanishing arbitrary functions one gets the second eq. (\ref{mnls22}) of the AKNS system. The first eq. (\ref{mnls11}) we will derive below.
 
Next, let us consider the linear system \footnote{Notice that the connections from (\ref{sys11})-(\ref{sys21}) and (\ref{sys1})-(\ref{sys2}) can be related as $\widetilde{{\cal A}}_{\mu} = \widetilde{{\cal P}} ( {\cal A}_{\mu} ) \,\,(\mu = \{x, t\}),\,\,\,\widetilde{a}_i =  \widetilde{{\cal P}}(a_i)\,\, (i=0,1)$, where $\widetilde{{\cal P}} $ is the parity transformation defined  in (\ref{par1}) and (\ref{aknsparity}).}
\br
\label{sys11}
\pa_t \widetilde{\Phi} &=& \widetilde{A}_t \widetilde{\Phi}\\
\label{sys21}
\pa_x \widetilde{\Phi} &=& \widetilde{A}_x \widetilde{\Phi}\\
\label{cA1}
\widetilde{A}_x &\equiv &  \zeta \pa_x q + 2 q \,\(\frac{\widetilde{a}_0 \zeta +\zeta^2 \widetilde{a}_1 }{2 \zeta q- i \pa_x q}\),\\
\label{cA2}
\widetilde{A}_t &\equiv &  -\zeta \, \int^{x}\, dx' \, \frac{\widetilde{b}_0 + \widetilde{b}_1 \zeta + \widetilde{b}_2 \zeta^2}{(2 \zeta q - i \pa_{x'}q)^2},
\er
with
\br
\nonumber
\widetilde{b}_0 &\equiv& \frac{1}{2} i \pa_x q^2 (2 \pa_t \widetilde{a}_0 - \pa_x V^{(1)} \pa_x q ) - i(\pa_x q)^2 (V^{(1)} \pa_x q -\pa^3_x q)-2 \widetilde{a}_0 (q^2 \pa_x V^{(1)} + \pa_x q \pa^2_x q - q \pa^3_x q)\\
\widetilde{b}_1 &\equiv& -2 q^2 [2 \pa_t \widetilde{a}_0 + \pa_x V^{(1)} (\widetilde{a}_1 - 2 \pa_x q)] - 2 \widetilde{a}_1 \pa_x q \pa^2_x q + q [2i \pa_t \widetilde{a}_1 \pa_x q + 4 V^{(1)} (\pa_x q)^2+2 (\widetilde{a}_1 - 2 \pa_x q )\pa^3_x q],\nonumber\\
\widetilde{b}_2 &\equiv& -8 q^2 [\pa_t \widetilde{a}_1 - i (q \pa_x V^{(1)} + V^{(1)} \pa_x q - \pa^3_x q)].\nonu
\er  
	 
The compatibility condition of the system of eqs. (\ref{sys11})-(\ref{sys21}); i.e. $\pa_t \pa_x (\widetilde{\Phi})-\pa_x \pa_t (\widetilde{\Phi})=0$, furnishes the next polynomial in powers of $\zeta$ 
\br
\label{pol11}
\{ 8 i q^2\pa_x(V^{(1)} q + i \pa_t q - \pa^2_x q)\} \zeta^2 -\\ 
\label{pol21}
 \{2 \widetilde{a}_1 [\pa_x q(\pa^2_x q-i \pa_t q) + \bar{q} (q \pa_x V^{(1)} + i \pa_x \pa_t q - \pa^3_x q) ] +2 \pa_x q^2 \pa_x[V^{(1)} q+ i \pa_t q - \pa^2_xq   ]   \} \zeta-\\
\{2 \widetilde{a}_0 [\pa_x  q (\pa^2_x q - i \pa_t q ) + q  (q  \pa_x V^{(1)} + i \pa_x \pa_t q - \pa^3_x q )] + i (\pa_x q )^2\pa_x (V^{(1)} q + i \pa_t q - \pa^2_x q)\} \equiv 0
\label{pol31}
\er 
Therefore, equating to zero the coefficient of $\zeta^2$ in (\ref{pol11}) provides the identity
\br
\pa_x [V^{(1)}q - i \pa_t q- \pa^2_x q] = 0.
\er
Next, replacing this identity into the coefficients of $\zeta$ and $\zeta^{0}$ in the eqs. (\ref{pol21}) and (\ref{pol31}), respectively,  one can get 
\br
\label{dakns2}
\widetilde{a}_i\, [i \pa_t q + \pa^2_x q - V^{(1)} q]  =0,\,\,\,\,\,i=0,1.
\er
Since the auxiliary fields $\widetilde{a}_i$ are nonvanishing arbitrary functions one gets the first eq. (\ref{mnls11}) of the AKNS system.

Therefore, the linear formulation (\ref{sys11})-(\ref{sys21}) is related to the first deformed AKNS eq.  (\ref{mnls11}), whereas the linear formulation  (\ref{sys1})-(\ref{sys2})  is related to the second deformed AKNS eq. (\ref{mnls22}). These separate formulations can be joined together  into just one linear system defined as
\br
\label{ss11}
\pa_x\(\begin{array}c
\Phi \\
\widetilde{\Phi} \end{array}\) &=& {\cal M} \, \(\begin{array}c
\Phi \\
\widetilde{\Phi} \end{array}\),\,\,\,\,\,\, {\cal M} \equiv \(\begin{array}{cc}
A_x  & 0\\
0  & \widetilde{A}_x \end{array}\) \\
\pa_t\(\begin{array}c
\Phi \\
\widetilde{\Phi} \end{array}\) &=& {\cal N}\, \(\begin{array}c
\Phi \\
\widetilde{\Phi} \end{array}\) ,\,\,\,\,\,\,{\cal N} \equiv  \(\begin{array}{cc}
A_t  & 0\\
0  & \widetilde{A}_t \end{array}\)\,.\label{ss22}
\er
So,  the compatibility condition of this system provides the zero-curvature eq.
\br
\label{zc1}
\pa_{t} {\cal M}  - \pa_{x} {\cal N} + \Big[{\cal M} ,\,  {\cal N} \Big] = 0. 
\er
Notice that  ${\cal M}$ and ${\cal N}$ are diagonal matrices, and so, one has $[{\cal M} ,\,  {\cal N} ]=0$; therefore, the linear formulation (\ref{ss11})-(\ref{ss22}) splits into the relevant formulations in (\ref{sys1})-(\ref{sys2}) and  (\ref{sys11})-(\ref{sys21}), respectively. Then, the eqs. of motion of the deformed AKNS model (\ref{mnls11})-(\ref{mnls22}) can be obtained from (\ref{zc1}).  

For completeness we provide a gauge transformation between the system (\ref{lin01})-(\ref{lin02}) and the  system  (\ref{sys1})-(\ref{sys2}). So, the gauge transformation  (\ref{gauge1})-(\ref{gauge3}) can be written as
\br
\phi &=& e^{-\Lambda} \Phi\\
A_x &=& {\cal A}_x + \pa_x \Lambda\\
A_t &=& {\cal A}_t + \pa_t \Lambda,
\er
where $\Omega \equiv \pa_x\Lambda$ satisfies the Riccati eq. 
\br
\pa_x \Omega =2 \Omega^2 -  (2i \zeta + 2 A_x - \frac{\pa_x \bar{q}}{\bar{q}} )\,  \Omega + i \zeta A_x + \frac{1}{2} A^2_x + \frac{1}{2} \bar{q}q + \frac{1}{2} \pa_x A_x - \frac{1}{2} A_x \frac{\pa_x \bar{q}}{\bar{q}}.
\er 
A similar construction can be performed for the gauge transformation of the sector with the connection $\(\widetilde{A}_x,\, \widetilde{A}_t\)$.  In fact,  in order to perform  that gauge transformation for  the full system  (\ref{ss11})-(\ref{ss22}) we can have
\br
\(\begin{array}c
\phi \\
\widetilde{\phi} \end{array}\) &=&g \,\(\begin{array}c
\Phi \\
\widetilde{\Phi} \end{array}\),\,\,\,\,\,\,g \equiv  \(\begin{array}{cc}
  e^{-\Lambda} & 0\\
0  & e^{-\widetilde{\Lambda}} \end{array}\) \\
A_{\mu} &=&  {\cal A}_{\mu} +  \pa_{\mu} \Lambda\\
\widetilde{A}_{\mu} &=& \widetilde{{\cal A}}_{\mu} + \pa_{\mu} \widetilde{\Lambda},\,\,\,\,\mu = \{x, t\}.
\er
Since the matrices ${\cal M} $\, , $ {\cal N}$\, and $g$ are diagonal  the connection components  from the both sectors do not couple in the process.

Actually, all of the constructions above can be reproduced for the modified NLS model (\ref{nlsd}) since one can perform the reduction process of the modified AKNS model ($ MAKNS \rightarrow MNLS$)  through the identifications (\ref{nlspsi}). 

\subsection{Infinite set of non-local conserved charges}
\label{sec:nonlocal}

For each of the linear systems in (\ref{sys1})-(\ref{sys2}) and (\ref{sys11})-(\ref{sys21}) it is possible to construct a set of non-local conserved charges. The construction of analogous  linear systems and their associated non-local charges have recently been performed for some deformations of the sine-Gordon and KdV models \cite{npb1, jhep33}. In fact, following an iterative technique developed by  Br\'ezin et.al. \cite{brezin}, the authors in \cite{npb1, jhep33} have uncovered infinite set of non-local conservation laws for the relevant linear systems associated to the  deformations of the SG and KdV models, respectively. So, in order to make this paper self-contained we summarize the main points of the construction since the procedure is quite similar to the ones undertaken for the SG and KdV models.  So, let us define the currents 
\br
J_{\mu}^{(n)} &=& \frac{\pa}{\pa x_\mu} \chi^{(n)},\,\,\,x_\mu \equiv x, t;\,\,\,\,n=0,1,2,...\\
d \chi^{(1)} &=& A_{\mu} dx_{\mu}\\
&\equiv& A_x dx + A_t dt ,\\
J_{\mu}^{(n+1)} &=& \frac{\pa}{\pa x_\mu} \chi^{(n)}-A_{\mu} \chi^{(n)};\,\,\,\,\,\chi^{(0)}=1,
\er

Next, an inductive procedure is used to show that the  (non-local) currents $J_{\mu}^{(n)}$ are conserved
\br
\label{nlcl}
\pa_{t} J^{(n)}_{t} - \pa_{x} J^{(n)}_{x} =0,\,\,\,\,n=1,2,3,...
\er
In fact, the first non-trivial  current becomes $J_{\mu}^{(1)}=(A_x, A_t)$ whose conservation law $\pa_t A - \pa_x A_t=0$  reproduces  the eq.  (\ref{eqm1}). The second  order current becomes $J_{\mu}^{(2)}=(A_x - A_x\chi^{(1)},A_t-A_t \chi^{(1)})$, and from the conservation law (\ref{nlcl}), using the first order conservation law (\ref{eqm1}), one gets
\br
\label{j2}
\pa_t [A_x \chi^{(1)}] - \pa_x [A_t\chi^{(1)}]=0.
\er
The third order current becomes $J_{\mu}^{(3)}=(\frac{\pa}{\pa x}\chi^{(2)}-A_x\chi^{(2)}, \frac{\pa}{\pa t}\chi^{(2)}-A_t\chi^{(2)})$. So, at this order, the conservation law  (\ref{nlcl}), upon using the first (\ref{eqm1}) and second (\ref{j2}) order conservation laws, can be written as 
\br
\label{j3}
\pa_t [A_x \chi^{(2)}] - \pa_x [A_t \chi^{(2)}]=0.
\er  
where
\br
\pa_ x \chi^{(2)}= A_x - A_x\chi^{(1)},\,\,\,\,\pa_ t \chi^{(2)}= A_t-A_t\chi^{(1)}.
\er
Then, one can write the infinite tower of non-local conservation laws as 
\br 
\label{nonl1}
\pa_t [ A_x \chi^{(1)} ] - \pa_x [ A_t \chi^{(1)} ] &=& 0,\\
\label{nonl2}
\pa_t [ A_x \chi^{(n)} ] - \pa_x [ A_t\chi^{(n)} ] &=&0,\,\,\,\,\,n=2,3,4,...\\
\label{nonl3}
\pa_x \chi^{(n)} &=& A_x - A_x \chi^{(n-1)},\,\,\,\,\,\,\pa_t \chi^{(n)} = A_t - A_t\chi^{(n-1)}.
\er 

A similar procedure can be performed for the sector with the gauge connection $\(\widetilde{A}_x,\, \widetilde{A}_t\)$ in (\ref{cA1})-(\ref{cA2}), giving rise to another tower of infinite number  of non-local conservation laws
\br 
\label{nonl11}
\pa_t [ \widetilde{A}_x \widetilde{\chi}^{(1)} ] - \pa_x [ \widetilde{A}_t \widetilde{\chi}^{(1)} ] &=& 0,\\
\label{nonl21}
\pa_t [ \widetilde{A}_x \widetilde{\chi}^{(n)} ] - \pa_x [ \widetilde{A}_t\widetilde{\chi}^{(n)} ] &=&0,\,\,\,\,\,n=2,3,4,...\\
\label{nonl31}
\pa_x \widetilde{\chi}^{(n)} &=&\widetilde{A}_x - \widetilde{A}_x \widetilde{\chi}^{(n-1)},\,\,\,\,\,\,\pa_t \widetilde{\chi}^{(n)} = \widetilde{A}_t - \widetilde{A}_t\widetilde{\chi}^{(n-1)}.
\er 

Due to the reduction process $ MAKNS \rightarrow MNLS$,  through the identification (\ref{nlspsi}),  the towers of non-local charges constructed above can directly be reproduced for the modified NLS model (\ref{nlsd}).  Moreover, additional reductions of the standard AKNS system have been reported which define some integrable non-local NLS, SG  and KdV  models \cite{ablowitz, reverse}. So, in the context of the modified AKNS the relevant NLS-type, SG-type and  KdV-type equations will appear for a convenient choice of the parameters $A, B$ and $C$, as well as the auxiliary fields like $r$ and $s$, in the MAKNS system  (\ref{ricc1})-(\ref{ricc2}). The suitable choices have been done in \cite{npb1} for the modified SG-like and in \cite{jhep33} for the modified KdV-like systems, respectively. So, our calculations and results above can be reproduced for the following reductions of the MAKNS system (\ref{ricc1})-(\ref{ricc2})
\br
\label{mnls44}
\bar{q}(x,t) & \equiv &\sigma \, q^{\star}(x, t), \,\,\,\,\,\,\,\,\, \sigma = \pm 1,\,\,\,\,\,\star \equiv  \mbox{complex conjugation}\\
\label{tmnls}
\bar{q}(x,t) & \equiv &\sigma \, q(x, -t), \\
\label{xmnls}
\bar{q}(x,t) & \equiv &\sigma \, q^{\star}(-x, t), \\
\label{xtmnls}
\bar{q}(x,t) & \equiv &\sigma \, q(-x, -t),  \,\,\,\,q\in \IC\\
\label{xtcmnls}
\bar{q}(x,t) & \equiv &\sigma \, q^{\star}(-x, -t),  \\
\label{xtrmnls}
\bar{q}(x,t) & \equiv &\sigma \, q(-x, -t),\,\,\,\,q\in \IR.
\er
The first reduction (\ref{mnls44}) is just the reduction $ MAKNS \rightarrow MNLS$  (\ref{nlspsi}) we have discussed in this paper. 
We expect that the second, third and fourth reductions above will give rise to non-local MNLS-type equations and the
last two of them to nonlocal modified KdV-type evolution equations. The construction of the anomalous charges and the symmetries satisfied by the relevant anomalies associated to the  NLS-type (\ref{tmnls})-(\ref{xtmnls}) and (real or complex) KdV-type (\ref{xtcmnls})-(\ref{xtrmnls}) reductions deserve careful analysis and we will postpone those important issues for future research. 

So far, the relevant deformations of the SG and KdV models have been considered in the literature \cite{npb1, jhep33}, and  they share a similar structure regarding their nonlocal conservation laws with the one of the AKNS-type models in (\ref{nonl1})-(\ref{nonl3}), since their linear formulations possess the same form as in (\ref{sys1})-(\ref{sys2}) or (\ref{sys11})-(\ref{sys21}). Since the algebra of conserved charges in certain two-dimensional integrable quantum field theories is also present in the classical theory as a Poisson-Hopf algebra \cite{mackay1, mackay2}, it would be interesting to search for those type of classical Yangian algebras related to the set of non-local currents and charges for the deformations of the integrable models. The non-local conserved charges, as in the non-linear $\sigma-$model, would be relevant at the quantum level and they would imply absence of particle production (see e.g. \cite{abdalla, luscher}).

The AKNS-type models are quite ubiquitous in the nonlinear science and it would be interesting to investigate the relevance and physical consequences of the infinite towers of infinitely many anomalous and non-local charges discussed in this paper. Some remarkable and profound relationships  between integrable models and gauge theories have been uncovered in recent years. For example, it has been proposed a kind of triality among gauge theories, integrable models and gravity theories in some UV regime. In particular,  the $(1+1)D$ nonlinear Schr\"odinger equation corresponds to the $2D\, {\cal N} = (2, 2)^{\star} U(N )$super  Yang-Mills theory (see \cite{nian} and references therein).  We will postpone those important issues and some relevant applications for a future work.
    
\section{Some conclusions and discussions}
 
\label{ap:conclu}

Quasi-integrability properties of certain deformations of the NLS model have been examined. New anomalous charges related to  infinite towers of infinitely many  quasi-conservation laws were uncovered.  By direct construction in sec. \ref{secmnlsano}, we have obtained novel towers of quasi-conservation laws such that the anomaly densities exhibit odd parities under the  special space-time  symmetry (\ref{par1})-(\ref{par2}). We have shown that each monomial or polynomial with homogeneous degree and even parity turns out to be the density of an anomalous charge;  since in each case one can construct a quasi-conservation law with odd parity anomaly density. In addition, for each monomial or polynomial of that type one can construct an infinite number of higher order quasi-conservation laws with anomalous charges and relevant anomalies possessing  successively increasing degrees. Moreover, an anomaly cancellation mechanism has been introduced in order to construct the exact conservation laws, since a convenient linear combination of a set of relevant anomalies identically vanishes.  

In sec. \ref{sec:stNLS} it has been shown that even the standard NLS model possesses infinite towers of infinitely many  anomalous conservation laws. Subsequently, we showed analytically the vanishing of the space-time integrated anomalies and then the quasi-conservation of the infinite tower of anomalous charges for $N-$soliton solution satisfying the special parity symmetry ${\cal C}{\cal P}_s{\cal T}_d$ in (\ref{cpstd1}). So, this is the first example, of an analytical, and not only numerical, demonstration of the vanishing of the space-time integrated anomalies associated to the whole quasi-conservation laws of an integrable system. Some towers of anomalous charges have also been discussed in the standard sine-Gordon \cite{jhep1} and KdV \cite{jhep33}  models; however, in the present paper we have examined the complete set of anomalies associated to the standard NLS model, and so, it is expected that these kind of properties will appear in the other integrable systems and their quasi-integrable deformations.   
   
Through numerical simulations of $2-$bright soliton collisions, in sec. \ref{simul}, we have checked the  quasi-conservation properties of the lowest order charges of the modified NLS  model, appearing in the towers of quasi-conservation laws defined in (\ref{q1d}), (\ref{q2d}), (\ref{q3d}) and (\ref{q4kd}), respectively. So, we computed the space and space-time integrals of their associated anomaly densities $\hat{\alpha}_1$,  $\hat{\beta}_2$,  $\hat{\gamma}_2$ and  $\hat{\delta}_1$, defined in (\ref{an1d}), (\ref{beta2d}), (\ref{gam33d}) and (\ref{ano4}), respectively, for three types of two-soliton collisions of the particular deformation of  NLS model (\ref{mnls}).  In our numerical simulations presented in the Figs 1-15 we have observed that the space and space-time integrals of the set of anomalies $\hat{\alpha}_1$,  $\hat{\beta}_2$,  $\hat{\gamma}_2$ and  $\hat{\delta}_1$ vanish with good accuracy. In fact, the space and space-time integrals of the anomaly densities vanish within the order of errors $10^{-4}$ and $10^{-6}$, respectively, and sometimes, within the order of errors  $10^{-11}$ and $10^{-15}$, respectively,  depending on the type of $2-$bright soliton configuration. So, our numerical simulations allow us to conclude that for 2-bright soliton configurations the relevant charges are asymptotically conserved and their collisions are elastic within numerical accuracy. 
 
We have performed the Riccati-type pseudo-potential approach to deformations of the AKNS model in sec. \ref{sec:riccati}, such that the modified NLS is obtained through a certain reduction. In this framework it has been  constructed infinite towers of quasi-conservation laws and discussed their properties and relationships with the MNLS model. This construction reproduced the tower of NLS-type quasi-conserved charges obtained  in the anomalous zero-curvature approach of \cite{jhep3}.  Moreover, in sec. \ref{sec:dual1} we have introduced a dual Riccati-type pseudo-potential approach and uncovered, in that framework, a novel set of infinite number of quasi-conservation laws, such that it encompasses the quasi-conservation laws obtained in sec.  \ref{secmnlsano} by a constructive method starting from the eqs. of motion. 

In the framework of the Riccati-type pseudo-potential approach we have constructed a couple of  linear systems of equations, (\ref{sys1})-(\ref{sys2}) and  (\ref{sys11})-(\ref{sys21}), whose relevant compatibility conditions give rise to the modified AKNS system of equations  (\ref{mnls11})-(\ref{mnls22}). The second system of linear eqs. (\ref{sys11})-(\ref{sys21}) is related to the first one  (\ref{sys1})-(\ref{sys2})  through the transformation  (\ref{par1}) and (\ref{aknsparity}). In subsection \ref{sec:nonlocal} we have constructed two towers of  infinite sets of non-local conservation laws associated to the linear formulations, respectively. These linear systems and their associated non-local charges deserve more careful considerations; in particular, regarding their relationships of their associated non-local currents with  the so-called classical Yangians \cite{mackay1, mackay2}. 

In view of the current results, on deformations of SG, KdV and in this paper on deformations of NLS, one can inquire about the non-local properties of the quasi-integrable systems studied in the literature, such as the deformations of the Bullough-Dodd, Toda and SUSY sine-Gordon systems \cite{jhep6, toda, epl}, and more specific structures, such as the complete list of the towers of infinite number of anomalous charges and the (non-local) exact conservation laws, as discussed in this paper. Moreover, it would be interesting to consider the general AKNS model and study their (non-local) reductions (\ref{tmnls})-(\ref{xtrmnls}) as proposed  in  \cite{reverse, ablowitz}, as well as their relevant deformations in the lines discussed above.

\section{Acknowledgments}

HB thanks FC-UNI (Lima-Per\'u) and FC-UNASAM (Huaraz-Per\'u) for hospitality during the initial stage of the work. MC thanks the Peruvian agency Concytec for partial financial support. LFdS thanks CEFET Celso Sukow da Fonseca-Rio de Janeiro-Brazil for kind support.
The authors thank S.Y. Lou  for enlightening comments about the Ref. \cite{lou1}. The authors thank A. C. R. do Bonfim, H. F. Callisaya, C. A. Aguirre, J. P. R. Campos, R. Q. Bellido, J.M.J. Monsalve and A. Vilela for useful discussions.

\appendix

\section{The $u_n'$s of the first set of charges}
\label{fsca1}

The $u_n'$s can be determined recursively by substituting the expansion (\ref{expan}) into (\ref{ricc1}). Then the first quantities become

\br
u_1 &=& -\frac{1}{2} i q\\
u_2 &=& \frac{1}{2} i \pa_x u_1\\
u_3 &=& \frac{1}{2} i \( - \bar{q} u_1^2 + \pa_x u_2\)\\
u_4 &=& \frac{1}{2} i \( - 2 \bar{q} u_1 u_2 + \pa_x u_3\)\\
u_5 &=& \frac{1}{2} i \( -  \bar{q} u_2^2 - 2 \bar{q} u_1 u_3 + \pa_x u_4\)\\
u_6 &=& \frac{1}{2} i \( -  2 \bar{q} (u_3 u_2 + u_1 u_4) + \pa_x u_5\)\\
&&............................\nonumber
\er
The above sequence can be written for any $n$ (even or odd) as follows
\br
u_n & = & \frac{1}{2} i \Big[ -2 \bar{q} \sum_{ \begin{array}c i_1+i_2 = n-1\\
i_1 \neq i_2 \end{array}}  u_{i_1}  u_{i_2}  +\pa_x u_{n-1} \Big],\,\,\,\,\,\,\,\,\,\,\,\,\,\,\,\,\,\,\,\,\,\,\,\,\,\,\,\,\,\,\,\,n = even\\
u_n & = & \frac{1}{2} i \Big[ - \bar{q} \,  u_{\frac{n-1}{2}}^2 -2 \bar{q} \sum_{\begin{array}c i_1+i_2 = n-1\\
i_1 \neq i_2 \end{array} } \, u_{i_1}  u_{i_2}  +\pa_x u_{n-1} \Big],\,\,\,\,\,\,n = odd.
\er  
In terms of the field components the first six $u_i\,(i=1,2,...6)$ become
\br
u_1 &=& -\frac{1}{2} i q\\
u_2 &=& \frac{1}{4}  \pa_x q\\
u_3 &=&\frac{1}{8} i \( \bar{q} q q + \pa^2_x q \)\\
u_4 &=&-\frac{1}{16}  [ 4 \bar{q} q \pa_x q + q q \pa_x \bar{q}  + \pa^3_x q ]\\
u_5 &=&-\frac{1}{32} i [ 2 (\bar{q} q)^2 q + 5 \bar{q} (\pa_x q)^2 + 6 q \( \pa_x q \pa_x \bar{q} + \bar{q}  \pa^2_x q\) + q^2 \pa^2_x \bar{q} + \pa^4_x q ]\\
u_6 &=& \frac{1}{64} \Big[6 q^3  \bar{q} \pa_x \bar{q}  +11 \pa_x \bar{q} (\pa_x q)^2  + 18  \bar{q} \pa_x q \pa^2_x q + 4 q ( 3 \pa^2_x q  \pa_x \bar{q} + 2 \pa_x q \pa^2_x \bar{q} +2 \bar{q} \pa^3_x q ) + \nonumber \\
&& q^2 (16 \bar{q} ^2 \pa_x q +  \pa^3_x \bar{q} ) + \pa^5_x q  \Big].
\er

\section{The $\chi$ components}

\label{ap:chi}
The components of the expansion of $\chi$ in a recursive form become
\br
\chi_1 &=& - i \bar{q}  u_1 X\\
\chi_2 &=& -i  \bar{q}  u_2 X  + \frac{1}{2} i \pa_x \chi_1 -   \frac{1}{2} i \chi_1 \frac{\pa_x \bar{q} }{\bar{q}} \\
\chi_3 &=& -i  \bar{q}  u_3 X  -  i  \bar{q}  u_1 \chi_1  + \frac{1}{2} i \pa_x \chi_2 -   \frac{1}{2} i \chi_2 \frac{\pa_x \bar{q} }{\bar{q}} \\
\chi_4 &=& -i  \bar{q}  u_4 X    -i  \bar{q}  u_2 \chi_1  -i  \bar{q}  u_1 \chi_2 + \frac{1}{2} i \pa_x \chi_3 -   \frac{1}{2} i \chi_3 \frac{\pa_x \bar{q} }{\bar{q}} \\
\chi_5 &=& -i  \bar{q}  u_5 X    -i  \bar{q}  u_3 \chi_1   - i  \bar{q}  u_2 \chi_2 - i  \bar{q}  u_1 \chi_3 + \frac{1}{2} i \pa_x \chi_4 -   \frac{1}{2} i \chi_4 \frac{\pa_x \bar{q} }{\bar{q}} \\
&&.............\nonumber
\er
The above sequence can be written for any $n$ as
\br
\chi_n &=& - i\,  \bar{q}  u_n X    -i \bar{q}  \sum_{i_1+i_2 = n-1} u_{i_1} \chi_{i_2}  + \frac{1}{2} i \,\pa_x \chi_{n-1} -   \frac{1}{2} i \chi_{n-1} \frac{\pa_x \bar{q} }{\bar{q}};\,\,\,\,\,\,\,\,\,  \chi_0\equiv 0,\,\,\,\,\,\,\,n=1,2,....
\er
The first five components become
\br  \nonumber
\chi_1 &=& -\frac{1}{2}  \bar{q} q X.\\ \nonumber
\chi_2 &=& -\frac{1}{4} i [2 \bar{q}  \pa_x q X +  \bar{q}  q \pa_x X].\\
\label{chis}
\chi_3 &=& \frac{1}{8}  [3 \bar{q}  \pa_x q \pa_x X + 3 \bar{q}  X ( \bar{q}  q^2 + \pa^2_x q) +  \bar{q}  q \pa^2_x X].\\
\nonumber
\chi_4 &=&\frac{i}{16} [  \bar{q} q^2\( 5 \bar{q} \pa_x X + 4 X \pa_x  \bar{q} \) + 6   \bar{q}  \pa^2_x q \pa_x X + 4 \bar{q} \pa_x q \pa^2_x X + 4 \bar{q} X \pa^3_x q + \bar{q} q \( 16 X \bar{q} \pa_x q + \pa^3_x X \) ]. 
\\ 
\nonumber
\chi_5 &=& -\frac{1}{32} \bar{q} \Big\{ 10  q^3 \bar{q}^2 X + q^2 \( 9 \pa_x X \pa_x  \bar{q} + 7 \bar{q} \pa^2_x X + 5 X \pa_x^2 \bar{q} \) + \\
\nonumber
&& 5  [ 2 \pa^2_x q \pa^2_x X + 2 \pa^3_x q \pa_x X +  \pa_x^3 X \pa_x q + X (5 \bar{q} (\pa_x q)^2 + \pa^4_x q)]+ \\
\nonumber
&& q [ 2 \pa_x q (17 \bar{q} \pa_x X + 15 X \pa_x  \bar{q} ) + 30 X  \bar{q} \pa^2_x q + \pa^4_x X ] \Big\}. 
\er

\section{The first $\bar{u}_n$ and $\bar{\chi}_n$}
\label{app:uchid}

Next we provide the lowest order components of the dual potentials $\bar{u}$ and $\bar{\chi}$
\br
\bar{u}_1 &=& \frac{1}{2} i \bar{q},\\
\bar{u}_2 &=& \frac{1}{4}  \pa_x \bar{q},\\
\bar{u}_3 &=&-\frac{1}{8} i \( q \bar{q}\bar{q} + \pa^2_x \bar{q} \),\\
\bar{u}_4 &=&-\frac{1}{16}  [ 4 q \bar{q} \pa_x \bar{q} + \bar{q} \bar{q} \pa_x q  + \pa^3_x \bar{q} ],\\
\bar{u}_5 &=&\frac{1}{32} i [ 2 (\bar{q}q )^2 \bar{q} + 5 q (\pa_x \bar{q})^2 + 6 \bar{q} \( \pa_x \bar{q} \pa_x q + q  \pa^2_x \bar{q}\) + \bar{q}^2 \pa^2_x q + \pa^4_x \bar{q} ] ,\\
\bar{u}_6 &=& \frac{1}{64} \Big[6 \bar{q}^3  q \pa_x q  +11 \pa_x q (\pa_x \bar{q})^2  + 18  q \pa_x \bar{q} \pa^2_x \bar{q} + 4 \bar{q} ( 3 \pa^2_x \bar{q}  \pa_x q + 2 \pa_x \bar{q} \pa^2_x q +2 q \pa^3_x \bar{q}) + \nonumber \\
&& \bar{q}^2 (16 q^2 \pa_x \bar{q} +  \pa^3_x q) + \pa^5_x \bar{q}  \Big]
\er
and
\br  \nonumber
\bar{\chi}_1 &=& -\frac{1}{2}  \bar{q} q X,\\ \nonumber
\bar{\chi}_2 &=& \frac{1}{4} i [2 q \pa_x \bar{q} X +  \bar{q}  q \pa_x X],\\
\label{chisd}
\bar{\chi}_3 &=& \frac{1}{8}  [3 q  \pa_x \bar{q} \pa_x X + 3 q  X ( q  \bar{q}^2 + \pa^2_x \bar{q}) +  \bar{q}  q \pa^2_x X],\\
\nonumber
\bar{\chi}_4 &=&-\frac{i}{16} [  q \bar{q}^2\( 5 q \pa_x X + 4 X \pa_x  q \) + 6  q  \pa^2_x \bar{q} \pa_x X + 4 q \pa_x \bar{q} \pa^2_x X + 4 q X \pa^3_x \bar{q} + \bar{q} q \( 16 X q \pa_x \bar{q} + \pa^3_x X \) ],\\
\\ 
\nonumber
\bar{\chi}_5 &=& -\frac{1}{32} q \Big\{ 10  \bar{q}^3 q^2 X + \bar{q}^2 \( 9 \pa_x X \pa_x  q + 7 q \pa^2_x X + 5 X \pa_x^2 q \) + \\
\nonumber
&& 5  [ 2 \pa^2_x \bar{q} \pa^2_x X + 2 \pa^3_x \bar{q} \pa_x X +  \pa_x^3 X \pa_x \bar{q} + X (5 q (\pa_x \bar{q})^2 + \pa^4_x \bar{q})]+ \\
\nonumber
&& \bar{q} [ 2 \pa_x \bar{q} (17 q \pa_x X + 15 X \pa_x  q ) + 30 X  q \pa^2_x \bar{q} + \pa^4_x X ] \Big\}. 
\er

\section{3-soliton parameters}
\label{ap:3sol}
The next equations have been used in order to construct the ${\cal C}{\cal P}_s{\cal T}_d$ symmetric $3-$soliton. The functions $f_3$ and $g_3$ become 
\br
f_3&=& e^{\xi_1}+e^{\xi_2} +e^{\xi_3}+e^{\xi_1+\xi_2+ \xi_4 +\theta_{12}+\theta_{14}+\theta_{24}} + e^{\xi_1+\xi_2+ \xi_5 +\theta_{12}+\theta_{15}+\theta_{25}}+e^{\xi_1+\xi_2+ \xi_6 +\theta_{12}+\theta_{16}+\theta_{26}} +\nonumber\\
&& e^{\xi_1+\xi_3+ \xi_4 +\theta_{13}+\theta_{14}+\theta_{34}} + e^{\xi_1+\xi_3+ \xi_5 +\theta_{13}+\theta_{15}+\theta_{35}}+e^{\xi_1+\xi_3+ \xi_6 +\theta_{13}+\theta_{16}+\theta_{36}}+\nonumber\\
&&e^{\xi_2+\xi_3+ \xi_4 +\theta_{23}+\theta_{24}+\theta_{34}} + e^{\xi_2+\xi_3+ \xi_5 +\theta_{23}+\theta_{25}+\theta_{35}}+e^{\xi_2+\xi_3+ \xi_6 +\theta_{23}+\theta_{26}+\theta_{36}}+\nonumber\\
&&e^{\sum_{j=1, j \neq 6}^6 \xi_j + \sum_{j<l (j,l \neq 6)}^6 \theta_{jl}}+e^{\sum_{j=1, j\neq 5}^6 \xi_j + \sum_{j<l (j,l \neq 5)}^6 \theta_{jl}}+e^{\sum_{j=1, j\neq 4}^6 \xi_j + \sum_{j<l (j,l \neq 4)}^6 \theta_{jl}}.\er
and
\br 
g_3 &=&1+ e^{\xi_1 + \xi_4 + \theta_{14}} +e^{\xi_1 + \xi_5 + \theta_{15}} + e^{\xi_1 + \xi_6 + \theta_{16}} +  e^{\xi_2 + \xi_4 + \theta_{24}} +e^{\xi_2 + \xi_5 + \theta_{25}} + e^{\xi_2 + \xi_6 + \theta_{26}} + \nonumber\\
&& e^{\xi_3 + \xi_4 + \theta_{34}} +e^{\xi_3 + \xi_5 + \theta_{35}} + e^{\xi_3 + \xi_6 + \theta_{36}} +\nonumber\\
&&e^{\xi_1+\xi_2 + \theta_{12}} [e^{\xi_4+\xi_5+\theta_{14}+ \theta_{15}+\theta_{24}+\theta_{25}+ \theta_{45}} + 
e^{\xi_4+\xi_6+\theta_{14}+\theta_{16}+ \theta_{24}+\theta_{26}+\theta_{46}}+\\
&&
e^{\xi_5+\xi_6+\theta_{15}+\theta_{16}+ \theta_{25}+\theta_{26}+\theta_{56}}]+\nonumber\\
&&e^{\xi_1+\xi_3 + \theta_{13}} [e^{\xi_4+\xi_5+\theta_{14}+ \theta_{15}+\theta_{34}+\theta_{35}+ \theta_{45}} + 
e^{\xi_4+\xi_6+\theta_{14}+\theta_{16}+ \theta_{34}+\theta_{36}+\theta_{46}}+\\
&& e^{\xi_5+\xi_6+\theta_{15}+\theta_{16}+ \theta_{35}+\theta_{36}+\theta_{56}}]+
\\
&&e^{\xi_2+\xi_3 + \theta_{23}} [e^{\xi_4+\xi_5+\theta_{24}+ \theta_{25}+\theta_{34}+\theta_{35}+ \theta_{45}} + 
e^{\xi_4+\xi_6+\theta_{24}+\theta_{26}+ \theta_{34}+\theta_{36}+\theta_{46}}+\\
&&e^{\xi_5+\xi_6+\theta_{25}+\theta_{26}+ \theta_{35}+\theta_{36} +\theta_{56}}]+\\
&& e^{\sum_{j}^6 \xi_j + \sum_{j<l (j=1,l=2)}^6 \theta_{jl}}.
\er
with
\br
\xi_1 &=& k_1  \widetilde{x}  + i k_1^2  \widetilde{t}  + \eta_{01} - \frac{\theta_{12}+\theta_{13}+\theta_{14}+ \theta_{15}+ \theta_{16}}{2} \equiv \eta_1 -   \frac{\theta_{12}+\theta_{13}+\theta_{14}+ \theta_{15}+ \theta_{16}}{2} ,\,\,\, \, \\
\xi_2 &=& k_2 \widetilde{x} + i k_2^2 \widetilde{t} + \eta_{02} - \frac{\theta_{12}+\theta_{23}+\theta_{24}+ \theta_{25}+ \theta_{26}}{2} \equiv \eta_2 - \frac{\theta_{12}+\theta_{23}+\theta_{24}+ \theta_{25}+ \theta_{26}}{2},\\
\xi_3 &=& k_3  \widetilde{x}  + i k_3^2  \widetilde{t}  + \eta_{03} - \frac{\theta_{13}+\theta_{23}+\theta_{34}+\theta_{35}+\theta_{36}}{2} \equiv \eta_3 - \frac{\theta_{13}+\theta_{23}+\theta_{34}+\theta_{35}+\theta_{36}}{2} ,\\
\xi_4 &=& \bar{\xi}_{1},\,\,\,\,\,\xi_5 = \bar{\xi}_{2},\,\,\,\,\,
\xi_6 = \bar{\xi}_{3},\,\,\,\,\,\,\,\widetilde{x} \equiv  x-x_{\Delta},\, \widetilde{t} \equiv t-t_{\Delta}.
\er
For $G_3$ one has 
\br
G_3 &\equiv& \cosh{[\frac{\sum_{j} \eta_j}{2}]} + e^{-\frac{\Theta_{14}}{2}}  \cosh{[\frac{ (\eta_1+\eta_4)-(\eta_2+\eta_5)-(\eta_3+\eta_6)}{2}]}+ \label{g31}\\
\nonumber
&&e^{-\frac{\Theta_{15}}{2}}  \cosh{[\frac{ (\eta_1-\eta_4)-(\eta_2-\eta_5)-(\eta_3+\eta_6)}{2}]}+e^{-\frac{\Theta_{16}}{2}}  \cosh{[\frac{ (\eta_1-\eta_4)-(\eta_2+\eta_5)-(\eta_3-\eta_6)}{2}]}+\nonumber\\
&&e^{-\frac{\Theta_{24}}{2}}  \cosh{[\frac{ (\eta_1-\eta_4)-(\eta_2-\eta_5)+(\eta_3+\eta_6)}{2}]}+e^{-\frac{\Theta_{25}}{2}}  \cosh{[\frac{ (\eta_1+\eta_4)-(\eta_2+\eta_5)+(\eta_3+\eta_6)}{2}]}+\nonumber\\
&&e^{-\frac{\Theta_{26}}{2}}  \cosh{[\frac{(\eta_1+\eta_4)-(\eta_2-\eta_5)+(\eta_3-\eta_6)}{2}]}+e^{-\frac{\Theta_{34}}{2}}  \cosh{[\frac{ (\eta_1-\eta_4)+(\eta_2+\eta_5)-(\eta_3-\eta_6)}{2}]}+\nonumber\\
&&e^{-\frac{\Theta_{35}}{2}}  \cosh{[\frac{ (\eta_1+\eta_4)+(\eta_2-\eta_5)-(\eta_3-\eta_6)}{2}]}+e^{-\frac{\Theta_{36}}{2}}  \cosh{[\frac{ (\eta_1+\eta_4)+(\eta_2+\eta_5)-(\eta_3+\eta_6)}{2}]}  ,\nonumber
\er
where the parameters $\Theta_{ij}$ are provided below and satisfy the properties
\br
\Theta_{14},\, \Theta_{25},\, \Theta_{36} \in \IR;\,\,\bar{\Theta}_{15}=\Theta_{24},\,\bar{\Theta}_{16}=\Theta_{34},\bar{\Theta}_{26}=\Theta_{35}.
\er
Similarly, for $F_3$ one has
\br  
\nonumber
F_3&\equiv& \g_{14} e^{\frac{-(\eta_1-\eta_4) + (\eta_2-\eta_{5}) + (\eta_3-\eta_6)}{2}} +\g_{25} e^{\frac{(\eta_1-\eta_4) - (\eta_2-\eta_{5}) + (\eta_3-\eta_6)}{2}}+\g_{36} e^{\frac{(\eta_1-\eta_4) + (\eta_2-\eta_{5}) - (\eta_3-\eta_6)}{2}} + \\
&&R_{14} e^{\frac{\eta_1-\eta_4-\theta_{14}}{2}}\cosh{[\frac{(\eta_2+ \eta_5)+(\eta_3+\eta_6)-i \a_{14}}{2}]} + \nonumber\\
&& R_{25} e^{\frac{\eta_2-\eta_5-\theta_{25}}{2}}\cosh{[\frac{(\eta_1+ \eta_4)+(\eta_3+\eta_6)-i \a_{25}}{2}]} +\nonumber\\
&&R_{36} e^{\frac{\eta_3-\eta_6-\theta_{36}}{2}}\cosh{[\frac{(\eta_1+ \eta_4)+(\eta_2+\eta_5)-i \a_{36}}{2}]}+\label{f31}\\
&& \nonumber A_1 \,r_{14} e^{\frac{\eta_1-\eta_4}{2}}\cosh{[\frac{(\eta_2+ \eta_5)-(\eta_3+\eta_6)+i \b_{14}}{2}]}+\\
&& A_2 \,r_{25} e^{\frac{\eta_2-\eta_5}{2}}\cosh{[\frac{(\eta_1+ \eta_4)-(\eta_3+\eta_6)+i \b_{25}}{2}]}+\nonumber\\
&& A_3\, r_{36} e^{\frac{\eta_3-\eta_6}{2}}\cosh{[\frac{(\eta_1+ \eta_4)-(\eta_2+\eta_5) + i \b_{36}}{2}]}.\nonumber
\er
 
The above parameters $\g_{j(3+j)}, R_{j(3+j)}, r_{j(3+j)}, A_j,  \a_{j(3+j)}, \b_{j(3+j)}\, (j=1,2,3)$ are real and they are provided below. Some of the parameters $\theta_{jk}$ satisfy
\br
\theta_{14}, \theta_{25}, \theta_{36} \in \IR,\,\,\,\theta_{45} = \bar{\theta}_{12},\,\theta_{46} = \bar{\theta}_{13},\,\theta_{56} = \bar{\theta}_{23}.
\er
The parameters $\Theta_{jk}$ are defined as
\br
\nonumber
\Theta_{14} &=& \theta_{12}+\theta_{13}+\theta_{15}+\theta_{16}+\theta_{24}+\theta_{34}+\theta_{45}+\theta_{46},\,\,
\Theta_{15} =\theta_{12}+\theta_{13}+\theta_{14}+\theta_{16}+\theta_{25}+\theta_{35}+\theta_{45}+\theta_{56},\\
\nonumber \Theta_{16} &=&\theta_{12}+\theta_{13}+\theta_{14}+\theta_{15}+\theta_{26}+\theta_{36}+\theta_{46}+\theta_{56},\,\,
\bar{\Theta}_{24} =\theta_{12}+\theta_{14}+\theta_{23}+\theta_{25}+\theta_{26}+\theta_{34}+\theta_{45}+\theta_{46},\\ 
\Theta_{25} &=&\theta_{12}+\theta_{15}+\theta_{23}+\theta_{24}+\theta_{26}+\theta_{35}+\theta_{45}+\theta_{56},\,\,
\nonumber\bar{\Theta}_{26} =\theta_{12}+\theta_{16}+\theta_{23}+\theta_{24}+\theta_{25}+\theta_{36}+\theta_{46}+\theta_{56},\\
\bar{\Theta}_{34} &=&\theta_{13}+\theta_{14}+\theta_{23}+\theta_{24}+\theta_{35}+\theta_{36}+\theta_{45}+\theta_{46},\,\,
\nonumber\Theta_{35} =\theta_{13}+\theta_{15}+\theta_{23}+\theta_{25}+\theta_{34}+\theta_{36}+\theta_{45}+\theta_{56}\\
\Theta_{36} &=&\theta_{13}+\theta_{16}+\theta_{23}+\theta_{26}+\theta_{34}+\theta_{35}+\theta_{46}+\theta_{56}.\label{Thetas}
\er
Finally, one has
\br
\g_{14} &\equiv& e^{-\frac{\theta_{12}+\theta_{13}+\theta_{14}+\theta_{25}+\theta_{26}+\theta_{35}+ \theta_{36}+\theta_{45}+\theta_{46}}{2}},\\
&=&\frac{1}{ (k_{1R} k_{2R} k_{3R})^2}  \Big|\frac{(k_1-k_2)(k_1-k_3)}{4 (k_2+k_3)} \Big|^4.\\
\g_{25} &\equiv& e^{-\frac{\theta_{12}+\theta_{14}+\theta_{16}+\theta_{23}+\theta_{25}+\theta_{34}+ \theta_{36}+\theta_{45}+\theta_{56}}{2}},\\
&=&\frac{1}{(k_{1R} k_{2R} k_{3R})^2}  \Big|\frac{(k_1-k_2)(k_2-k_3)}{4 (k_1+k_3)} \Big|^4.\\
\g_{36} &\equiv& e^{-\frac{\theta_{13}+\theta_{14}+\theta_{15}+\theta_{23}+\theta_{24}+\theta_{25}+ \theta_{36}+\theta_{46}+\theta_{56}}{2}},\\
&=&\frac{1}{(k_{1R} k_{2R} k_{3R})^2} \Big|\frac{(k_1-k_3)(k_2-k_3)}{4 (k_1+k_2)}\Big|^4.\\
R_{14} e^{-i \frac{\a_{14}}{2}} &\equiv& e^{-\frac{\theta_{24}+\theta_{34}+\theta_{45}+\theta_{46}}{2}},\\
R_{25} e^{-i \frac{\a_{25}}{2}} &\equiv& e^{-\frac{\theta_{15}+\theta_{35}+\theta_{45}+\theta_{56}}{2}}, \\
R_{36} e^{-i \frac{\a_{36}}{2}} &\equiv& e^{-\frac{\theta_{16}+\theta_{26}+\theta_{46}+\theta_{56}}{2}},\\
r_{14} e^{-i \frac{\b_{14}}{2}} &\equiv& e^{-\frac{\theta_{12}+\theta_{15}+\theta_{34}+\theta_{46}}{2}}, \\
r_{25} e^{-i \frac{\b_{25}}{2}} &\equiv& e^{-\frac{\theta_{12}+\theta_{24}+\theta_{35}+\theta_{56}}{2}}, \\
r_{36} e^{-i \frac{\b_{36}}{2}} &\equiv& e^{-\frac{\theta_{13}+\theta_{26}+\theta_{34}+\theta_{56}}{2}},\\
A_1 &=& e^{\frac{\theta_{23} +\theta_{56}+\theta_{14}+\theta_{35}+\theta_{26}}{2}},\\
A_2 &=& e^{\frac{\theta_{13} +\theta_{46}+\theta_{25}+\theta_{34}+\theta_{16}}{2}},\\
A_3 &=& e^{\frac{\theta_{12} +\theta_{45}+\theta_{24}+\theta_{15}+\theta_{36}}{2}}.
\er

\end{document}